\documentclass{aa}

\usepackage{graphicx}

\usepackage{lipsum}
\usepackage{widetext}

\usepackage[varg]{txfonts}

\usepackage{natbib}
\bibpunct{(}{)}{;}{a}{}{,} 

\begin{document}



 \title{On the propagation of Alfv\'en waves in the dusty interstellar medium}
 \subtitle{Can we neglect dust inertia in molecular clouds?}

  \author{  Patrick Hennebelle \inst{\ref{inst1}} 
   and Ugo Lebreuilly \inst{\ref{inst1}} 
    }

%
%

   \institute{ AIM, CEA, CNRS, Université Paris-Saclay, Université Paris Diderot, Sorbonne Paris Cité, F-91191 Gif-sur-Yvette, France,
      \label{inst1} 
      }

\date{}

\abstract
{Alfv\'en waves are fundamental magnetized modes which play an important role in the dynamics of magnetized flows such as the interstellar medium (ISM).
}
{In weakly ionised medium, their propagation critically depends on the ionisation rate but also on the charge carriers which, depending 
on gas density can be ions, electrons or dust grains. The latter in particular are well known to have a drastic influence on the magnetic resistivities
in the dense ISM such as collapsing dense cores. Yet, in most calculations, for numerical reasons, the grain inertia is usually neglected.}
{We investigate analytically the propagation of Alfv\'en waves both in a single-size  and multi-size grain medium such as the ISM and
we obtain exact expressions giving wavenumbers as a function of  wave frequencies. These 
expressions are then solved analytically or numerically taking into account or neglecting grain inertia.}
{Whereas at large wavelengths, neglecting  grain inertia is a very good approximation, the situation 
is rather different  for wavelengths shorter than a critical value, which broadly scales as  $1/n$, $n$ being the gas density. More precisely, when inertia is neglected the waves do not propagate at short wavelengths or, due to 
the Hall effect, develop for one circular polarisation only, a whistler mode such that ${\cal R} _e (\omega) \propto k^2$, whereas the other polarisation presents a zero group velocity, i.e. ${\cal R} _e (\omega) \propto k^0$.
 When grain inertia is accounted for, the propagation of  the two polarisations tend to be more symmetrical and the whistler mode is 
 only present at density higher than $\simeq 10^8$ cm$^{-3}$. At lower density it is replaced by a mode having ${\cal R} _e (\omega) \propto k^{\simeq 1.2}$. Interestingly, one of the polarisation presents a
 distribution, instead of a single $\omega$ value. Importantly, for short wavelengths, wave damping is considerably reduced when inertia  is properly accounted for. }
{To properly handle the propagation of Alfv\'en waves at short wavelengths, it is necessary to treat self-consistently grain inertia.
We discuss the possible consequences this may have in the context of diffuse and dense molecular gas regarding turbulence, magnetic braking and protoplanetary disk formation as well as cosmic rays propagation in the dense ISM. }

\keywords{ physical processes: wave --- physical processes: magnetohydrodynamics (MHD) --- physical processes: turbulence --- stars: formation --- ISM: dust,extinction  }

\maketitle


\section{Introduction}

It is now well recognized that magnetic field is playing a fundamental role in many astrophysical contexts from galaxies to stars and planets. 
 In a weakly ionised medium such as the interstellar medium (ISM), 
its impact on the gas depends on its coupling with the neutrals. This coupling is related  not only to the ionisation degree but also to the charge carriers namely ions and electrons as well as 
on the dust grains.  One difficulty which arises with these latter is that unlike ions and electrons, they present a distribution of sizes that span
several order of magnitude. 

Whereas few studies have addressed this problem in fully ionised plasma \citep{tripathi1996,cramer2002}, 
so far most of the works in the context of the dense ISM have not accounted for the grain dynamics but have taken into account 
their impact on the magnetic resistivities. This includes the works of \citet{nishi1991}, \citet{nakano2002}, \citet{kunz2009}, \citet{zhao2016}, \citet{marchand2016}, \citet{wurster2016},
\citet{guillet2020}, \citet{marchand2021}, \citet{tsukamoto2022}. In all these works, it has been found that the magnetic resistivities in the dense ISM are controlled by 
the grain distributions, particularly by the small ones, i.e. of size equal to a few nm. Ambipolar diffusion and Hall effect, respectively due to the decoupling 
of neutrals and ions with magnetic field, have been consistently found to play a significant role  and are dominant over Ohmic dissipation up to densities 
of about  $\simeq 10^{11}$ cm$^{-3}$. 

Apart from the context of protoplanetary disks where grains are playing a fundamental role regarding planet formation and have intensively 
been studied since several decades
\citep[e.g.][]{drazkowska2022}, 
 the dynamics of the grains has been recognized to be significant  in context such as ISM shocks \citep{ciolek2002,guillet2007}, collapsing dense cores
\citep{lebreuilly2019} or diffuse ISM \citep{hopkins2016}. These works constitute, particularly when  the magnetic field is taken into account, 
important attempts towards fully consistent ISM studies. 

Indeed, grains follow their own dynamics and since they carry a fraction of the electric charges, they influence the propagation of magnetic 
waves and more generally the coupling between the gas and the magnetic field. Therefore a fully consistent approach should  eventually 
be considered. In the present paper, we aim at progressing along this line, by addressing self-consistently the very fundamental magnetic waves, namely the Alfv\'en waves. 
Indeed Alfv\'en waves  are certainly the most fundamental novelty that magnetic field introduces compared to an unmagnetized plasma. This is because
Alfv\'en waves are transverse, incompressible modes that carries momentum and even angular momentum in rotating and magnetized flows. 

For instance, Alfv\'en waves are likely responsible of the magnetic braking \citep{mouschovias1979,joos2012,hirano2020} which appears to play a fundamental role in the formation of 
protoplanetary disks \citep{allen2003,hennebelle2016,hetal2020,wurster2018,zhao2020,maury2022}.
A detailed analysis of Alfv\'en waves propagating in the dense ISM, that is to say accounting for the proper resistivities \citep[e.g.][]{nakano2002},  has been conducted by \citet{wardle1999}.
In particular, they stress the role of the Hall effect that deeply modifies the propagation of Alfv\'en waves and introduce a strong difference
between the two circularly polarised modes. Whereas one of them present, at short wavelengths, $\lambda$,  a dispersion relation such that ${\cal R}_e(\omega) \propto \lambda^{-2}$
($\omega$ being the frequency  of the wave),
which is typical of  whistler modes ; the other circular polarisation is such that  ${\cal R}_e(\omega) \propto \lambda^{0}$. This latter mode therefore has a zero group velocity 
(ZGV) and therefore is not transporting momentum and energy. This strong effect has turned out to have consequences regarding the formation of protoplanetary disks since 
aligned and anti-aligned rotators are not subject to the same magnetic braking \citep[e.g.][]{braiding2012,tsukamoto2015,wurster2016}. These analysis so far 
have neglected grain inertia and the question arises to what extend this may modify this effect and more generally what is the impact of this approximation on the wave propagation?

Interestingly,  \citet{kulsrud1969} \citep[see also][]{roberge2007,soler2013} have shown that in a partially ionised medium, when the inertia of ions is accounted for, Alfv\'en waves whose propagation is impossible at 
intermediate wavelengths, can propagate at short wavelengths. This is because the ion-neutral friction is negligible in the long and short wavelength limit but may become 
dominant at intermediate wavelengths. Since the dense ISM is very weakly ionised, with ion abundances below $10^{-7}$, the ion fluid mass is rather low compared to the neutral one and 
we may expect the regime for which ions dynamics play a role, to occur at very short wavelengths. Since they carry electric 
charges, similar processes are expected for dust grains.
However, the total mass of dust is about hundred times below the mass of the neutrals. Whereas this still is a small fraction, it is nevertheless much larger than 
the total mass of the ions. In particular, it is worth remembering that since the Alfv\'en velocity is proportional to the square-root of the density, it is typically {\it only} ten times larger for waves
propagating in the dust fluid than in the neutral one. 
Moreover due to dynamical effects, dust enrichment by a factor of a few, has been obtained in collapse calculations \citep{lebreuilly2020}. 
Therefore it seems important to investigate the role that dust dynamics may have on Alfv\'en wave propagation.

The paper is organised as follows. In the second part the equations and the physics of grains in the ISM are discussed and presented.
The dispersion relation for Alfv\'en waves propagation in a multi-size grain medium is obtained. The third part is dedicated to the simpler 
but enlightening  case of a single-size grain medium. The physics of the various waves is discussed as well as the possible consequences in term
of momentum transport by the waves. In the fourth part, a multi-size grain medium is considered. Several densities, ionisation and grain 
distributions are explored and discussed. In the fifth part, we discuss the possible consequences of our results regarding the development 
of magnetized turbulence and the magnetic braking occurring in dense cores and controlling the formation of protoplanetary disks. The sixth part
concludes the paper.

\section{Equations and grain properties}

\subsection{The fluid equations}
 The equations describing the dynamics of a magnetized fluid are well known.
The situation is  rendered significantly more complicated when a continuum of  dust grains is considered.
To simplify the notations and take advantage of the similarities between the various equations describing 
charged particles, we note $X_{i,e,a}$ any quantity like density or velocity associated to respectively ions, electrons
or charged grains of size $a$. By convention, we will note 
\begin{eqnarray}
 \int da {d \rho_{i,e,a} \over d a}  X_{i,e,a} = \int da {d \rho_{ia,} \over d a}  X_{a} + \rho_i X_i + \rho_e X_e
\end{eqnarray} 



The momentum equation for neutrals is
\begin{eqnarray}
\rho \left(  \partial _t {\bf V} + {\bf V} \cdot \nabla {\bf V} \right) = -C_s^2 \nabla \rho  
+ 
\rho \int da {d \rho_{i,e,a,} \over d a}   K_{i,e,a} ({\bf V_{i,e,a}} - {\bf V} ),
\label{V_mg}
\end{eqnarray}
where $\rho$,  $\rho_{i,e,a} $ are the neutral and charged particle density, ${\bf V} $, 
 ${\bf V_{i,e,a}} $, the velocities, $C_s$ is the sound speed and $K_{i,e,a} $
are the friction coefficients of the charged particles.

The momentum equation for charged particles is
\begin{eqnarray}
\nonumber
  \partial _t {\bf V_{i,e,a}} + {\bf V_{i,e,a}}  \cdot \nabla  {\bf V_{i,e,a}}  = {Z_{i,e,a}  e \over m_{i,e,a} }  ( {\bf E} + { {\bf  V_{i,e,a}} \over c } \times {\bf B} ) \\
 + \rho  K_{i,e,a} ({\bf V} - {\bf V_{i,e,a}} ), 
\label{Vd_mg}
\end{eqnarray}
where ${\bf E}$ and  ${\bf B}$ are the electric and magnetic fields respectively. $Z_{i,e,a} $, $ m_{i,e,a} $ are the  charge and mass
of the charged particles and $a$
represents the grain size. In practice  since there is a continuity of grain size, $\rho_{a}$ and ${\bf V}_{a}$ are distributions.
 Note that we do not consider a population of neutral grains because the time scale for charge fluctuation is shorter than the 
grain Larmor time scale $1/ \omega$ \citep[see for instance Fig. 5 of][]{guillet2020}, meaning that all grains possess an effective mean charge.

The electric current is given by
\begin{eqnarray}
 {\bf j} =  \int da {d \rho _{i,e,a} \over da} { Z_{i,e,a} e \over m_{i,e,a} }{\bf V_{i,e,a}} , 
\label{j_mg}
\end{eqnarray}
and the Maxwell-Faraday equation
\begin{eqnarray}
\partial _t {\bf B} + c \nabla \times  {\bf E} = 0.
\label{E}
\end{eqnarray}

The Maxwell-Amp\`ere equation is
\begin{eqnarray}
{\bf \nabla \times B} = { 4 \pi \over c} {\bf j}.
\label{j}
\end{eqnarray}

We have not considered the continuity equations at this stage since Alfv\'en waves are incompressible.

\subsection{Grain and ionisation}

\subsubsection{Grain properties}
The grain characteristics vary a lot with the environment.
 For instance, in the diffuse interstellar medium, the size typically goes from a few nanometers to a few hundreds of nanometers 
and roughly follows the so-called MRN distribution \citep{mathis1977} such like $d n / da \propto a^{\lambda_{mrn}}$, where $a$ is the grain
 size and $\lambda_{mrn} \simeq -3.5$. The reality is however likely more complex \citep{jones2017} particularly in
  the dense interstellar medium which is investigated here.  In dense environment, the grain size distribution is less known
 but it appears that the grains
coagulate and  grow in mass and size, possibly becoming as big as hundreds of micrometers or possibly even larger \citep{valeska2019}.
How this coagulation exactly proceeds is still a matter of investigation but several teams have recently been stressing 
the active role of the charged particles-neutral slip induced by the magnetic field \citep{guillet2020,silsbee2020,marchand2021,kawasaki2022,lebreuilly2023} 
together with the turbulent processes \citep{ormel2007}. The resulting grain distributions appear to be complicated and to depend 
on the magnetic intensity and the collapsing time. Therefore to keep the discussion simple enough at this stage, we will restrict the investigation to 
few configurations. For the single size grain models, two sizes are being used namely $a=3 $nm and $a=30$ nm. For multiple size grains, 
we  chose $a_{min}= 1$nm and  $a_{max}= 100$nm or $a_{min}= 10$nm and  $a_{max}= 1000$nm. This latter case, is typically used at high density when 
grain growth is expected to have occurred. 
Except for one case, we always assume $\lambda_{mrn} =-3.5$.

The details of grain binning and various properties, which  are all standard, are given in Appendix~\ref{grainion}.

\subsubsection{Ionisation models}
\label{ionisation_sec}

\setlength{\unitlength}{1cm}
\begin{figure}
\begin{picture} (0,8)
\put(0,0){\includegraphics[width=8cm]{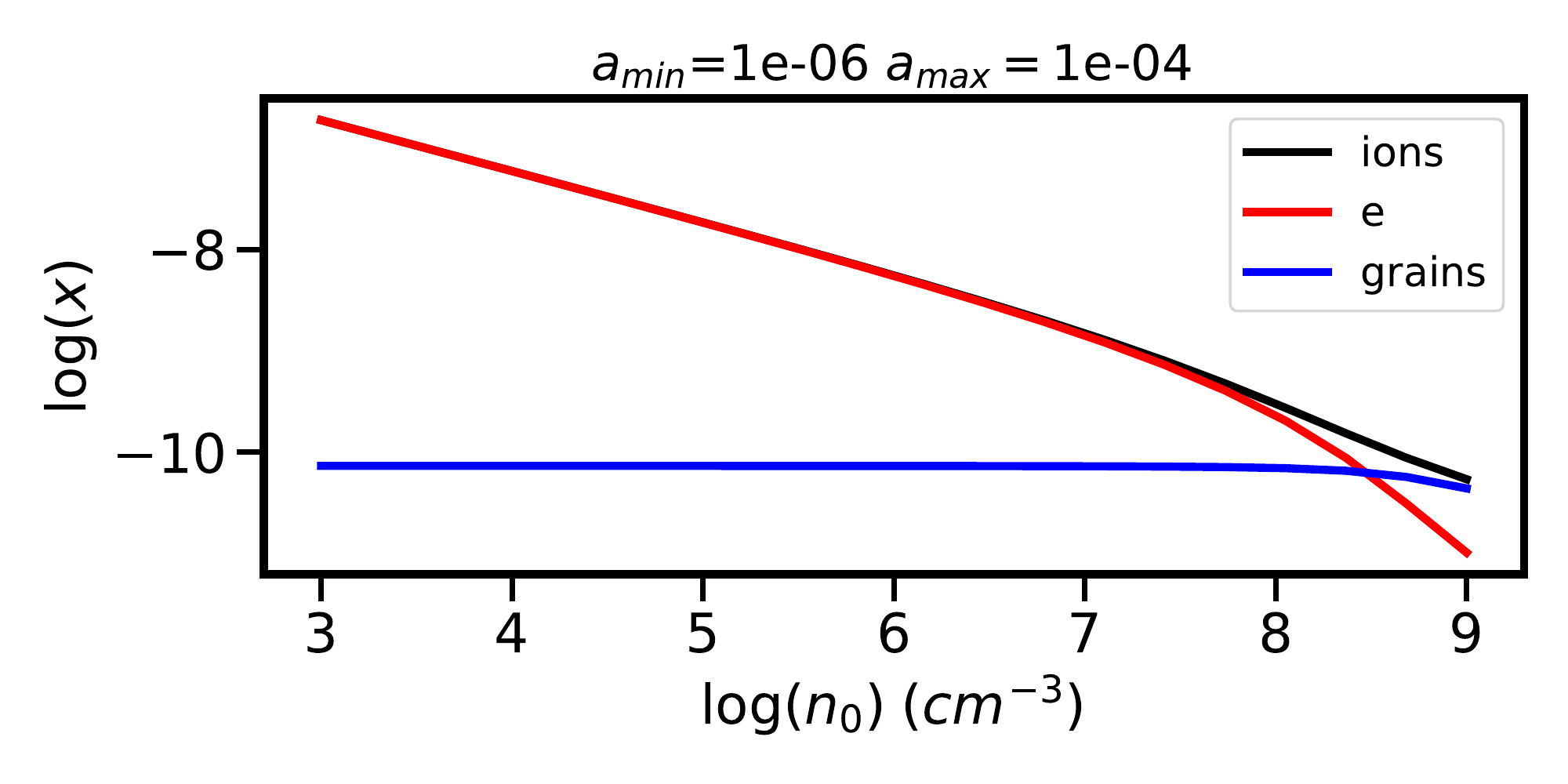}}  
\put(0,4){\includegraphics[width=8cm]{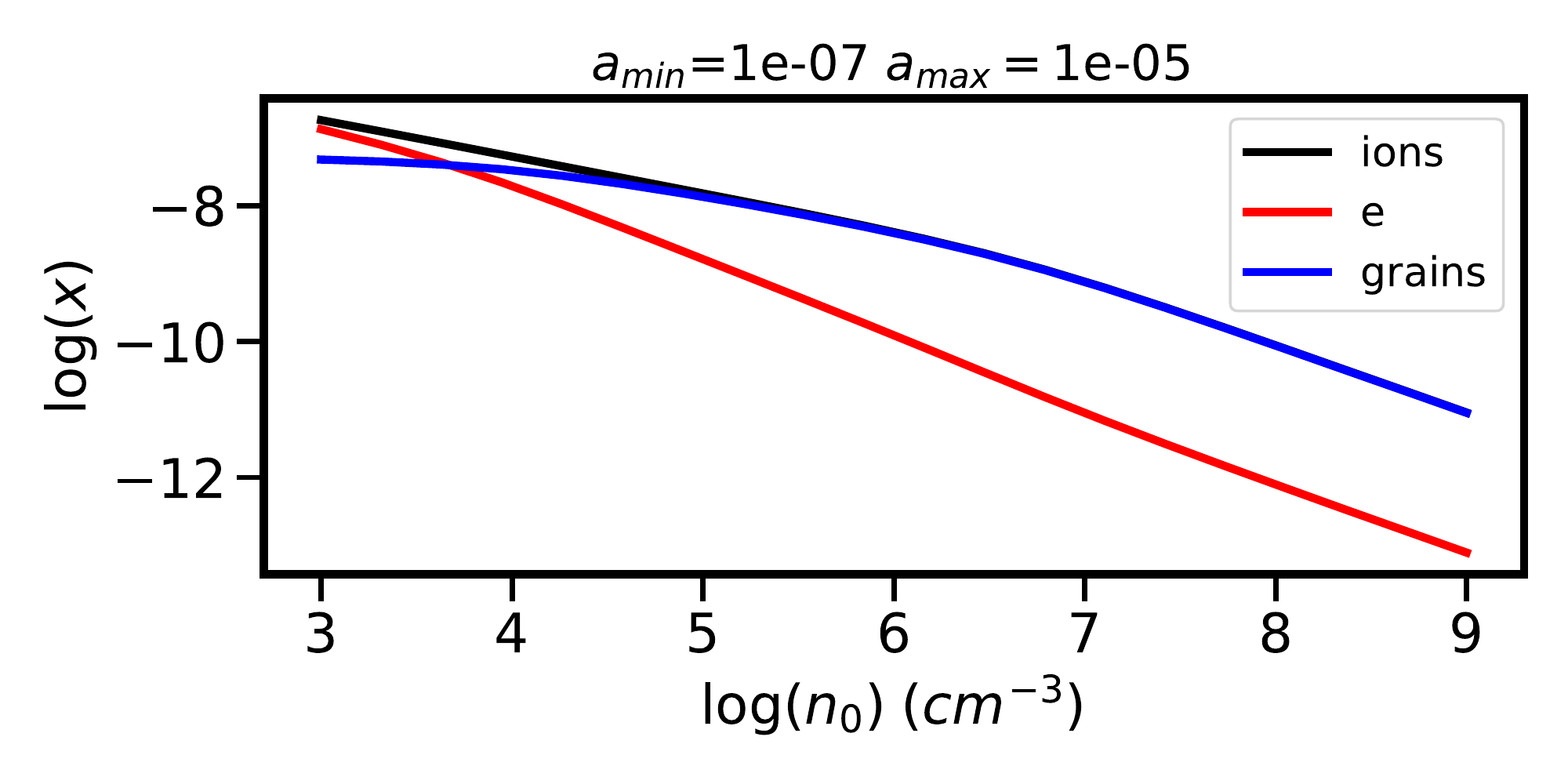}}  
\end{picture}
\caption{ Abundances of  ions, electrons and charged grains as a function of gas density for two grain distributions. 
 The size grain distribution is assumed to be a power law, $d n / da \propto a^{\lambda_{mrn}}$ with $\lambda _{mrn} = -3.5$ and 
grains of size between $a_{min}$ and $a_{max}$ are considered. The ionisation rate is assumed to be  $5 \, 10^{-17}$ s$^{-1}$. }  
\label{ionisation}
\end{figure} 

For the ionisation, we have made two assumptions. When a single size grain is investigated (Sect.~\ref{ssg}) we simply 
assume $x_i ^{ref} = 10^{-7} (n / 10^3 {\rm cm^{-3}} )^{-1/2}$ \citep{shu1987}. 
{Since the number of charges carried on small grains cannot be larger than one, we assume that the numbers
of electrons is such that $n_e = \max( n_i - n_g, 0)$. 

When grain distribution is studied 
(Sect.~\ref{msg}), the ionisation is self-consistently calculated using the model presented in 
\citet{marchand2021}, which is based on the model of \citet{draine1987}.  In this model, the charge carried by the grains varies with 
their size and  typically increases with it. 
 
 Figure~\ref{ionisation} shows the ion, electron and charged grain abundances as a function of density
and for the two different grain distribution, $a_{min}= 1$nm and  $a_{max}= 100$nm (top panel) or $a_{min}= 10$nm and  $a_{max}= 1000$nm
(bottom panel). In the first case, the negative charges are essentially carried by the grains except at low density whereas in the second case
they are carried by the electrons except at high density.

\subsection{Alfv\'en waves propagation}
We  consider the propagation of Alfven waves along a uniform magnetic field  ${\bf B}=B_0 {\bf k}$. 
Introducing for variables the two polarisation modes respectively denoted by "+" and "-", 
\begin{eqnarray}
\nonumber
V_\pm = V_x \pm i V_y, \\
 B_\pm = B_x \pm i B_y, 
 \end{eqnarray}
and considering the following perturbations

\begin{eqnarray}
\nonumber
\nonumber
 \delta v_\pm =  V^\pm  \exp( i ( \omega  t - k  z ) ), \\
 \delta v_{i,e,a,\pm} =  V_{i,e,a} ^\pm \exp( i ( \omega  t - k  z ) ), \\
 \delta b_\pm =  b ^\pm  \exp( i ( \omega  t - k z ) ), 
\nonumber
\end{eqnarray}
we obtain for Eqs.~(\ref{V_mg},\ref{Vd_mg},\ref{j_mg})

\begin{eqnarray}
i \omega    V_\pm  = 
 \int da {d \rho_{i,e,a} \over d a}   K_{i,e,a} ({V_{i,e,a, \pm }} -  V_\pm ),
\label{V_mgpm}
\end{eqnarray}

Momentum equation for charge species

\begin{eqnarray}
  i \omega V_{i,e,a, \pm}  = {Z_{i,e,a}  e \over m_{i,e,a} }  (  E _\pm  -  \pm i  B_0 {{ V_{i,e,a, \pm}} \over c } )
 + \rho  K_{i,e,a} ( V_\pm - V_{i,e,a,\pm} ),
\label{Vd_mgpm}
\end{eqnarray}

Electric current
\begin{eqnarray}
 j_ \pm =  \int da {d \rho _{i,e,a} \over da} { Z_{i,e,a} e \over m_{i,e,a} } V_{i,e,a, \pm}
\label{j_mgpm}
\end{eqnarray}

Maxwell-Faraday equation
\begin{eqnarray}
\omega  B_\pm =   \pm i c k   E_ \pm,
\label{Epm}
\end{eqnarray}

Maxwell-Amp\`ere equation
\begin{eqnarray}
 \pm k  B_\pm = {4 \pi \over c}  j_\pm,
\label{jpm}
\end{eqnarray}

from which we get

\begin{eqnarray}
\left( i \omega    +    \int da {d \rho_{i,e,a} \over d a}   K_{i,e,a}  \right)   V_\pm  = 
 \int da {d \rho_{i,e,a} \over d a}   K_{i,e,a} {V_{i,e,a, \pm }} ,
\label{V2_mgpm}
\end{eqnarray}

\begin{eqnarray}
\nonumber
\left(   i \omega    \pm i   {Z_{i,e,a}  e \over m_{i,e,a}  }  { B_0 \over c}   +  \rho  K_{i,e,a}    \right)   V_{i,e,a, \pm}  =  {Z_{i,e,a}  e \over m_{i,e,a} }   E _\pm  
 + \rho  K_{i,e,a}  V_\pm  \\
  =    \pm i {\omega \over c k }  {Z_{i,e,a}  e \over m_{i,e,a} }   B _\pm   + \rho  K_{i,e,a}  V_\pm ,
\label{Vd2_mgpm}
\end{eqnarray}

or equivalently

\begin{eqnarray}
\nonumber
V_{i,e,a, \pm}  &=&   \pm i {\omega \over c k }     \left(   {    {Z_{i,e,a}  e \over m_{i,e,a} }   \over   i \omega    \pm i   {Z_{i,e,a}  e \over m_{i,e,a}  }  { B_0 \over c}   +  \rho  K_{i,e,a}    } \right)   B _\pm    \\
\nonumber
 &+&   \left(   {  \rho  K_{i,e,a}    \over   i \omega    \pm i   {Z_{i,e,a}  e \over m_{i,e,a}  }  { B_0 \over c}   +  \rho  K_{i,e,a}    } \right)     V_\pm,     \\
 &=&   \pm i {\omega \over c k }  A^{\pm} _{B, \; i,e,a}   B _\pm   +  A^{\pm} _{N, \; i,e,a}   V_\pm  
\label{Vd2_mgpm}
\end{eqnarray}

\begin{eqnarray}
- \pm k  B_\pm = {4 \pi \over c}   \int da {d \rho _{i,e,a} \over da} { Z_{i,e,a} e \over m_{i,e,a} } V_{i,e,a, \pm},
\label{j2pm}
\end{eqnarray}

and thus 

\begin{widetext}

\begin{eqnarray}
\left( i \omega    +    \int da {d \rho_{i,e,a} \over d a}   K_{i,e,a}   - \int da {d \rho_{i,e,a} \over d a}   K_{i,e,a}   A^{\pm} _{N, \; i,e,a}     \right)   V_\pm  = 
    \pm i {\omega \over c k }   \left(   \int da {d \rho_{i,e,a} \over d a}   K_{i,e,a}    A ^{\pm}_{B, \; i,e,a}    \right)  B _\pm      ,
\label{V3 _mgpm}
\end{eqnarray}

\begin{eqnarray}
\pm \left( -  k   -  i {\omega \over c k }    {4 \pi \over c}   \int da {d \rho _{i,e,a} \over da} { Z_{i,e,a} e \over m_{i,e,a} }   A^{\pm} _{B, \; i,e,a}    \right) B_\pm = {4 \pi \over c}   \int da {d \rho _{i,e,a} \over da} { Z_{i,e,a} e \over m_{i,e,a} }    A^{\pm} _{N, \; i,e,a}   V_\pm   
\label{j3pm}
\end{eqnarray}

The dispersion relation follows

\begin{eqnarray}
\nonumber
\left( i \omega    +    \int da {d \rho_{i,e,a} \over d a}   K_{i,e,a}   - \int da {d \rho_{i,e,a} \over d a}   K_{i,e,a}   A^{\pm} _{N, \; i,e,a}     \right)  \left(  k   +  i {\omega \over c k }    {4 \pi \over c}   \int da {d \rho _{i,e,a} \over da} { Z_{i,e,a} e \over m_{i,e,a} }   A^{\pm} _{B, \; i,e,a}    \right)    = \\
   - i  {4 \pi \over c}    {\omega \over c k }   \left(   \int da {d \rho_{i,e,a} \over d a}   K_{i,e,a}    A ^{\pm}_{B, \; i,e,a}    \right)    \int da {d \rho _{i,e,a} \over da} { Z_{i,e,a} e \over m_{i,e,a} }    A^{\pm} _{N, \; i,e,a}   ,
\label{disp_rel1}
\end{eqnarray}

It can also be written as 

\begin{eqnarray}
\nonumber
k ^2  \left( i \omega    +    \int da {d \rho_{i,e,a} \over d a}   K_{i,e,a}   (1 -  A^{\pm} _{N, \; i,e,a} )     \right)   = \\
    -i  {4 \pi \omega \over c^2 }   \left(   \int da {d \rho_{i,e,a} \over d a}   K_{i,e,a}    A ^{\pm}_{B, \; i,e,a}     \int da {d \rho _{i,e,a} \over da} { Z_{i,e,a} e \over m_{i,e,a} }    A^{\pm} _{N, \; i,e,a}   + \int da {d \rho _{i,e,a} \over da} { Z_{i,e,a} e \over m_{i,e,a} }   A^{\pm} _{B, \; i,e,a}     \left( i \omega    +    \int da {d \rho_{i,e,a} \over d a}   K_{i,e,a}   (1 -  A^{\pm} _{N, \; i,e,a} )     \right)   \right)   ,
\label{disp_rel2}
\end{eqnarray}

where

\begin{eqnarray}
\nonumber
 A^{\pm}_{N,i,e,a}  &=&
\left( \rho K_{i,e,a}     \over i \omega + \rho K_{i,e,a}  \pm i {B_0 Z_{i,e,a} e \over c m_{i,e,a} }   \right), \\
A^{\pm}_{B,i,e,a} &=& 
  \left(  { {Z_{i,e,a}  e \over m_{i,e,a} } \over  i \omega + \rho K_{i,e,a}  \pm i {B_0 Z_{i,e,a} e \over c m_{i,e,a}}   } \right).
\label{AAnia}
\end{eqnarray}

\end{widetext}

Equations~(\ref{disp_rel1}) and (\ref{AAnia}) constitute a complex dispersion relation. Due to the presence of $\omega$
at the denominator of $A_{N,i,e,a}$ and $A_{B,i,e,a}$ which are part of integrals over the grain size distribution,
the dispersion relation  is, strictly speaking, not a polynomial expression, although because of the finite number of grain size eventually considered, it is 
effectively a high order polynomials. 
Whereas Eq.~(\ref{disp_rel2}) explicitly gives $k$ as a function of $\omega$, getting $\omega$ as a function of $k$ (that is to say a real $k$)
is obviously more complex as will be seen below. 
 We stress that the only difference between  the "+" and "-" modes  is the term $i( \omega \pm B_0 Z_{i,e,a} e /  (c m_{i,e,a} ) ) $ that 
represents the Hall effect.
We therefore expect that these two modes should present important differences for frequencies close to the grain gyro frequencies $ B_0 Z_{i,e,a} e /  (c m_{i,e,a}) $.
In particular, since the grain are negatively charged,  we see that for the "+" mode  a kind of resonance is expected  to occur (whose amplitude is limited by the friction).





\subsection{Dispersion relation for a single size grain}
Equations~(\ref{disp_rel1}) and (\ref{AAnia}) are rather complex and it is therefore
enlightening to start considering a system that is as simple as possible.
This will allow verifying that simple regimes and behaviours are indeed recovered. 
We therefore study the single size grain case and 
since our main goal here is to gain insight on the problem,
we further neglect the ion inertia and ion-neutral friction. 
 This case is therefore equivalent to the standard  electron-ion-neutral case, 
originally considered by \citet{kulsrud1969}, where the electrons and the ions have been 
replaced by the ions and charged grains respectivelly.

The derivation has been obtained in Appendix~\ref{singlesizegrain} by a simplification 
of Eqs.~(\ref{rel_disp}) and (\ref{CCnia}). 
The corresponding  dispersion relation (single size grain and $K_i=0$) is




\begin{eqnarray}
\nonumber
   \omega^3
+ \left(   \pm  k^2     {c B_0 \over 4 \pi   n_i Z_i e} 
  - i  (\rho + \rho_{a}) K_{a}  \right)   \omega ^2 
    -  \\
     k^2   \left( \pm  i (\rho + \rho_{a})  K_{a} {c B_0 \over 4 \pi  n_i Z_i e } + {B_0^2  \over 4 \pi \rho _{a}} \right) \omega 
      + i k ^2 {B_0^2 \over 4 \pi}  K_{a} = 0,
\label{rel_disp_simpl8}
\end{eqnarray}
which is a third degree polynomials meaning that three roots are to be found. 

This relation can also be expressed as

\begin{eqnarray}
\nonumber
k^2  = \\
 {   \omega^3  - i  (\rho + \rho_{a}) K_{a} \omega^2  \over \left(  -\pm  \omega^2     {c B_0 \over 4 \pi   n_i Z_i e} + \left( \pm  i (\rho + \rho_{a})  K_{a} {c B_0 \over 4 \pi  n_i Z_i e } + {B_0^2  \over 4 \pi \rho _{a}} \right) \omega 
      - i {B_0^2 \over 4 \pi}  K_{a}   \right) },
\label{k_omeg4}
\end{eqnarray}
which implies that for a given $\omega$, there are two possible $k$ solutions of Eq.~(\ref{k_omeg4}).

\section{The single size grain case}
\label{ssg}

As the multi-size grain case appears to be obviously more complex, we start with a single size grain fluid for which simple analysis 
can be enlighteningly performed.

\subsection{Some simple asymptotic behaviours}

\subsubsection{Neglecting  grain inertia}
It is worth discussing first the behaviour obtained when the grain intertia is being neglected. This latter approximation is accomplished by
removing in Eqs.~(\ref{CCnia2_ap}) the terms $i \omega$  
at the numerator of Eq.~(\ref{CCnia2_ap}). This leads to replace Eq.~(\ref{rel_disp_simpl8})
by

\begin{eqnarray}
\nonumber
 \pm  i k^2    \omega {c B_0 \over 4 \pi}  
\left(  \rho  K_{a}  \pm i {B_0 Z_{a} e \over c m_{a}}  \right)
  + 
  i \omega^2   n_i Z_i e  
  \left(   \rho  K_{a}  \right)
   + \\  
   \left(   k^2    {c B_0 \over 4 \pi}  \right)
   \left(  -  i {B_0 n_i Z_i e \over c }  K_{a}      \right)  =  0,
\label{rel_disp_noiner}
\end{eqnarray}
and this leads to (using  $n_i Z_i = - n_{a} Z_{a}$)

\begin{eqnarray}
 \omega^2  
 +  k^2    {  B_0 \over 4 \pi}  
\left( \pm {c \over n_i Z_i e } - i {B_0 \over \rho K_{a} \rho_{a}}  \right)  \omega
    - \left(   k^2    { B_0^2 \over 4 \pi \rho}  \right)
    =  0.
\label{rel_disp_noiner2}
\end{eqnarray}

The two solutions describing waves propagating along positive z are

\begin{eqnarray}
\nonumber
\omega_{\pm} = {1 \over 2 } \left(   k^2    {  B_0 \over 4 \pi}  
\left(  {c \over n_i Z_i e } + i {B_0 \over \rho K_{a} \rho_{a}}  \right)  \right. \\
\left. \pm \left(   k^4    {  B_0 ^2 \over 16 \pi^2}  
\left( {c \over n_i Z_i e } + i {B_0 \over \rho K_{a} \rho_{a}}  \right)^2  + 4 
k^2    { B_0^2 \over 4 \pi \rho} 
\right) ^{1/2} \right)
\label{omeg_noiner2}
\end{eqnarray}

Note that the other solutions of Eq.~(\ref{rel_disp_noiner2}) are physically equivalent 
to the two solutions $\omega _{\pm}$ as they simply  correspond to propagation toward
negative z.

$\bullet$ In the limit of long wavelengths, we get
\begin{eqnarray}
\label{limk0}
k \rightarrow 0, \; \omega \rightarrow {B_0 \over \sqrt{4 \pi \rho}  }k + i k^2 { B_0^2 \over 8 \pi \rho  } { 1 \over K_{a} \rho _{a}},
\end{eqnarray}
the two modes propagate at the Alfv\'en speed. The dust friction leads to a dissipation proportional to $k^2$. \\

$\bullet$  In the limit of short wavelengths, two different modes are found \citep{wardle1999}
\begin{eqnarray}
\label{limkinf0}
k \rightarrow \infty, \; \omega \rightarrow k^2 {B_0 c \over 4 \pi n_i Z_i e}  + i k^2 { B_0^2 \over 8 \pi \rho  } { 1 \over K_{a} \rho _{a}},
\end{eqnarray}
and
\begin{eqnarray}
\label{limkinf1}
k \rightarrow \infty, \; \omega \rightarrow {B_0^2 \over  \rho } { -  
{  c B_0 \over n_i Z_i e } + i { B_0^2 \over \rho K_{a} \rho_{a} }    \over  \left({  c B_0 \over n_i Z_i e }\right)^2 + \left( { B_0^2 \over \rho K_{a} \rho_{a} }  \right)^2 }.
\end{eqnarray}
The first one presents a frequency proportional to $k^2$ which is characteristic of the Hall effect. These modes are known as the whistler modes.
For the second one, the frequency tends toward a constant, meaning that the group velocity $\partial  _k {\cal R}_e (\omega) =0$ and 
so these waves do no not transport energy.

\subsubsection{Effect of the inertia}
A first and immediate difference between Eq.~(\ref{rel_disp_simpl8}) and  Eq.~(\ref{rel_disp_noiner2}), which respectively includes and 
excludes grain inertia is that the former is of degree three whereas the latter is of degree two. As expected there are less degrees of freedom and 
therefore less modes when grain inertia is neglected. Again it is worth, studying the asymptotic behaviours. \\ \\

$\bullet$  In the limit $\omega \rightarrow 0$, $k \rightarrow 0$, we have

\begin{eqnarray}
\label{limk0_in}
k \rightarrow 0, \; \omega \rightarrow {B_0 \over \sqrt{4 \pi (\rho+\rho_{a}) }  }k + i k^2 { B_0^2 \over 8 \pi (\rho+\rho_{a})  } { 1 \over K_{a} \rho _{a}},
\end{eqnarray}
which is almost identical to the limit with no inertie as stated by Eq.~(\ref{limk0}), except for a minor correction 
in the densities. \\ \\

$\bullet$  In the limit $k \rightarrow \infty$, we see that if $\omega$
remains finite, it must satisfy a second order equation that is independent of 
$k$, namely

\begin{eqnarray}
\nonumber
   \omega ^2    -  
      \left(   i (\rho + \rho_{a})  K_{a}  \pm {B_0   n_i Z_i e \over c  \rho _{a} }  \right) \omega 
      \pm i  {B_0 n_i Z_i e \over c }  K_{a} = 0.
\label{mode_omeg2}
\end{eqnarray}
We thus get two modes which in the limit $k \rightarrow \infty$ tend to a constant $\omega$ and 
therefore have a vanishing group velocity. \\ \\

$\bullet$  In the limit $\omega \rightarrow \infty$, $k \rightarrow \infty$, we see that Eq.~(\ref{rel_disp_simpl8}) becomes

\begin{eqnarray}
\nonumber
   \omega^2
+ \left(   \pm  k^2     {c B_0 \over 4 \pi   n_i Z_i e} 
  - i  (\rho + \rho_{a}) K_{a}  \right)   \omega  
    -  \\
     k^2   \left( \pm  i (\rho + \rho_{a})  K_{a} {c B_0 \over 4 \pi  n_i Z_i e } + {B_0^2  \over 4 \pi \rho _{a}} \right)  = 0,
\label{rel_disp_highomeg}
\end{eqnarray}
which admits for relevant solutions
\begin{eqnarray}
\omega = {1 \over 2} \left(    k^2     {c B_0 \over 4 \pi   n_i Z_i e} 
  + i  (\rho + \rho_{a}) K_{a}  \right. \\
  \nonumber
  \left. 
 \left( \left(    k^2     {c B_0 \over 4 \pi   n_i Z_i e} 
  + i  (\rho + \rho_{a}) K_{a} 
   \right)^2 + \right. \right.  \\
   \left. \left.
   4 k^2   \left(   i (\rho + \rho_{a})  K_{a} {c B_0 \over 4 \pi  n_i Z_i e } + {B_0^2  \over 4 \pi \rho _{a}} \right) \right)^{1/2}
   \right),
   \nonumber
\end{eqnarray}
and thus
\begin{eqnarray}
k \rightarrow \infty, \;  \omega \rightarrow    k^2     {c B_0 \over 4 \pi   n_i Z_i e} 
  + i  (\rho + \rho_{a}) K_{a}.
  \label{limkinf_iner}
  \end{eqnarray}
Comparing this expression with Eq.~(\ref{limkinf0}), i.e. the asymptotic behaviour when grain inertia is neglected, 
we see a fundamental difference. Whereas the real parts are identical, the imaginary ones are different. 
More precisely, the dissipation is proportional to $k^2$ when inertia is neglected and to $k^0$ when it is taken 
into account. This is a major difference indicating that inertia cannot be neglected at short wavelengths.

To summarize, we see that in the long wavelength limit, the grain inertia does not play a significant role, it cannot however be neglected 
at scale sufficiently small.

\subsection{Orders of magnitude}
The asymptotic analysis allows us to anticipate the most important regimes regarding the Alfv\'en waves. 
We have three modes of propagation 
\begin{eqnarray}
\label{propagation}
\omega_{AW,n} = { B_0 \over \sqrt{4 \pi \rho} } k, \\
\nonumber
\omega_{AW,d} = { B_0 \over \sqrt{4 \pi \rho_d} } k, \\
\nonumber
\omega_{Hall} = {c B_0  \over 4 \pi n_i Z_i e} k^2, 
\end{eqnarray}
which respectively correspond to the Alfv\'en velocity for the neutrals, the Alfv\'en velocity for the grains (about 10 times larger than
$\omega_{AW,n}$) and the velocity of the whistler mode. 

There are two modes of dissipation
\begin{eqnarray}
\label{diss}
\omega_{diss,n} = {B_0 ^2  \over 8 \pi \rho K_d \rho_d } k^2, \\
\nonumber
\omega_{diss,d} = K_d \rho,
\end{eqnarray}
where the  latter is the relevant dissipation at large $k$  whereas the former is valid at small $k$.

By comparing the various expressions, we can infer different cases. First, we have the transition between the  neutral Alfv\'enic 
regime and the dissipation dominated mode above which Alfv\'en  waves do not propagate. This occurs when $\omega _{diss,n} \simeq \omega_{AW,n}$
at a wavenumber $k_{AW,diss,n}$
\begin{eqnarray}
\label{AWdissn}
k_{AW,diss,n} = { 2 \sqrt{4 \pi} K_d \rho_d \sqrt{\rho} \over B_0 }.
\end{eqnarray}
Then the dust Alfv\'enic regime and the dissipation dominated mode below which Alfv\'en waves do not propagate,
i.e. $\omega _{diss,d} \simeq \omega_{AW,d}$
 giving 
 \begin{eqnarray}
\label{AWdissd}
k_{AW,diss,d} = {  \sqrt{4 \pi} K_d \rho \sqrt{\rho_d} \over B_0 }.
\end{eqnarray}
Another fundamental regime comes from the transition between Alfv\'en and whistler propagation. This leads to two wavenumbers
\begin{eqnarray}
\label{AWhalln}
k_{AW,Hall,n} = {  \sqrt{4 \pi}  n_i Z_i e  \over   c  \sqrt{\rho}  },
\end{eqnarray}
and 
\begin{eqnarray}
\label{AWhalld}
k_{AW,Hall,d} = {  \sqrt{4 \pi}  n_i Z_i e  \over   c  \sqrt{\rho_d}  }.
\end{eqnarray}

For convenience, writing $x_i^{ref}= 10^{-7} (n / 10^3 {\rm cm}^{-3})^{-1/2}$ and $\beta _{mag} = (3/2 k_B T n) / (B_0^2 / (8 \pi))$, we obtain 

\begin{eqnarray}
\label{AWdissn_num}
\lambda _{AW,diss,n} = \\
\nonumber
7.2 \times 10^3 \, {\rm au} \; \beta _{mag}^{-1/2} \left( { n \over 10^{3} {\rm cm}^{-3}  } \right) ^{-1} {a \over 10^{-7} {\rm cm}}
 \left( {  \rho _d \over \rho } \times 100 \right)^{-1},
\end{eqnarray}

\begin{eqnarray}
\label{AWdissd_num}
\lambda _{AW,diss,d} = \\
\nonumber
1.4 \times 10^3 \, {\rm au} \; \beta _{mag}^{-1/2} \left( { n \over 10^{3} {\rm cm}^{-3}  } \right) ^{-1} {a \over 10^{-7} {\rm cm}}  
\left( {  \rho _d \over \rho } \times 100 \right)^{-1/2},
\end{eqnarray}

\begin{eqnarray}
\label{AWdhalln_num}
\lambda _{AW,hall,n} = 4.2 \, {\rm au},
\end{eqnarray}

\begin{eqnarray}
\label{AWdhalld_num}
\lambda _{AW,hall,d} = 0.42 \, {\rm au}.
\end{eqnarray}

Several interesting trends can already be inferred. First, the dissipation wavelengths 
scale with the density as $n^{-1}$ and therefore vary significantly during the collapse. 
In particular, given than the Jeans length varies as $n^{-1/2}$, 
the dissipation length  becomes continuously smaller with respect to the Jeans length. 
The dissipation wavelengths also increase with the grain size. 
Second the Hall wavelengths do not depend on density nor on the grain size (assuming that the ionisation 
is as stated by $x_i^{ref}$).  The consequence this may have in the context of collapsing 
cloud are discussed in Sect.~\ref{disk_form}.

\subsection{ Solving for the dispersion relation $\omega$ vs $\lambda$}

\setlength{\unitlength}{1cm}
\begin{figure}
\begin{picture} (0,8)
\put(0,4){\includegraphics[width=8cm]{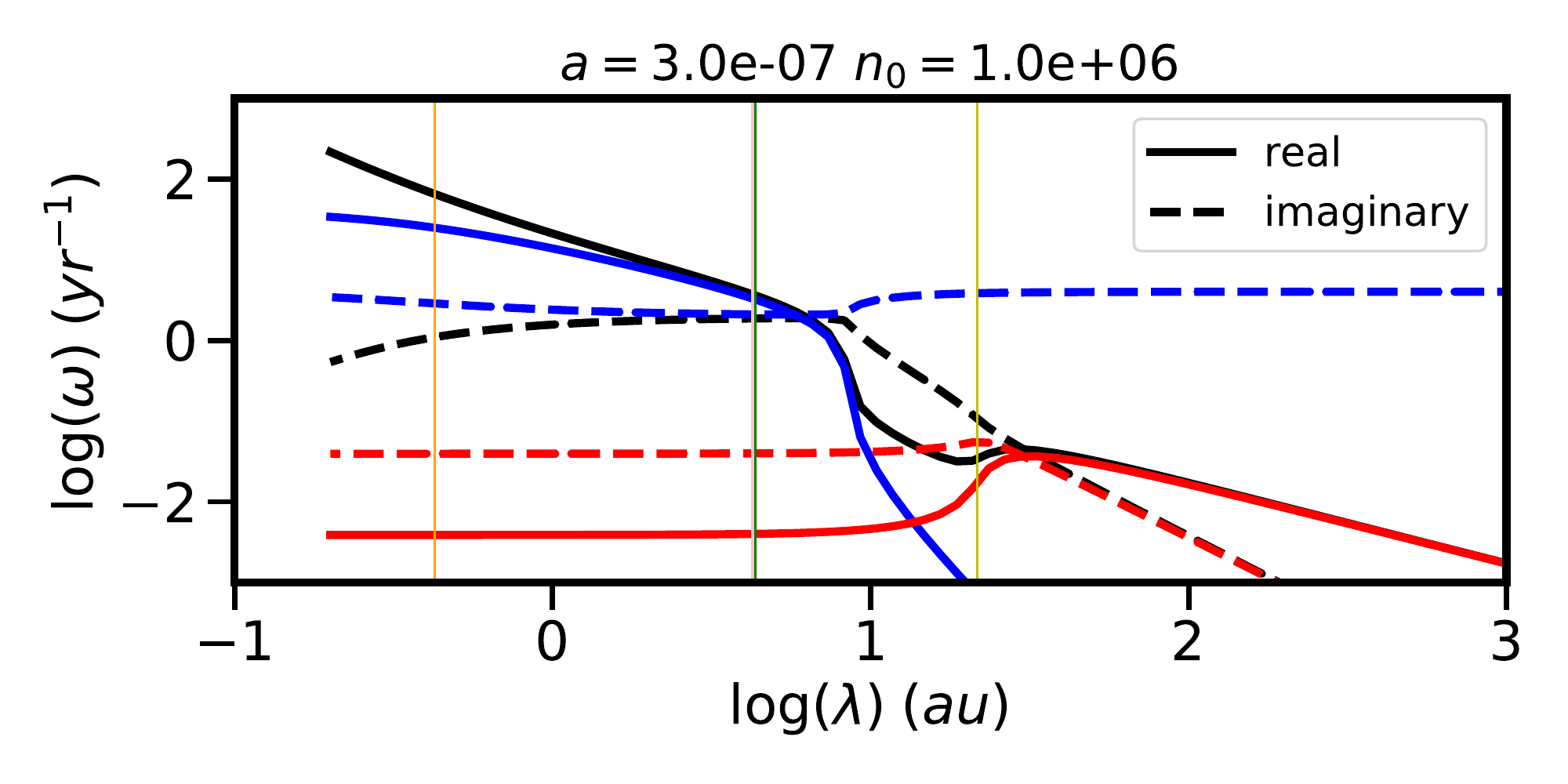}}  
\put(0,0){\includegraphics[width=8cm]{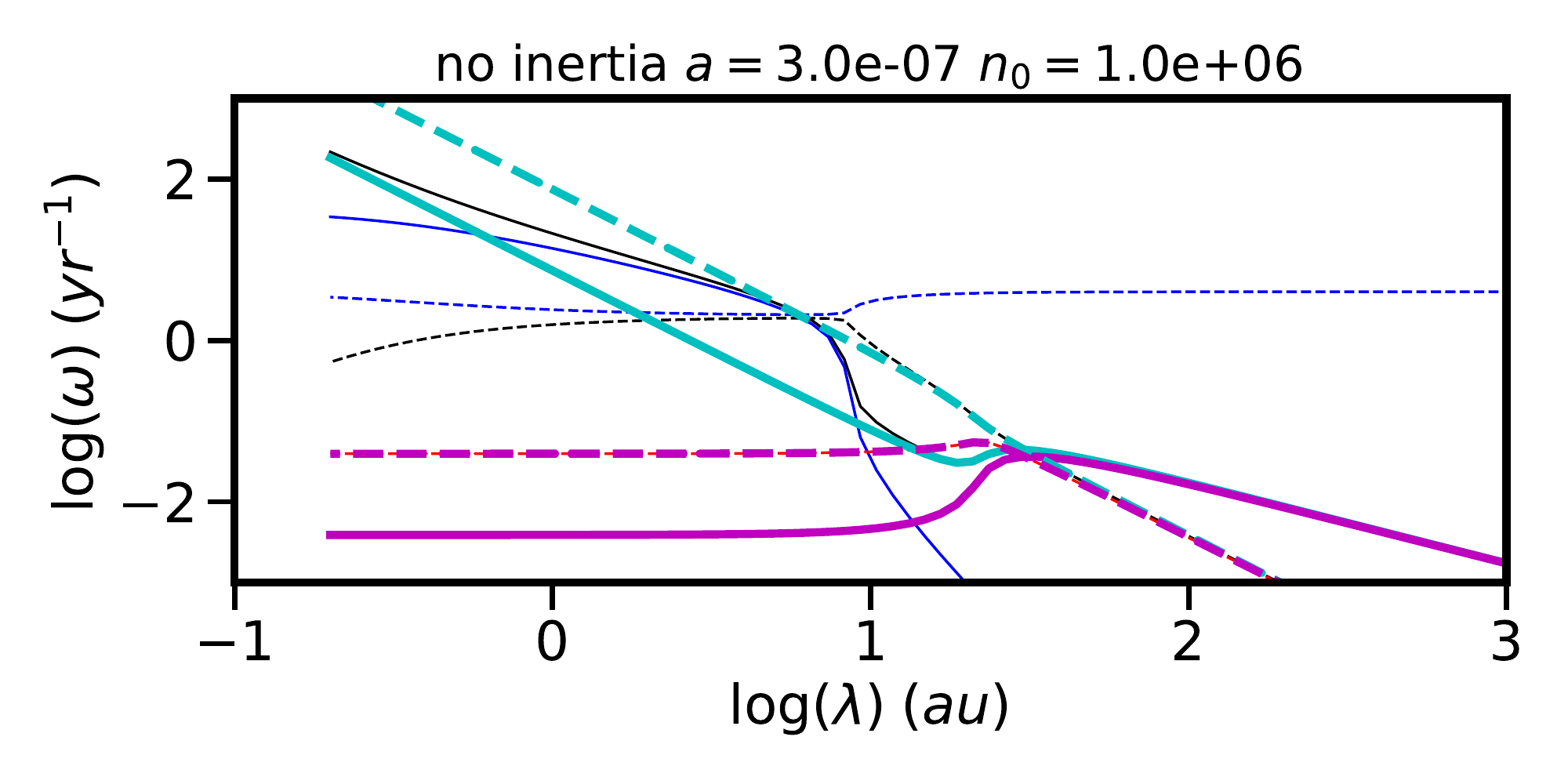}}  
\end{picture}
\caption{  Wave frequency as a function of wavelengths for $n_0=10^6$ cm$^{-3}$ and a single grain size of $a=3 \times 10^{-7}$ cm. 
In top panel, the grain inertia is considered 
and the three modes are displayed (dark, blue and red lines). Both the real (solid lines) and the imaginary (dashed lines) parts are shown.
In bottom panels, the grain inertia  is neglected and the two corresponding modes are displayed (cyan and purple).
The modes of top panel have been drawn in the bottom panel (thin lines) to make comparison easier.  
At large wavelengths, the inertia does not play any role. However at scale smaller 
than about 10 au, the inertia drastically modify the Alfv\'en wave propagation.
The yellow, green, pink and orange vertical lines represent $\lambda _{AW,diss,n}$, $\lambda _{AW,diss,d}$, $\lambda _{AW,hall,n}$, $\lambda _{AW,hall,d}$ respectively
(see Eqs.~\ref{AWdissn_num}-\ref{AWdhalld_num}). Note that in this particular case $\lambda _{AW,diss,d} = \lambda _{AW,hall,n}$ and so
the pink line is not visible. }
\label{3e-7n1e6}
\end{figure} 

\setlength{\unitlength}{1cm}
\begin{figure}
\begin{picture} (0,8)
\put(0,4){\includegraphics[width=8cm]{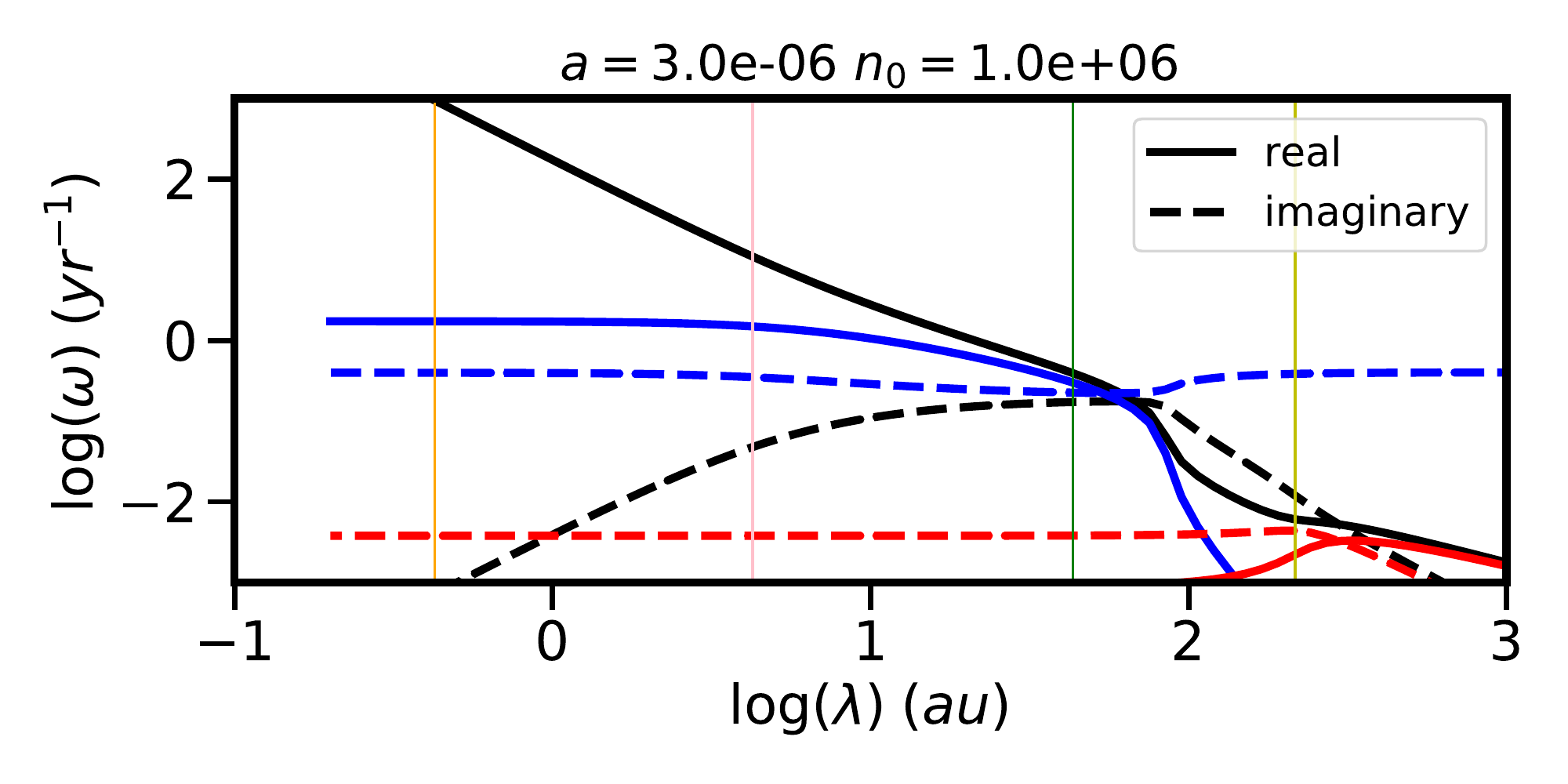}}  
\put(0,0){\includegraphics[width=8cm]{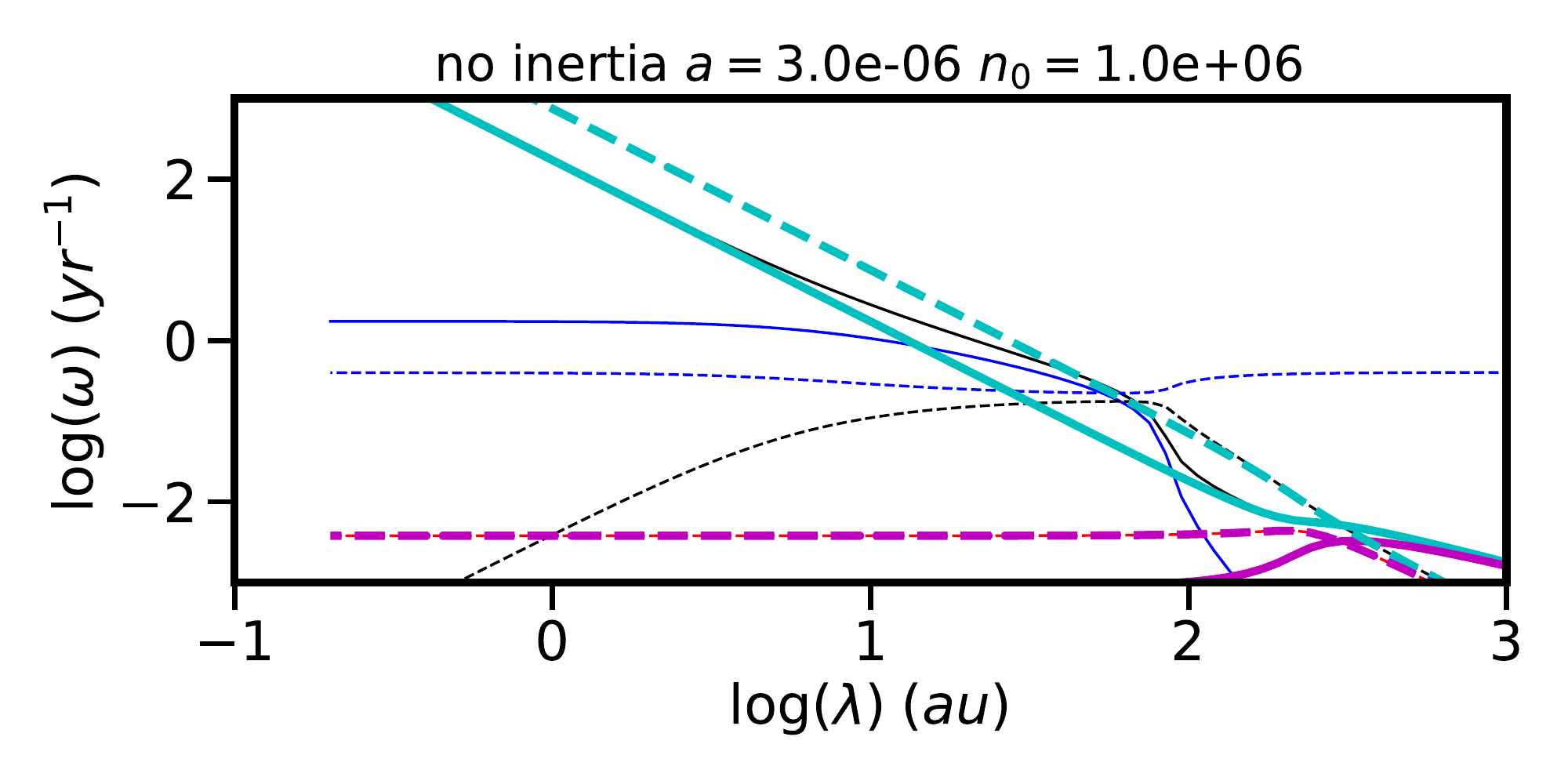}}  
\end{picture}
\caption{ Same as Fig.~\ref{3e-7n1e6}
for $n_0=10^6$ cm$^{-3}$ and $a=3 \times 10^{-6}$ cm. The difference 
between the inertia and no inertia cases is important below 100 au.}
\label{3e-6n1e6}
\end{figure}

\setlength{\unitlength}{1cm}
\begin{figure}
\begin{picture} (0,8)
\put(0,4){\includegraphics[width=8cm]{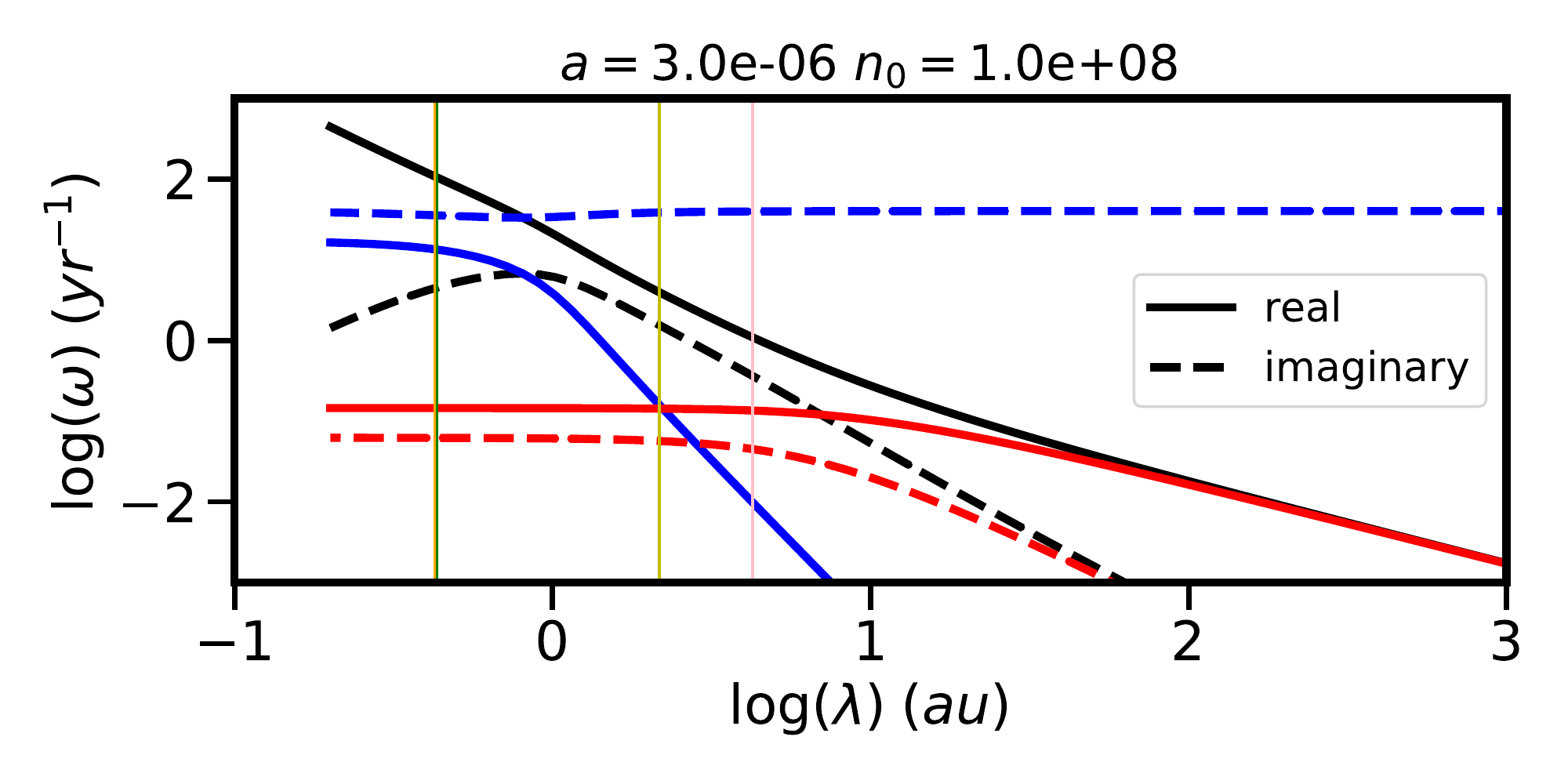}}  
\put(0,0){\includegraphics[width=8cm]{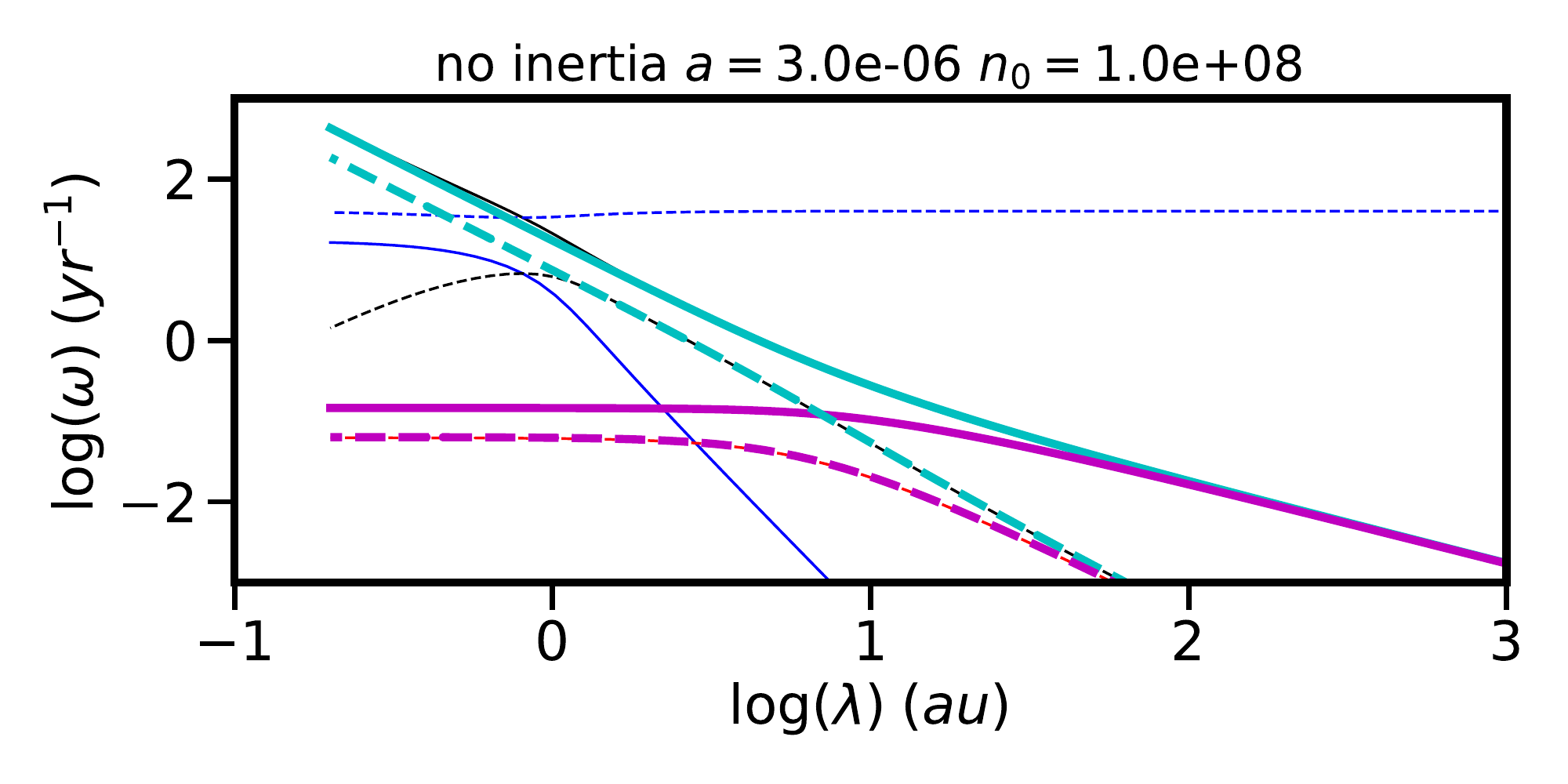}}  
\end{picture}
\caption{ Same as Fig.~\ref{3e-7n1e6}  for $n_0=10^8$ cm$^{-3}$ and $a=3 \times 10^{-6}$ cm. 
The difference 
between the inertia and no inertia cases is important below 1 au.}
\label{3e-6n1e8}
\end{figure} 

We now solve the dispersion relations both with and without grain inertia, as stated by Eqs.~(\ref{rel_disp_simpl8}) and~(\ref{rel_disp_noiner2}).
This is achieved by simply finding the roots of the second and third degree polynomials. 

 We assume a temperature of 10 K and a magnetic $\beta _{mag}= 0.1$, which define the value of $B_0$. We also recall that the ionisation is 
prescribed as explained in section.~\ref{ionisation_sec}. 
For $a= 3 \times 10^{-7}$ cm, the number of grains is large and thus the number of electrons is assumed to be 0. For
$a= 3 \times 10^{-6}$ cm, the number of grains is much smaller and thus $n_e = n_i - n_g > 0$. However, since in  section 3 for simplicity reason, both the inertia and the friction 
of ions and electrons are neglected, the only consequence it has on  Eq.~(\ref{rel_disp_simpl8}) is to replace $n_i$ by $n_i-n_e=n_g$ which also 
means that when  $a= 3 \times 10^{-6}$ cm the ionisation is given by $n_g / n_0$.

\subsubsection{$n_0=10^6$ cm$^{-3}$ and $a= 3 \times 10^{-7}$ cm}
Figure~\ref{3e-7n1e6} displays results for $n_0=10^6$ cm$^{-3}$ and $a= 3 \times 10^{-7}$ cm. 
The black, blue and red lines of top panel correspond to the three roots of Eqs.~(\ref{rel_disp_simpl8}), the solid lines
represent the real parts and dashed lines the imaginary ones.  
The yellow, green, pink and orange vertical lines stand for $\lambda _{AW,diss,n}$, $\lambda _{AW,diss,d}$, $\lambda _{AW,hall,n}$, $\lambda _{AW,hall,d}$ respectively
(as stated by Eqs.~\ref{AWdissn_num}-\ref{AWdhalld_num}).

For $\lambda > \lambda _{AW,diss,n}$ au, the red and blue lines are identical. They correspond to the two usual Alfv\'en modes and they behave 
as stated by Eq.~(\ref{limk0_in}). The black lines represent a very dissipative mode which has a very low group velocity. 
This corresponds to a mode for which the dust is decoupled from the neutral and is very dissipative. 
At about $\lambda = \lambda _{AW,diss,n}$ (yellow line),  the dissipation is very intense and the wave propagation is 
significantly reduced. In particular we see that the group velocity changes sign. At about 
 $\lambda \simeq \lambda _{AW,diss,d}$ (green line), the waves propagate as classical Alfv\'en waves however at the Alfv\'en speed of the dust, 
 which is ten times the Alfv\'en speed of the neutrals. The two Alfv\'en waves,  which we recall, represent the two circularly polarised modes, 
 behave similarly.  At $\lambda \simeq \lambda _{AW,hall,d}$ (orange line), we see that the two branches split due to the Hall effect as predicted 
 by Eqs.~(\ref{mode_omeg2}) and (\ref{limkinf_iner}). 
 
 Bottom panel of Fig.~\ref{3e-7n1e6} shows the solutions obtained when grain inertia is neglected as described 
 by Eq.~(\ref{rel_disp_noiner2}). The cyan and purple lines display the two roots. For convenience the 
 thin lines reproduce the results of top panel. For $\lambda > \lambda _{AW,diss,n}$ au, the solutions with and without inertia 
 are essentially identical. However, for $\lambda < 10$ au, major differences between the cases with and without inertia appear.
 First of all, we see that only the cyan mode has a non-vanishing group velocity. Indeed the purple root, is identical to the red one
 and has a vanishing group velocity. It therefore do not transport energy. Thus for 1 au < $\lambda$ < 10 au, 
neglecting inertia predicts an unphysically strong difference between the two circularly polarised modes.
 Second, the real part of the cyan mode has a slope much stiffer than the black and blue ones. It is 
 a whistler wave. The reason is that since the grain inertia is neglected, the Alfv\'enic grain modes do not exist. We also see 
 that the imaginary part of the cyan root is much higher than for the blue and black ones. This is because the 
 grain-neutral friction is considerably over-estimated. 
 
 \subsubsection{$n_0=10^6$ cm$^{-3}$ and $a= 3 \times 10^{-6}$ cm}
  
 Figure~\ref{3e-6n1e6} portrays the case $n_0=10^6$ cm$^{-3}$ and $a= 3 \times 10^{-6}$ cm.
 Because of the grain size being 10 times bigger than the case of Fig.~\ref{3e-7n1e6}, 
  $\lambda _{AW,diss,n}$ is correspondingly 10 times larger than in Fig.~\ref{3e-6n1e6}
  and overall the dispersion relations are similar except that they are shifter toward 
  longer wavelength when $a= 3 \times 10^{-6}$ cm.
  
  The difference with the waves corresponding to the no inertia case (bottom panel) is again very pronounced. 
  Again the cyan mode is a whistler wave which is completely asymmetrical with the 
  other polarisation represented by the purple one. Whereas at small wavelengths, the real part agrees well with the case with inertia, the 
  imaginary one is far too high, implying that the waves dissipate too rapidly if inertia is neglected. 
  More physical interpretations will be given for this case in Sect.~\ref{velocity} where the velocities of the various species are analysed.

\subsubsection{$n_0=10^8$ cm$^{-3}$ and $a= 3 \times 10^{-6}$ cm}
Figure~\ref{3e-6n1e8} portrays the case $n_0=10^8$ cm$^{-3}$ and $a= 3 \times 10^{-6}$ cm.
Due to the high density, $\lambda _{AW,diss,n}$ is 100 times smaller than in Fig.~\ref{3e-6n1e6}.
Consequently, it is smaller than $\lambda _{AW,hall,n}$ and thus we have a transition 
between two Alfv\'en waves and a whistler mode together with a zero group velocity wave (red mode). 
As expected, bottom panel reveals that the no inertia wave constitutes a good approximation of the 
exact case, except at small scale below 1 au where deviations between the two cases are significant.


%

\subsection{$\lambda$ vs $\omega$}

\setlength{\unitlength}{1cm}
\begin{figure}
\begin{picture} (0,8)
\put(0,4){\includegraphics[width=8cm]{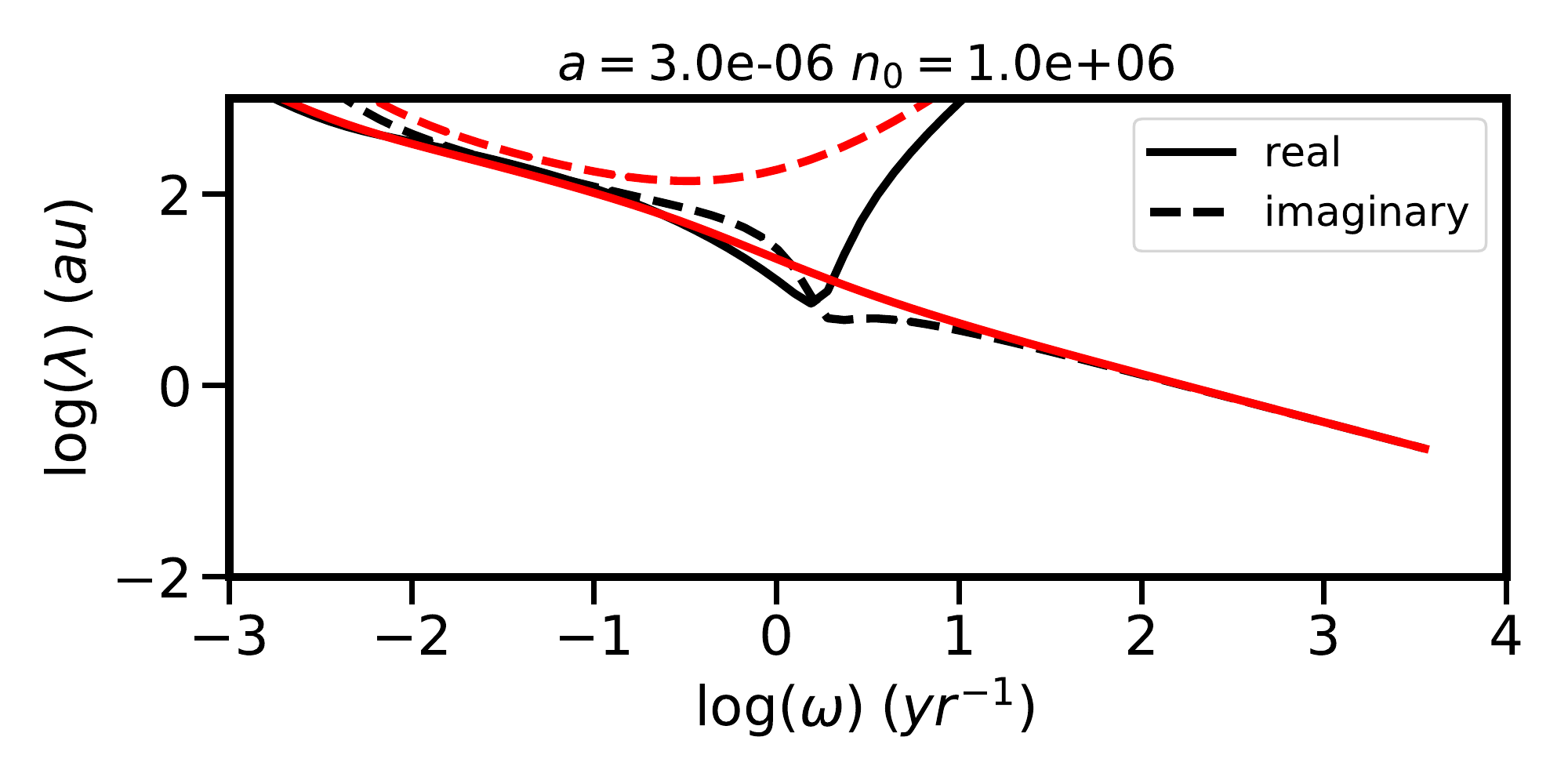}}  
\put(0,0){\includegraphics[width=8cm]{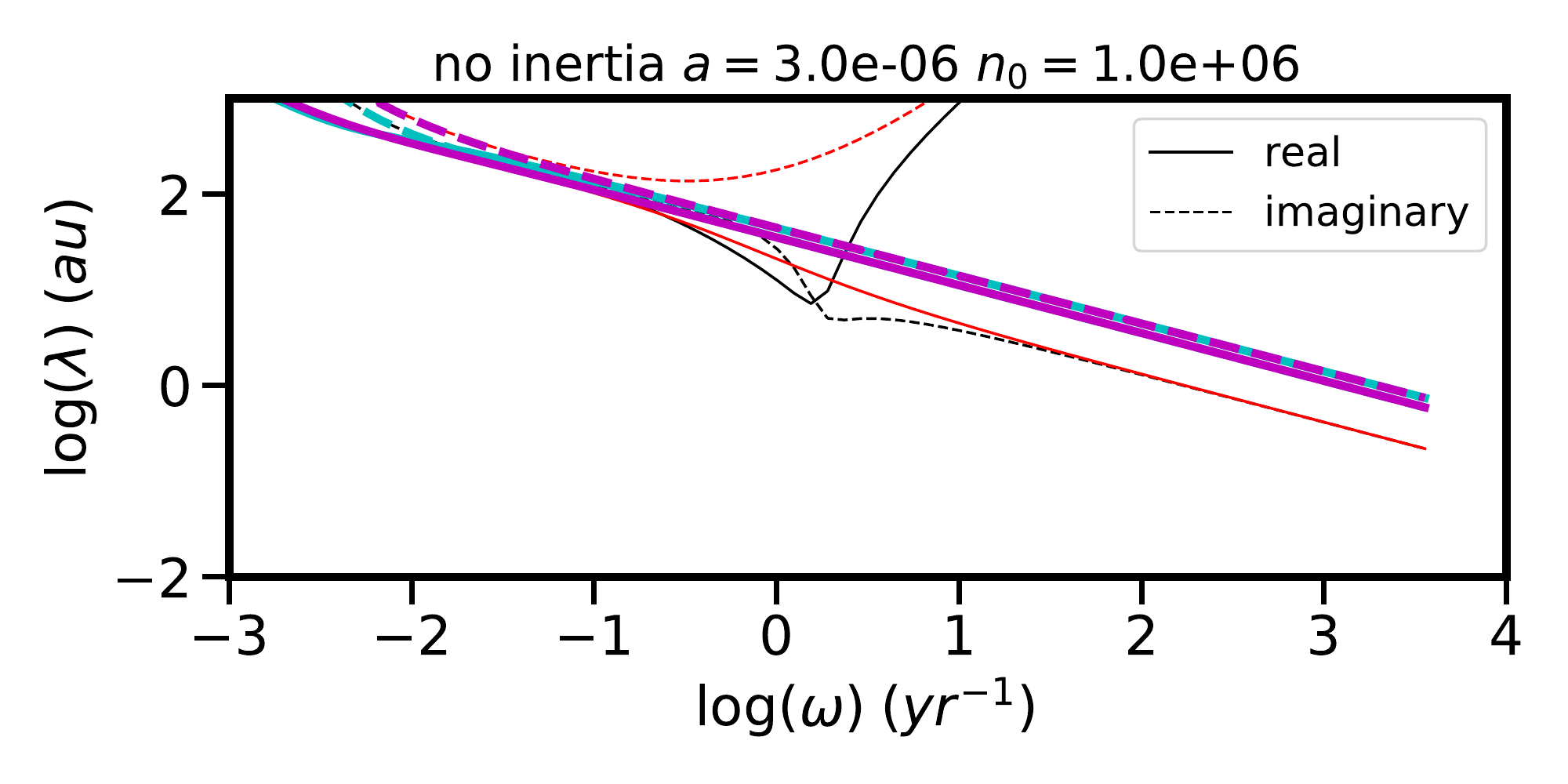}}  
\end{picture}
\caption{  Wavelength as a function of wave frequency for $n_0=10^6$ and a grain size $a=3 \times 10^{-6}$ cm
as stated by Eq.~(\ref{k_omeg4}). On top panel, the red and black lines represent the two solutions 
with grain inertia accounted for whereas on bottom panel the cyan and purple lines (which cannot be distinguished 
in this particular case) represent the solutions without grain inertia. To make comparison easier the the solutions of 
top panel have been reported (thin lines) on bottom panel.}
\label{lambomeg6}
\end{figure} 

A complementary description  to the dispersion  relation $\omega(\lambda)$, is provided by the 
$\lambda(\omega)$ relation as stated by Eq.~(\ref{k_omeg4}), which we see is a simple second order 
polynomials. The difference between the two approaches is that in the classical $\omega(\lambda)$, 
$\lambda$ is kept real while it is $\omega$ for the $\lambda(\omega)$ relation. Physically, this could 
 represent the waves generated by a perturbation at a fix spatial position and at a real frequency. The imaginary part of $k$
 then indicates the length over which the waves propagate. 
 
 Figure~\ref{lambomeg6} shows $\lambda$ as a function of $\omega$ for $a=3 \times 10^{-6}$ cm and $n_0=10^6$ cm$^{-3}$. 
 Top panel shows the two modes, '+'  is represented by dark lines whereas '-' is by red lines.
 Interestingly, a major difference, already noted by \citet{soler2013}, with the $\omega(\lambda)$ description, can be seen.
 The waves propagates at any wavelength including inbetween   $\lambda _{AW,diss,d} \simeq 120 $ au and  $\lambda _{AW,diss,n} \simeq 300$ au unlike 
 what happens in the $\omega(\lambda)$ description (strictly speaking the '+' mode propagates but its group velocity appears to be low).
 There is no contradiction, since the two descriptions correspond to two different physical situations.
 We see however that for $\lambda \simeq 200 $ au, the derivative $\partial _\omega \lambda$ is shallower. 
 Also in this region as expected, the dissipation is more significative as revealed by the imaginary part 
 which is comparable to the real one. At $\lambda \simeq \lambda _{AW,hall,n} \simeq 10 $ au, the two modes 
 split and for the '-'  polarisation we get $\lambda \propto \omega^{1/2}$ whereas for the '+' one (dark lines),  the waves are quickly damped. 
 
 Bottom panel of Fig.~\ref{lambomeg6} displays the no inertia case (purple line). For comparison, the results of the top panel 
 have been reproduced (thin lines). Neglecting inertia appears to be a good approximation for $\lambda < 100$ au. At smaller scales, 
 we see that the purple lines strongly differ from black and red ones. In particular, at a given $\omega$, the real part of 
 $\lambda$ is lower when the inertia is accounted for (except for high values of $\omega$ and for the '+' case).
 This indicates for instance that neglecting grain inertia leads to underestimate  the group velocity.


\subsection{Ion and grain velocities}
\label{velocity}

\setlength{\unitlength}{1cm}
\begin{figure}
\begin{picture} (0,8)
\put(0,4){\includegraphics[width=8cm]{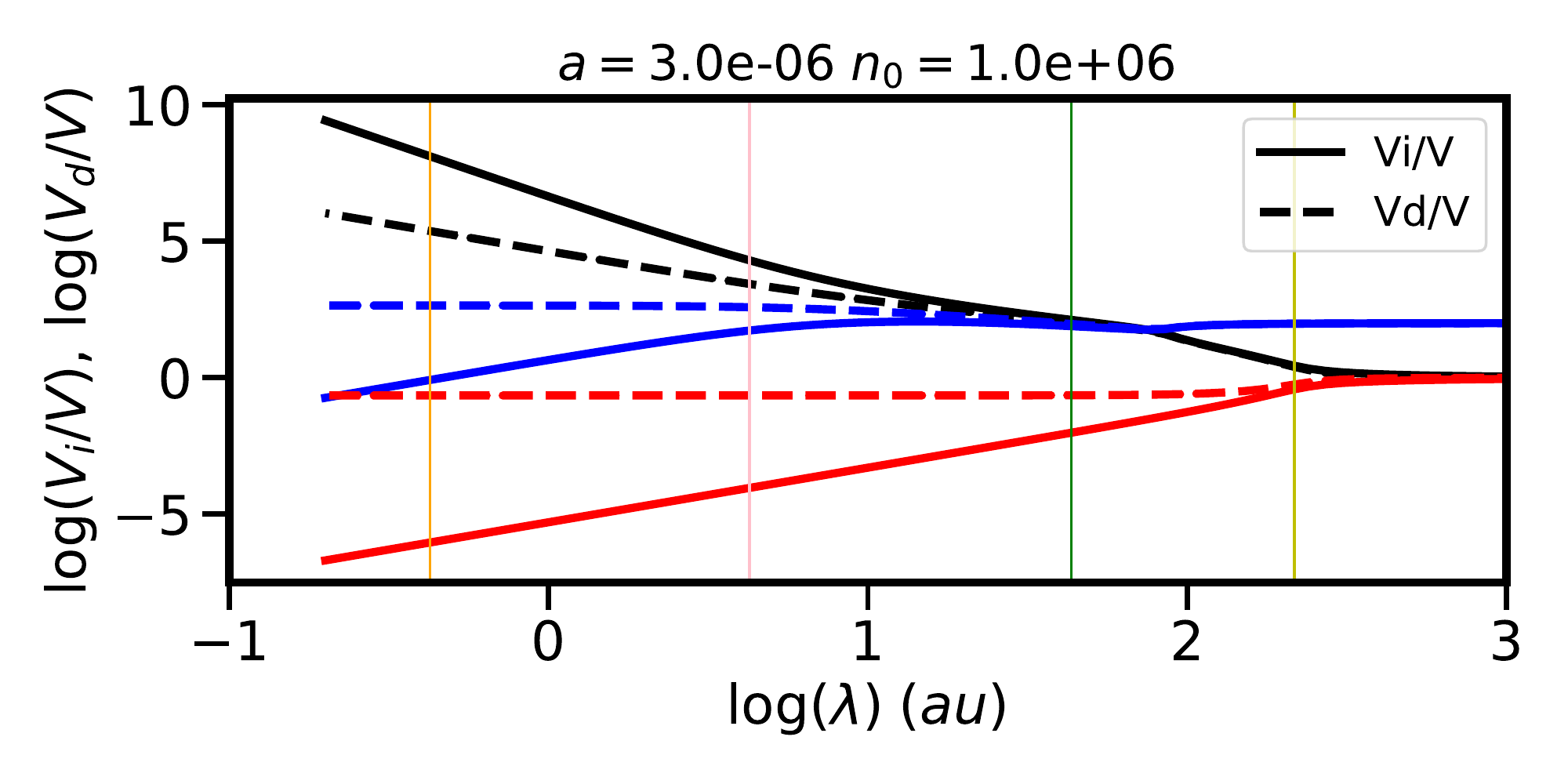}}  
\put(0,0){\includegraphics[width=8cm]{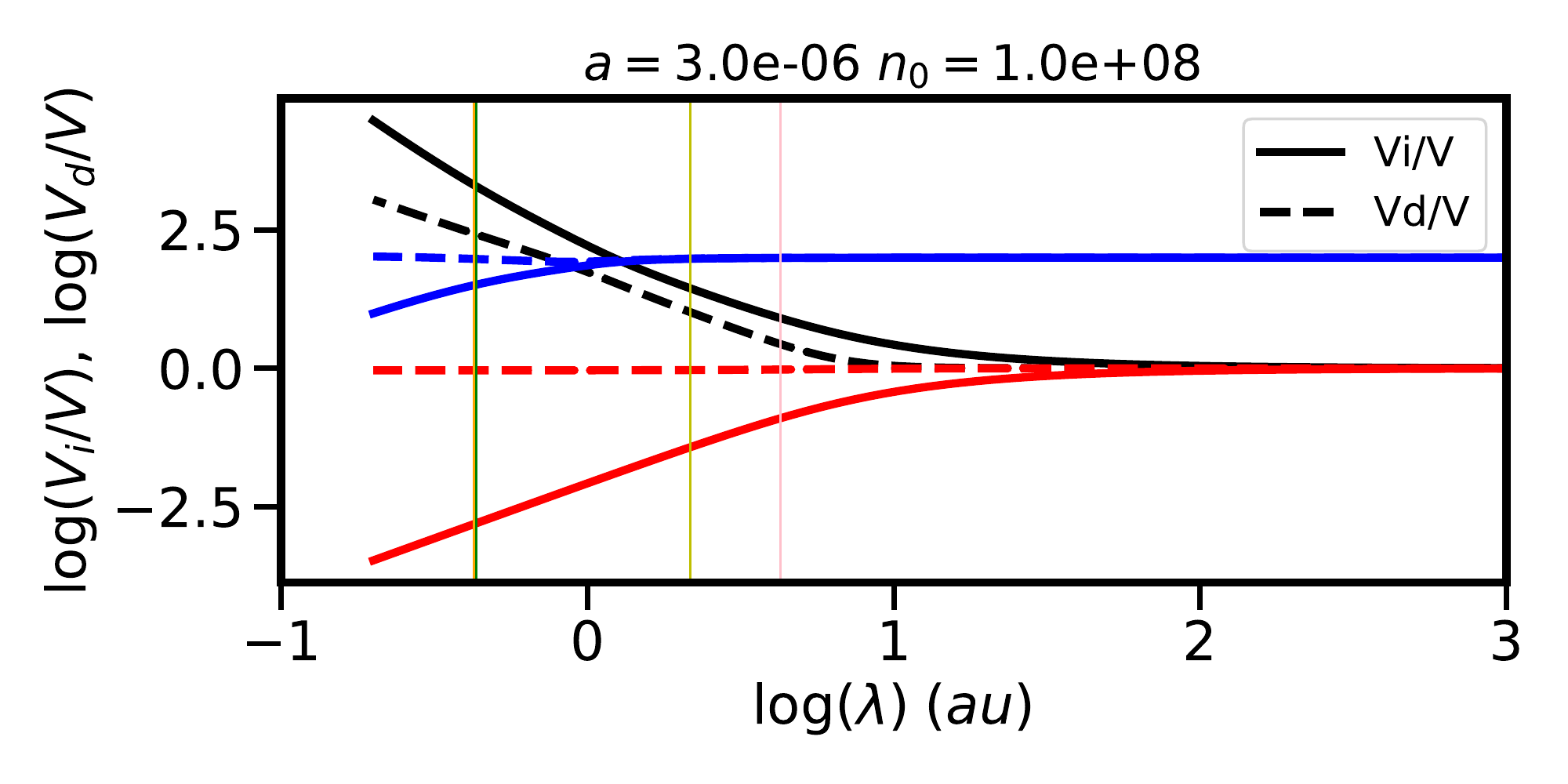}}  
\end{picture}
\caption{ Ion (solid lines) and grain (dashed line) velocities over neutral velocities as a function of wavelength.
Top panel displays the case of Fig.~\ref{3e-6n1e6}, i.e. $n_0=10^6$ cm$^{-3}$ and $a= 3 \times 10^{-6}$ cm. 
Bottom panel shows the one of Fig.~\ref{3e-6n1e8}, i.e. $n_0=10^8$ cm$^{-3}$ and $a= 3 \times 10^{-6}$ cm. 
The color coding is the same than the one used in  Fig.~\ref{3e-6n1e6} and Fig.~\ref{3e-6n1e8}. 
Whereas as large wavelength, the grains and the gas are dynamically coupled, at small scale, they become 
substantially different.}
\label{ViVd_V}
\end{figure} 

To get more insight regarding the physical meaning of the various waves, we look 
at the velocities of the various species. 
Figure~\ref{ViVd_V} portrays the ratio of the ion over neutral velocities, $V_i / V$ (solid lines) and 
grain over neutral velocities, $V_d/V$ (dashed lines). The 2 cases displayed in Figs.~\ref{3e-6n1e6} and \ref{3e-6n1e8} are shown with identical color coding.

Top panel corresponds to $n_0=10^6$ cm$^{-3}$ and $a= 3 \times 10^{-6}$ cm.  For $\lambda > \lambda_{AW,diss,n}$ (vertical yellow line) and for the two Alfv\'en modes (dark and red lines), the grains, the ions and the neutral have the same velocities since $V_i/V$ and $V_d/V$ are both equal to 1. The blue lines reveal that the corresponding mode is of different nature and why it is so dissipative. 
Whereas $V_i$ and $V_d$ are roughly equal, $V_d/V$ is about 100. This means that this modes correspond to dust and ions oscillating in almost static neutrals, therefore leading to heavy dissipation. At scale smaller than $\lambda_{AW,diss,n}$, whereas $V_i$ and $V_d$ are almost equal, 
$V_d / V$ is not equal to 1 any more. The neutrals and the dust are not dynamically coupled.  For the  dark and blue modes, the neutral velocity 
is much smaller than the dust one meaning that the Alfv\'en waves are carried by the dust grains whereas the neutrals are almost static. As seen from
Fig.~\ref{3e-6n1e6}, the dissipation is nevertheless limited (and drops with density) because the grains and the neutral are not dynamically well coupled. 
For the red mode, the neutrals are moving whereas the dust and the ions remain nearly static. Since the former do not feel directly the magnetic field, 
it is clear that no propagation is possible and thus the group velocity vanishes. At $ \lambda < \lambda _{AW,hall,d} $, the blue and dark modes (Alfv\'en waves carried by the dust), split. The ions and neutrals  decouple and  acquire different velocities.

In bottom panel, the case $n_0=10^8$ cm$^{-3}$ and $a= 3 \times 10^{-6}$ cm is displayed. Interestingly in these conditions, 
$\lambda _{AW,hall,n} > \lambda_{AW,diss,n}$. Thus the ion and dust velocities are different whereas in this range of densities the Alfv\'en waves are still
 carried by the neutrals that is to say the Alfv\'en speed is still given by $B / \sqrt{4 \pi \rho}$. At $\lambda \simeq \lambda _{AW,hall,d} $, the black and 
 blue modes eventually  split due to the Hall effect. 
 The velocity ratios remain much smaller than in the case presented in top panel (which has $n_0=10^6$ cm$^{-3}$) because 
 at a density of 10$^8$ cm$^{-3}$, the dust grains are much better coupled to the gas. 
 
 Altogether, we see that whereas at long wavelengths the dust grains, ions and neutrals are strongly coupled and have same velocities,  at short wavelengths,  they have completely different dynamics. More precisely, we expect that at wavelengths lower than the maximum between
 $\lambda_{AW,diss,n}$  and $\lambda _{AW,hall,n} $, the ions and the grains would have a velocity dispersion that is larger than the one of the neutrals.

\subsection{Group velocity and flux of momentum}

\setlength{\unitlength}{1cm}
\begin{figure}
\begin{picture} (0,8)
\put(0,4){\includegraphics[width=8cm]{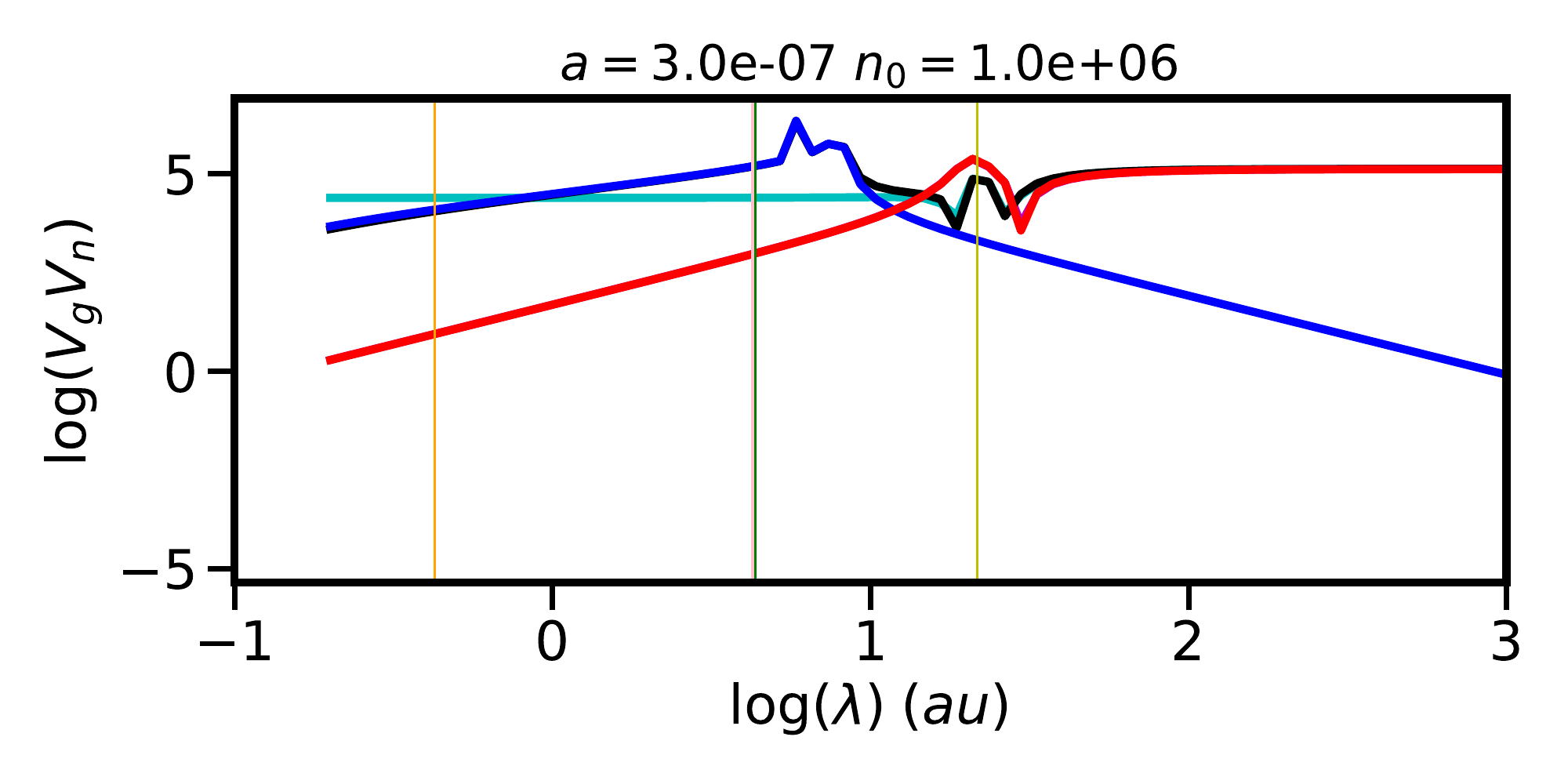}}  
\put(0,0){\includegraphics[width=8cm]{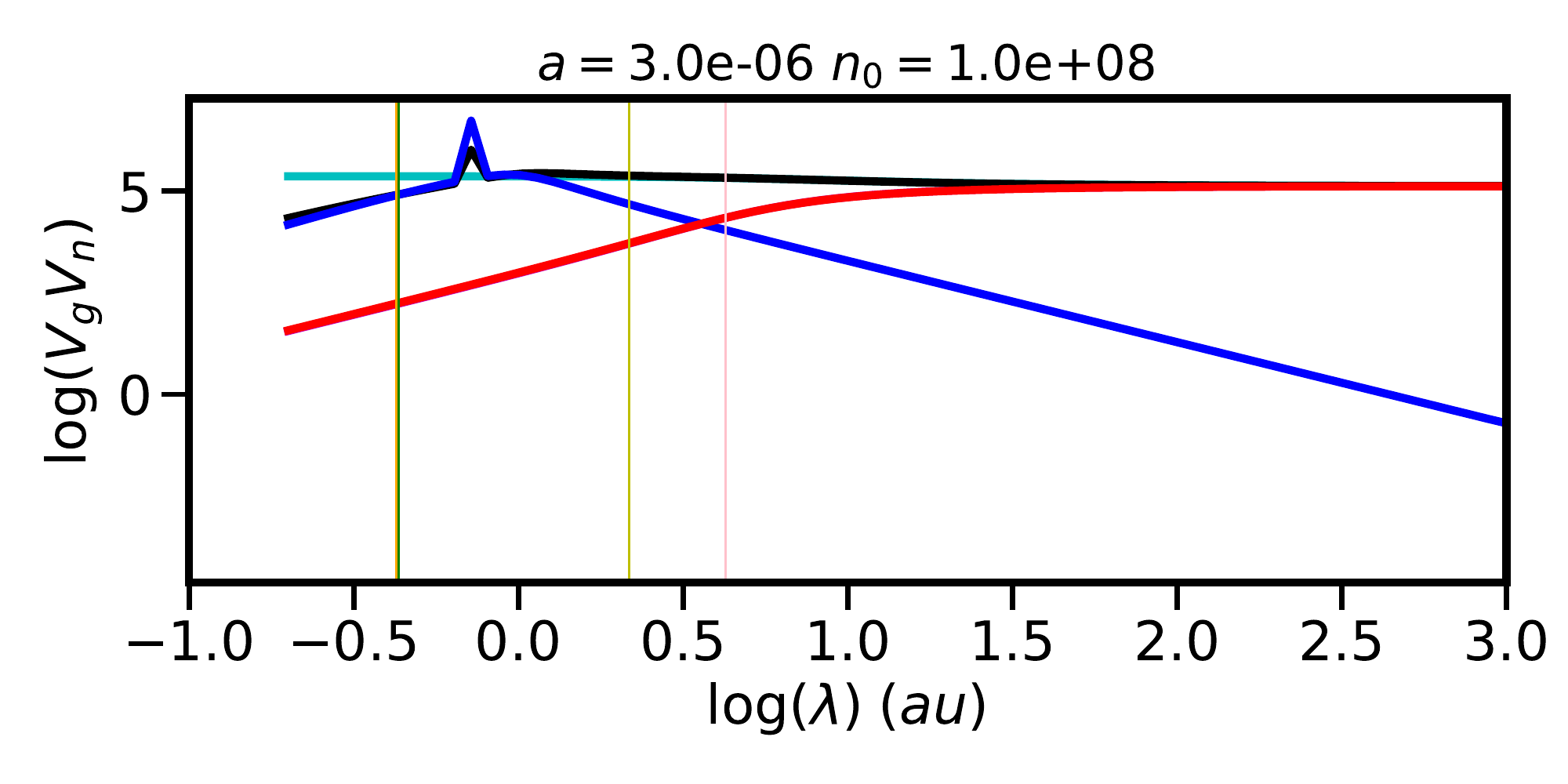}}  
\end{picture}
\caption{ Group velocity times neutral velocity (assuming a magnetic field $b=1$) as a function of wavelength. 
This quantity represents the flux of momentum transported by the wave.}
\label{VdVg_V}
\end{figure}

As recall in the introduction, in the context of protoplanetary disk formation, magnetic braking is believed to play a fundamental role \citep{zhao2020}.
In essence, magnetic braking is related to the flux of angular momentum that is carried away by Alfv\'en waves. 
A question, in particular, requires to be carefully investigated.
Whereas at large wavelengths, the neutrals and the dust grains are well coupled, it is not the case at small scales, as illustrated in Sect.~\ref{velocity} where
the neutrals have velocities which become  gradually smaller  compared to the grains and ions ones, as the wavelength decreases. On the other hand, at small scales the waves  present 
a group velocity, $v_g$,  which is higher than its value at large scales. This is because at small wavelengths, the waves propagate either at the Alfv\'en speed of the dust, 
i.e. $B_0 / \sqrt{4 \pi \rho_d}$ or at the velocity of whistler modes, namely $c B_0 / (2 \pi n_i Z_i e) k$, which increase with $k$. 
Since the flux of momentum is proportional to $v_g v_n$, it is therefore not straightforward to anticipate the actual trend.

Figure~\ref{VdVg_V} shows the product $v_g v_n$ as a function of $\lambda$ for the two cases displayed in Figs.~\ref{3e-7n1e6} and \ref{3e-6n1e8} .
To calculate $v_n$, it is assumed that $b=1$ which is therefore assumed to be a reference. 
The color coding is identical as the one of Figs.~\ref{3e-7n1e6} and \ref{3e-6n1e8}.
The dark and deep blue correspond to the case with inertia while the cyan and purple (not seen because identical to the red mode) ones to the case without inertia. 
For $n_0=10^6$ cm$^{-3}$ (top panel), for $\lambda < \lambda_{AW,diss,d}$ (green vertical lines) 
the flux of momentum, $v_g v_n$, increases with $\lambda$, meaning that magnetic braking is likely more efficient at larger scales.
Interestingly, we see that at wavelengths between 3 and 10 au, the flux of momentum, $v_g v_n$, is slightly larger than 
its value at larger wavelengths.  Note that the spikes seen for instance in top panel of Fig~\ref{ViVd_V}  around 50 au are due to the stiff variations of 
$\omega$ with $k$.  

The cases with and without inertia  are again quite different for $\lambda < \lambda_{AW,diss,d} $. 
First, we see that whereas the dark and deep blue modes have similar $v_g v_n$ (for $\lambda > \lambda_{AW,hall,d} $), the behaviour 
of the cyan  one is very different. Also at $\lambda \simeq 7$ au, the momentum flux, $v_g v_n$, is more than an order of magnitude smaller when 
inertia is neglected than when it is taken into account. 
This may indicate that when inertia is taken into account, as it should, the flux of angular momentum is more symmetrical for the two polarisations, therefore possibly leading to 
 less asymmetry for the disk formed in the aligned and anti-aligned configurations \citep[e.g.][]{tsukamoto2015,wurster2016,leeyn2021}.

The situation for $n_0=10^8$ cm$^{-3}$ (bottom panel) is different. For $\lambda > \lambda_{AW,hall,d} $ (orange line), the deep blue (with inertia) and cyan (without intertia)
modes are almost identical. One difference is that for $\lambda < \lambda_{AW,diss,n} $ (yellow line), the black mode (which corresponds to waves carried by dust), 
is not negligible, implying here again, that 2 modes  are contributing to the transport of momentum when inertia is accounted for, instead of one when inertia 
is not taken into account.

\section{Alfv\'en waves propagation in the dense ISM}
\label{msg}
We now investigate the more complex but more realistic multi-size grain ISM.

\subsection{Method}
The dispersion  relation for the multi-grain Alfv\'en waves is given by Eqs.~(\ref{disp_rel1})
 and~(\ref{AAnia}) which are complex equations. In particular, whereas from Eq.~(\ref{disp_rel2}), it is straightforward 
 to compute $k$ once $\omega$ is specified, getting $\omega$ for a given $k$ is considerably more complex. Indeed, 
 Eqs.~(\ref{disp_rel1}) is  an integral equation. When the number of bin of grains is finite,  Eqs.~(\ref{disp_rel1}) is 
 formally a high order polynomials. 
 To get the dispersion relation, we use an iterative method. First, we specify a real part for $\omega$ and a series of 200 
  imaginary parts logarithmically  distributed between ${\cal R}e (\omega) \times 10^{-4}$ and ${\cal R}e (\omega) \times 10^{4}$. For all these values, 
  we calculate $k$ using Eq.~(\ref{disp_rel2}). As the goal is to obtain a dispersion relation, we seek for real values of $k$.
  Thus we select the $\omega$ for which the imaginary part of $k$ is changing sign between two values of ${\cal I}m (\omega)$. We then iterate using a simple dichotomy
  iterative method until convergence is reached.    We have varied the number of bins of imaginary parts and found that increasing their number above 200 leads
  to identical results. We stress in particular, that whereas the total number of existing roots is proportional to the number of grain bins, most of these roots are likely not found because
  they are very dissipative weakly propagative waves and are therefore outside the search interval ($ \| \log ({\cal R}e (\omega) /  {\cal I}m (\omega) ) \| < 4$).

  In spite of the complexity of Eq.~(\ref{disp_rel1}), in many cases we found a unique solution (for each sign "+/-"). In particular we do not find 
  the highly dissipative zero group velocity waves, that were identified with single size grain. Likely enough this is because the range of search for ${\cal I}m(\omega)$
  is too restricted. It seems likely that there are many of this type of modes when a large number of size of grains is considered.

  Various configurations are explored below. We have used 1000 bins of grains and asked for a convergence on 
  ${\cal I}m(\omega)$ better than $10^{-3}$. In Sect~\ref{num_methode}, the influence of these parameters is explored. 
To assess the method, we have checked that for single size grain, 
  Eqs.~(\ref{disp_rel1}) and Eq.~(\ref{rel_disp_simpl8}) give indisguishable results.
   
\subsection{A fiducial case: $n_0=10^6$ cm$^{-3}$, $a_{min}=10^{-7}$ cm, $a_{max}=10^{-5}$ cm }   

\setlength{\unitlength}{1cm}
\begin{figure}
\begin{picture} (0,16)
\put(0,12){\includegraphics[width=8cm]{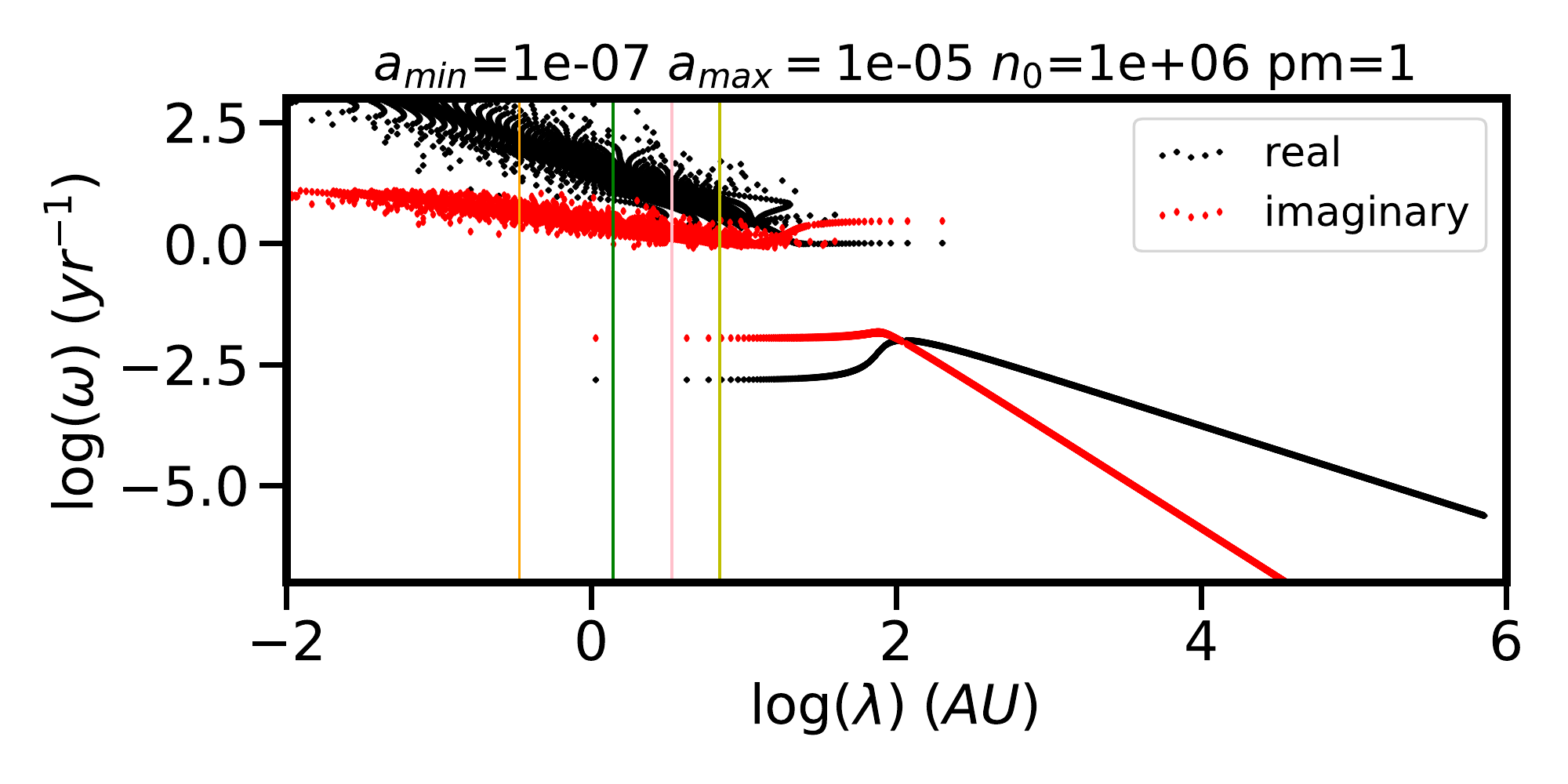}}
\put(0,8.3){\includegraphics[width=8cm]{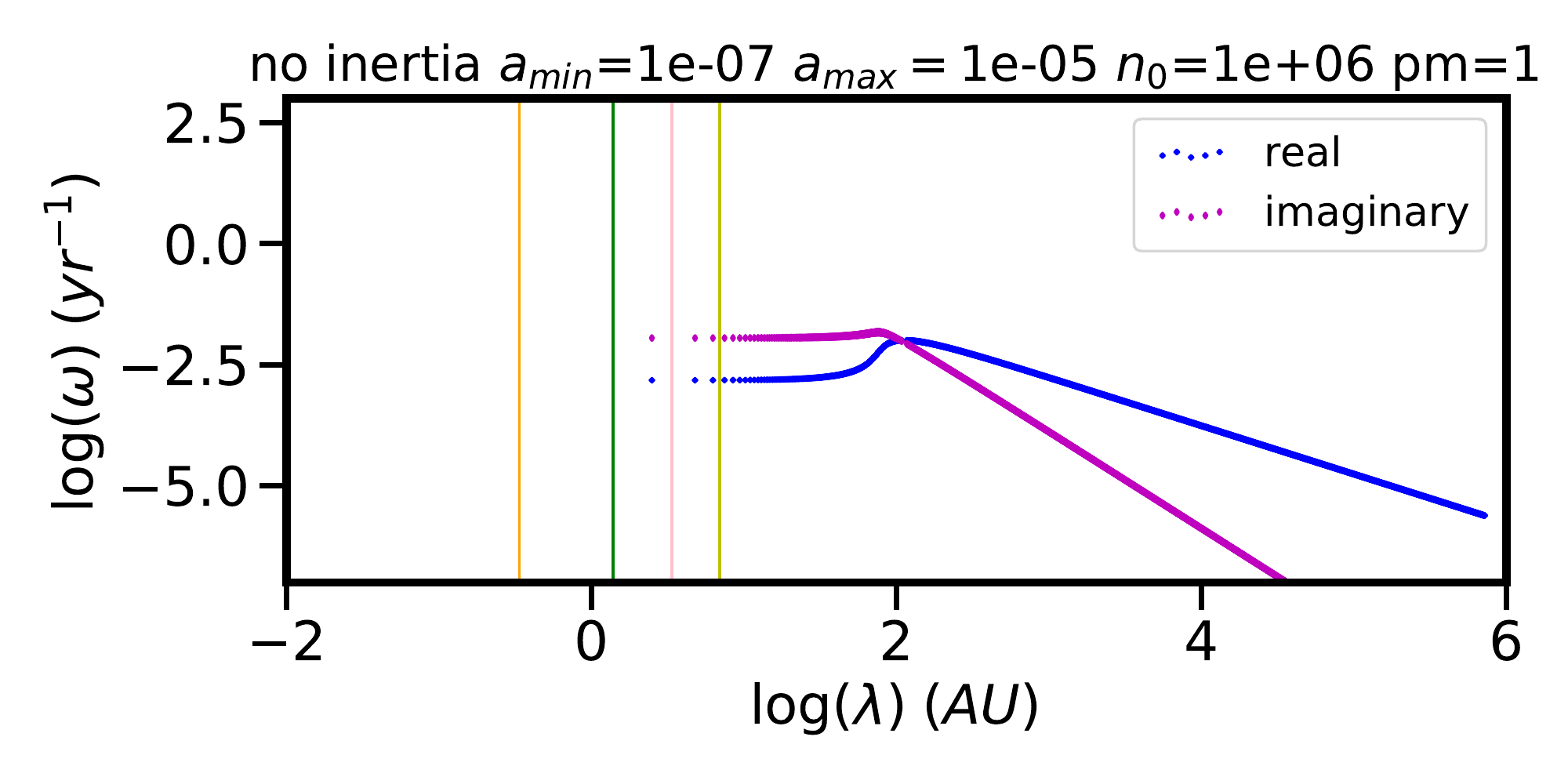}}
\put(0,3.7){\includegraphics[width=8cm]{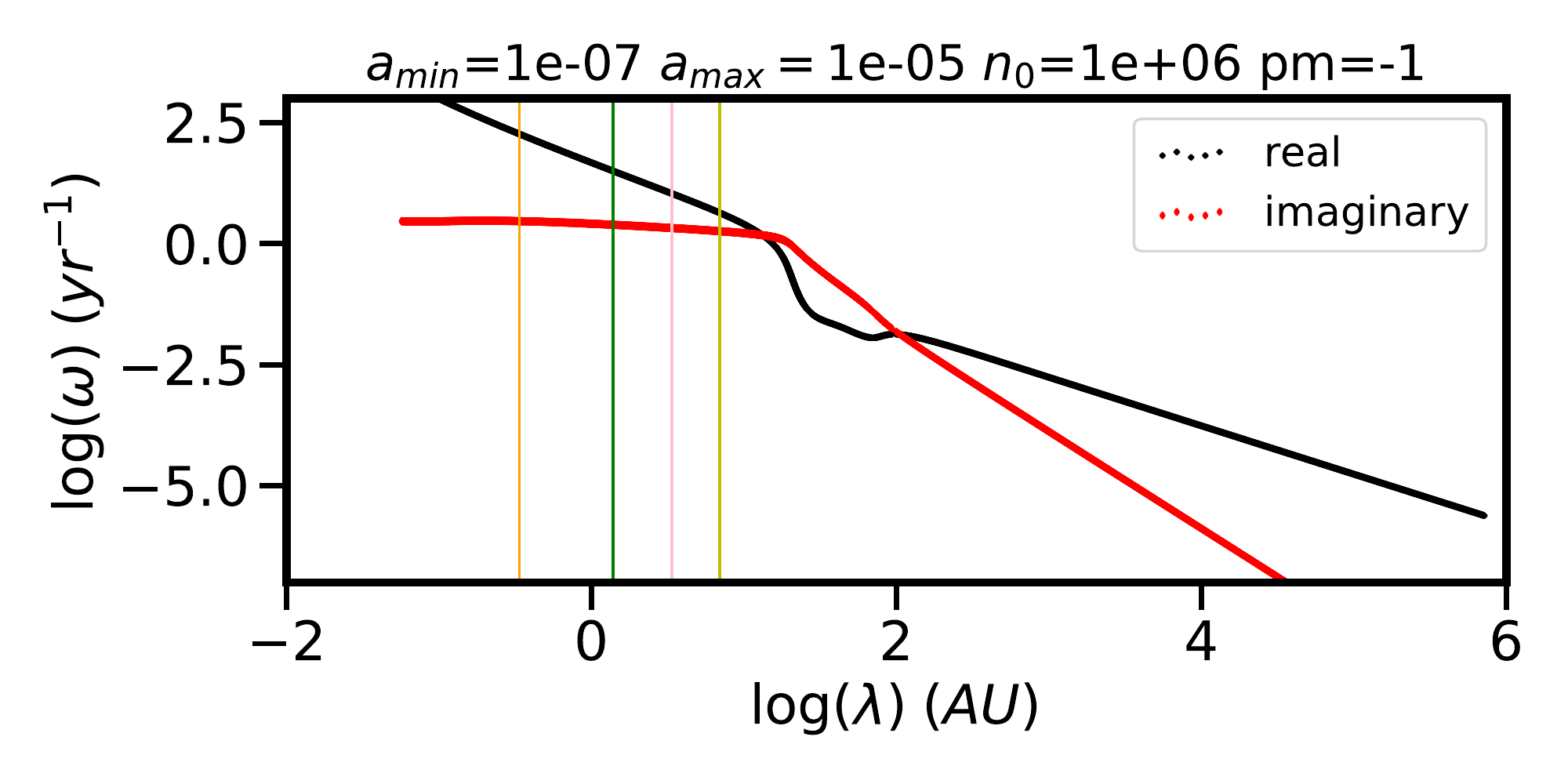}}
\put(0,0){\includegraphics[width=8cm]{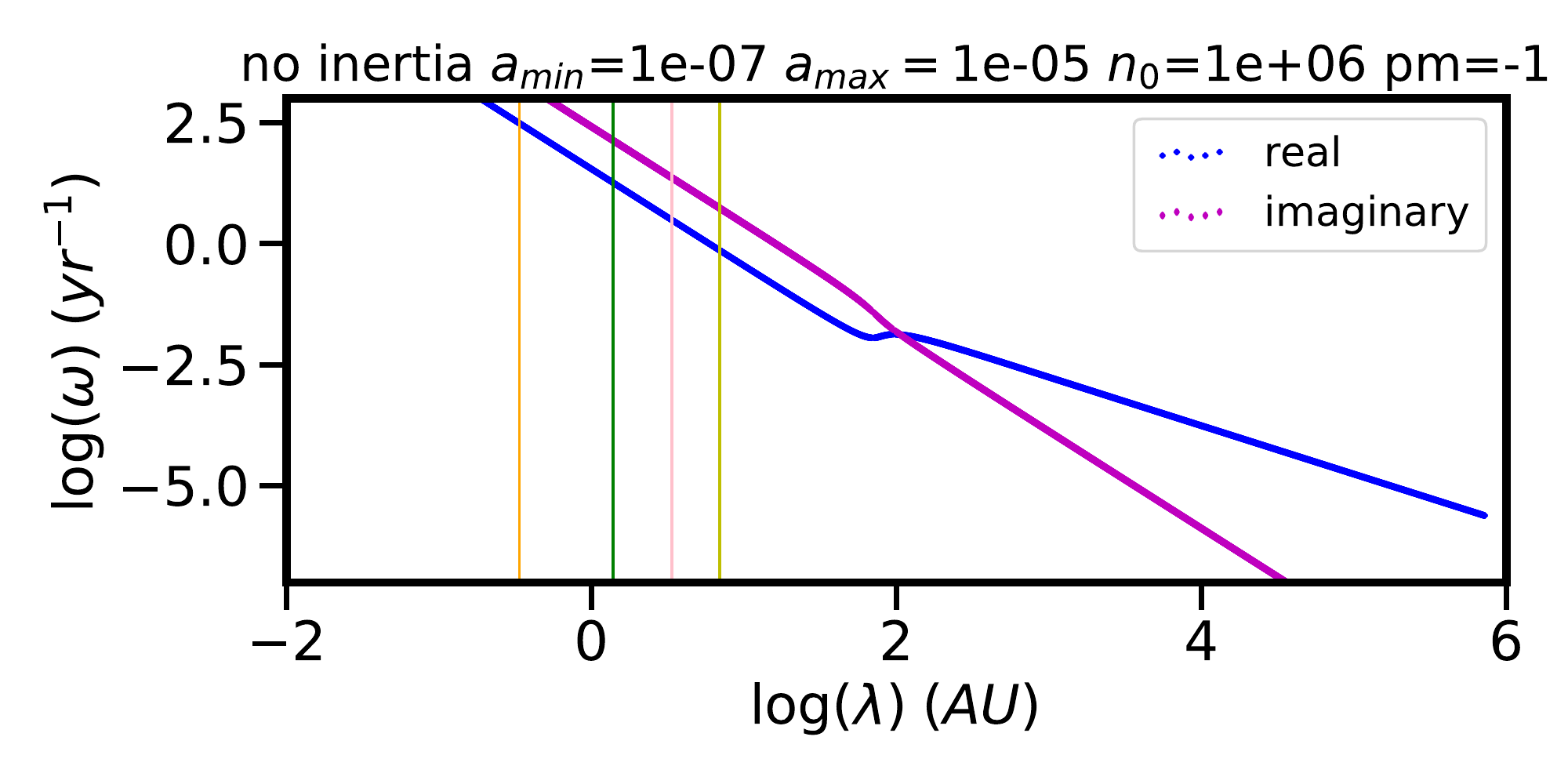}}
\end{picture}
\caption{  Wave frequency as a function of wavelengths for $n_0=10^6$ cm$^{-3}$ and $a_{min}=1 \times 10^{-7}$ cm $a_{max}=1 \times 10^{-5}$ cm.
First panel displays the real and imaginary part of the frequency for the "+" circularly polarised mode. Second panel shows the result when 
grain inertia is neglected. 
Third and fourth panels same as for first and second but displaying the "-" circularly polarised mode. }
\label{multn1e6}
\end{figure} 

We start by discussing in detail a case that we consider as being sufficiently typical. The parameters are 
$n_0=10^6$ cm$^{-3}$, $a_{min}=10^{-7}$ cm, $a_{max}=10^{-5}$ cm, $\lambda _{mrn}=-3.5$, $\beta _{mag}=0.1$ and $\zeta= 5 \times 10^{-17}$.

 Figure~\ref{multn1e6} portrays $\omega$ as a function of $\lambda$. 
 First-top panel is for the "+" mode (pm=1) and grain inertia is accounted for, that is to say Eqs.~(\ref{disp_rel1})-(\ref{AAnia}) are being solved.
 Second-top panel is identical to first-top except that grain inertia is neglected, which mathematically means that in 
 Eq.~(\ref{AAnia}) the terms $i \omega$ in $A^\pm_{n,a}$ and $A^\pm_{i,a}$ have been dropped.
 
 At large wavelengths, i.e. $\lambda  > 100$ au, we have a standard Alfv\'en wave, almost identical to the single size grain case. 
 Neglecting the inertia here is entirely justified, the real and imaginary parts are almost identical with and without inertia.
 
 At scale $\lambda < 100 $ au, we see in top panel a new behaviour. A branch which presents a  dispersion relation ($\omega$ vs $\lambda$) 
 broadly
 similar to an Alfv\'en wave carried by the grains (i.e. $v_a = B_0 / \sqrt{4 \pi \rho_d}$) is found. However, at a given $\lambda$, 
 there seems to be many, likely a continnum, of $\omega$ values. Mathematically, this is a clear consequence of the non-polynomial
 nature of  Eqs.~(\ref{disp_rel1})-(\ref{AAnia}) which allows a large number of solutions. Physically, this clearly comes from the 
 grain distribution. Each bin of grains has its own Alfv\'en velocity and coupling with the gas. At large wavelengths, 
 the grains are synchronized by their coupling with the neutrals but at small wavelengths, particularly for 
 $\lambda < \lambda_{AW,diss,n}$ we do not expect this to be necessarily the case.
 As what has been found with single size grains (Fig.~\ref{3e-7n1e6}), the discrepancy with the no inertia case is obvious. No solution, apart for the ZGV ones, are found for $\lambda < 100$ au. 
 This is because when grain inertia is neglected, the grains cannot carry waves independently of the neutrals.
 As for single size grain, this implies that at small scales, waves can propagate and therefore angular momentum for instance 
 could possibly be transported whereas this is not possible if inertia is neglected. 
 
 Because of the grain size distribution, the meaning of $\lambda_{AW,diss,n}$ is less obvious. The yellow line represents
 the value of   $\lambda_{AW,diss,n}$  obtained for $a_{min}$. The agreement between its value and the 
 $\lambda$ at which the group velocity vanishes, is much less good than in Fig.~\ref{3e-7n1e6} for instance, as 
 $\lambda_{AW,diss,n}$ is about 10 times lower. This is because larger grains have a larger  $\lambda_{AW,diss,n}$.
 
 Even more importantly, we do not see a sign of a change of behaviour around 
  $\lambda_{AW,hall,n}$ or  $\lambda_{AW,hall,d}$. Although the distribution of $\omega$ values precludes 
  a clear measurement, the mean $\omega$ appears to be broadly proportional to $\lambda^{-1.1}$-$\lambda^{-1.2}$, that 
  seems to be intermediate between the standard Alfv\'en and whistler  waves. 
  
  The third-top panel portrays the "-" mode (pm=-1). For $\lambda$ > 100 au, it behaves as the "+" mode. 
  The behaviour for $\lambda$ < 100 au, differs from  the "+" mode (first-top panel).  For all 
  $\lambda$, including below 100 au, there is now a unique   $\omega$. 
  Interestingly, we find that whereas for large wavelengths, ${\cal R}_e (\omega) \propto \lambda^{-1}$, at
  short wavelength we approximately get  ${\cal R}_e (\omega) \propto \lambda^{-1.2}$, which again 
  indicates a behaviour inbetween standard Alfv\'en and whistler modes.
 
  On the other-hand, bottom panel, which displays the case without grain inertia, reveals that
  at short wavelength, $\omega \propto \lambda^{-2}$ as expected for  whistler modes.
  This clearly shows that at short wavelengths,  neglecting inertia leads to 
  incorrect predictions.

\subsection{The effect of density}

\setlength{\unitlength}{1cm}
\begin{figure}
\begin{picture} (0,16)
\put(0,12){\includegraphics[width=8cm]{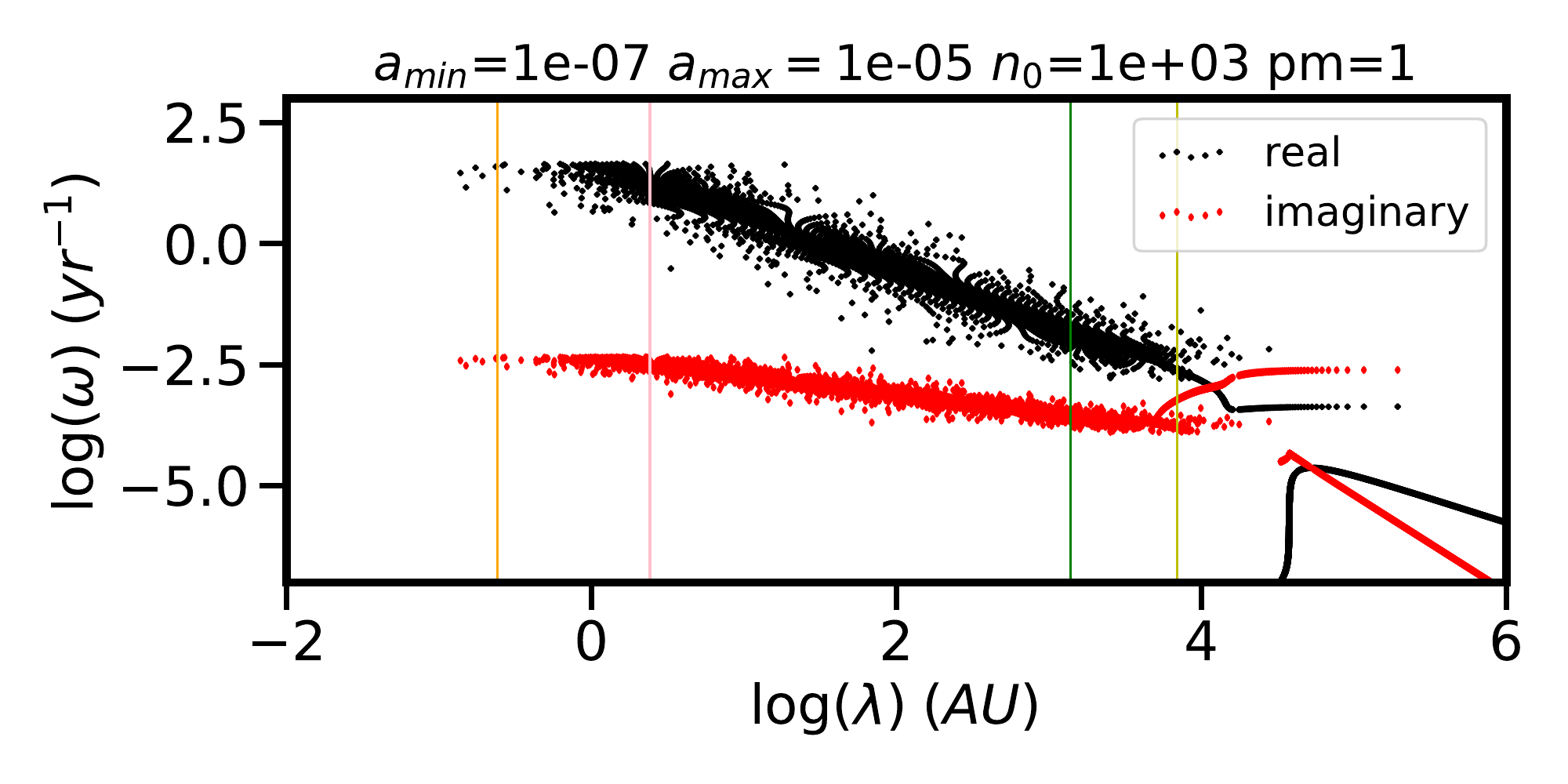}}
\put(0,8.3){\includegraphics[width=8cm]{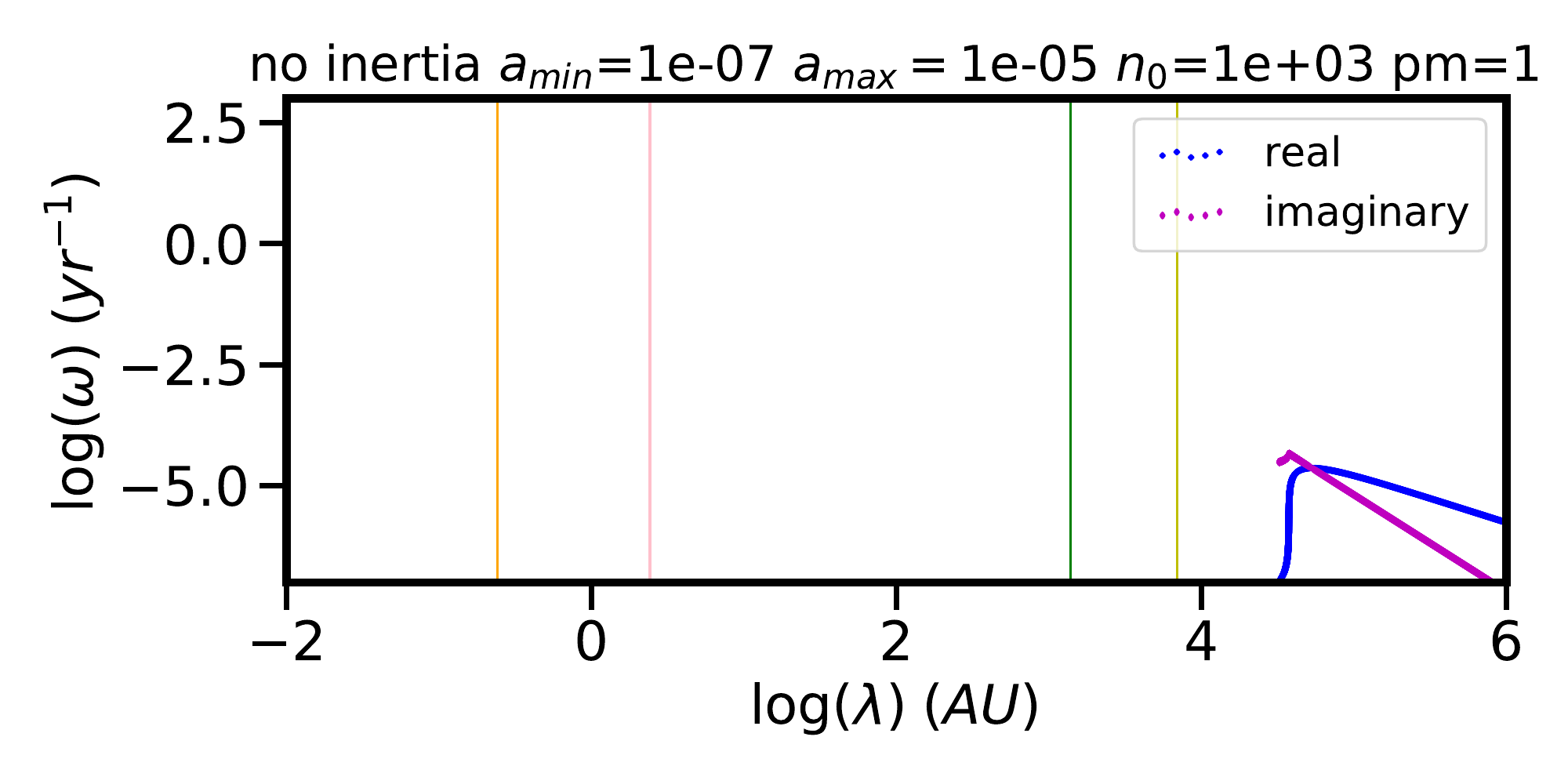}}
\put(0,3.7){\includegraphics[width=8cm]{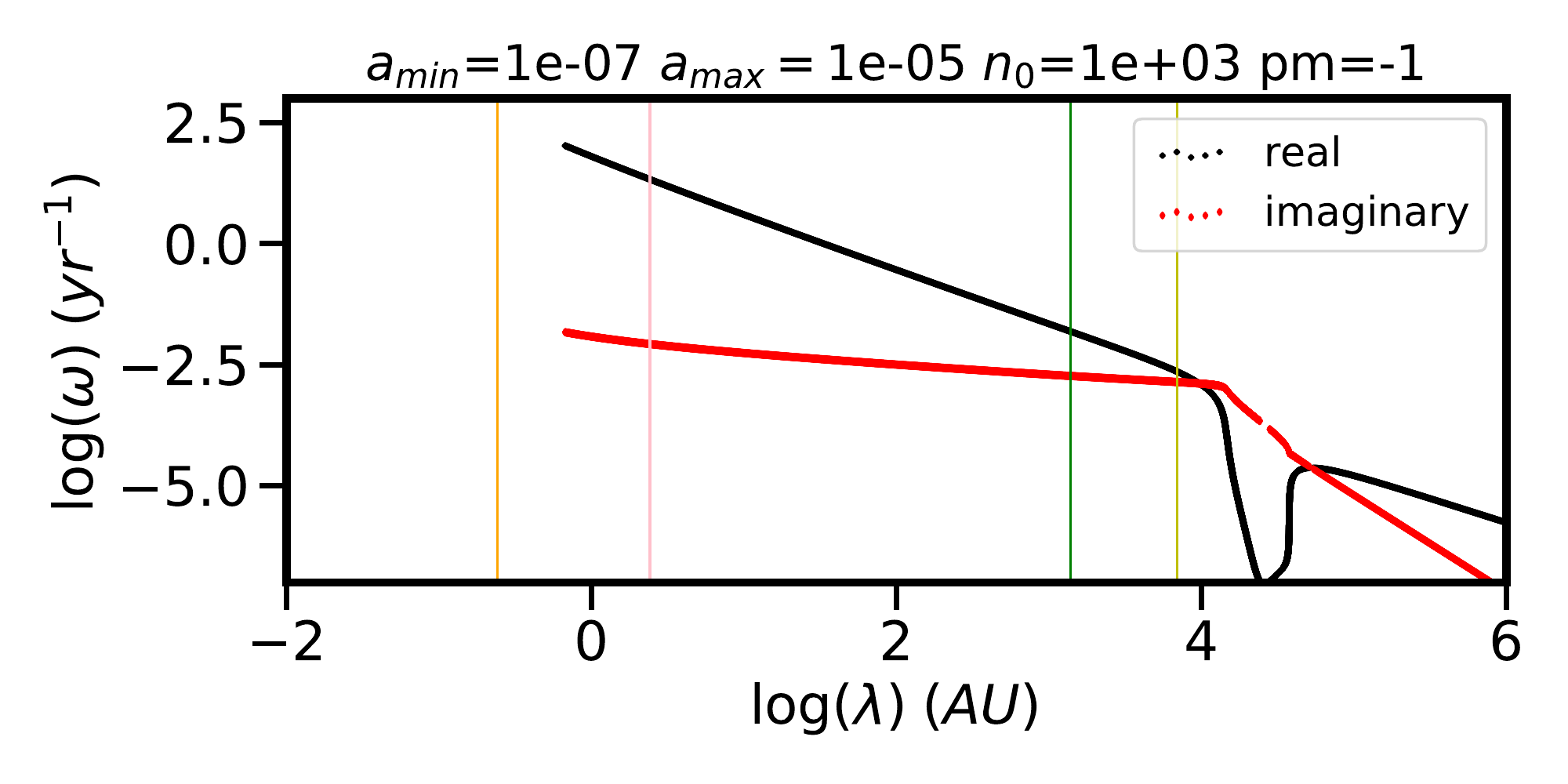}}
\put(0,0){\includegraphics[width=8cm]{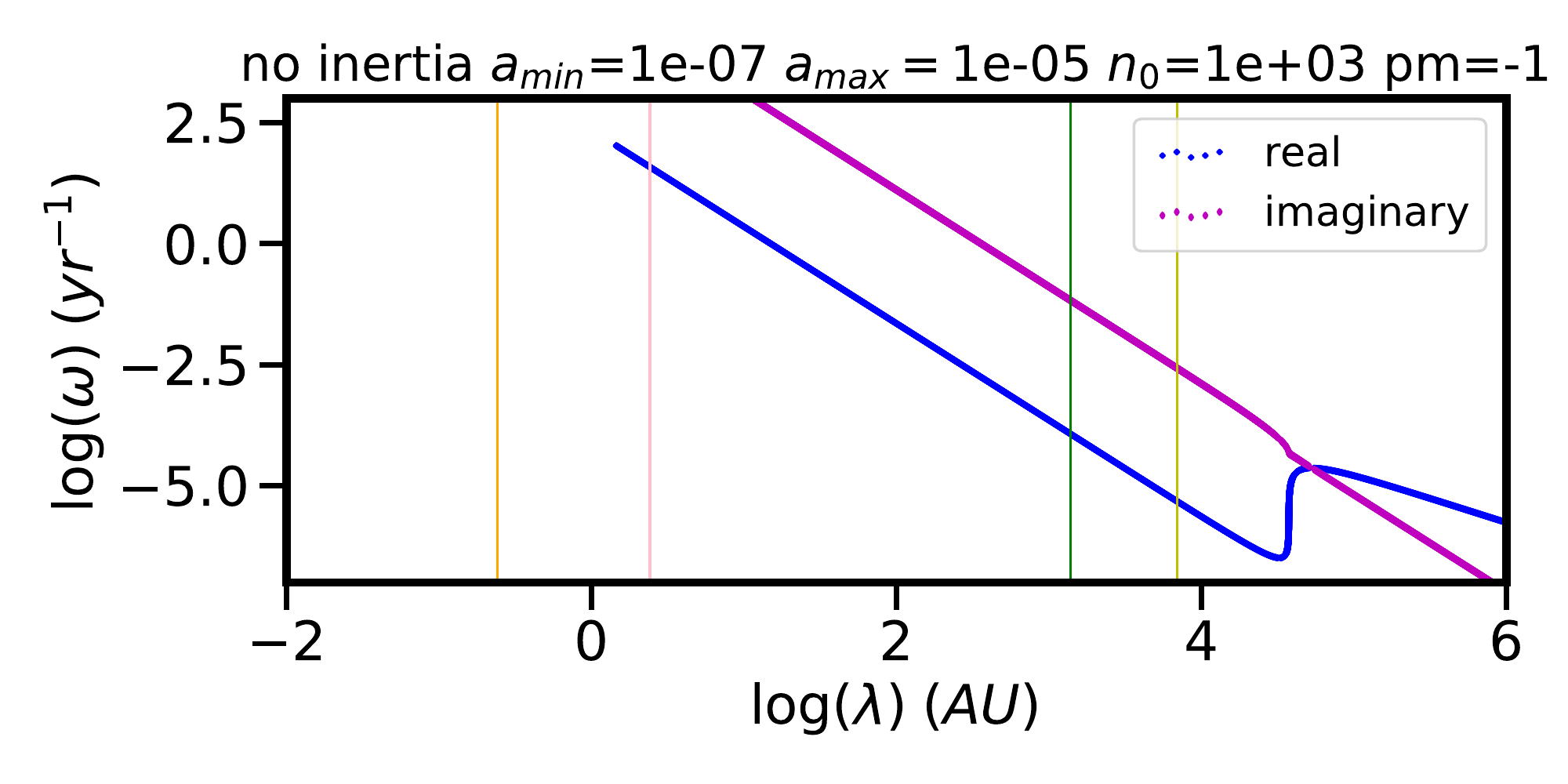}}
\end{picture}
\caption{  Same as Fig.~\ref{multn1e6}
for $n_0=10^3$ cm$^{-3}$ and $a_{min}=1 \times 10^{-7}$ cm $a_{max}=1 \times 10^{-5}$ cm.  }
\label{multn1e3}
\end{figure}

\setlength{\unitlength}{1cm}
\begin{figure}
\begin{picture} (0,16)
\put(0,12){\includegraphics[width=8cm]{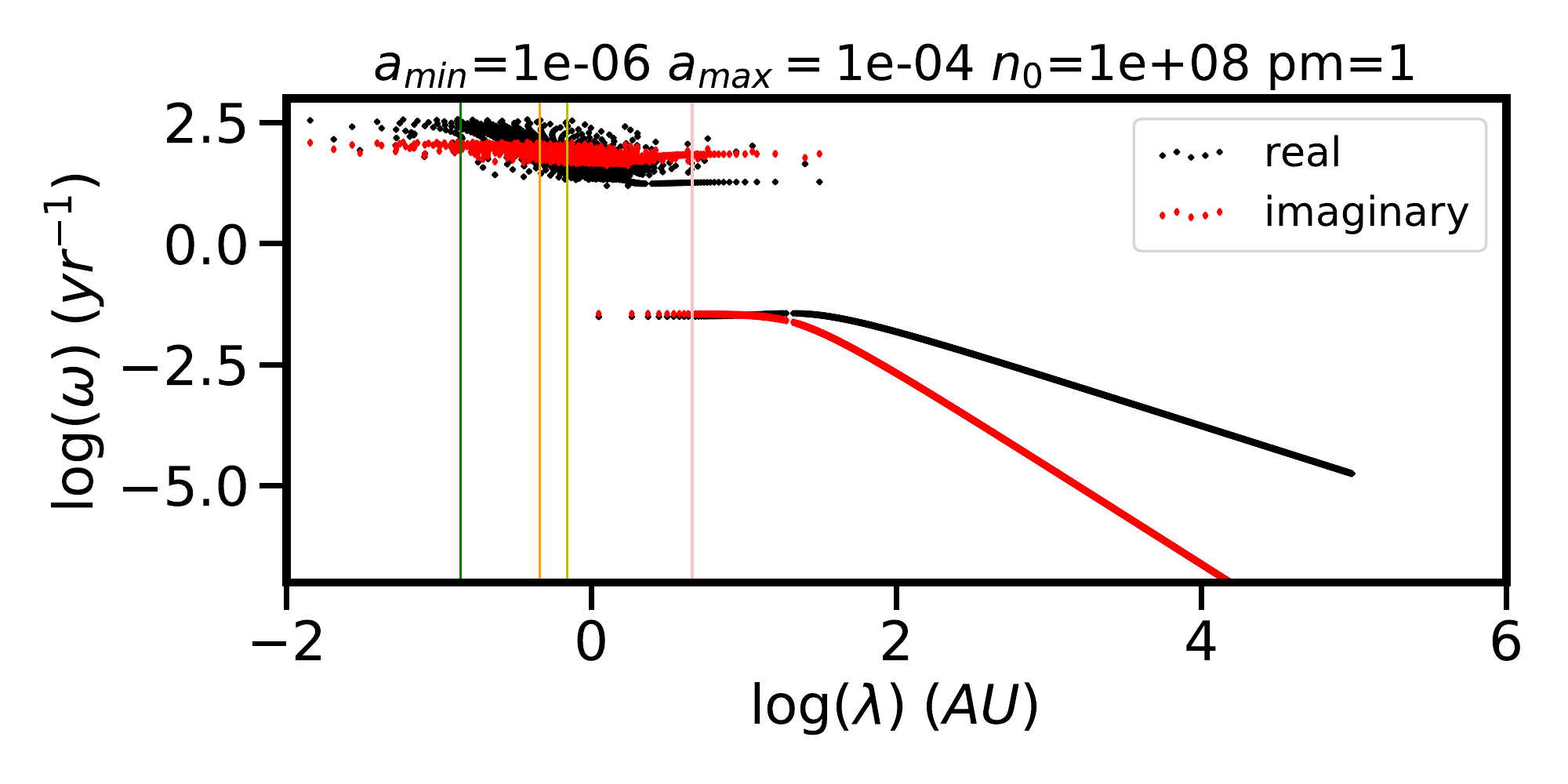}}
\put(0,8.3){\includegraphics[width=8cm]{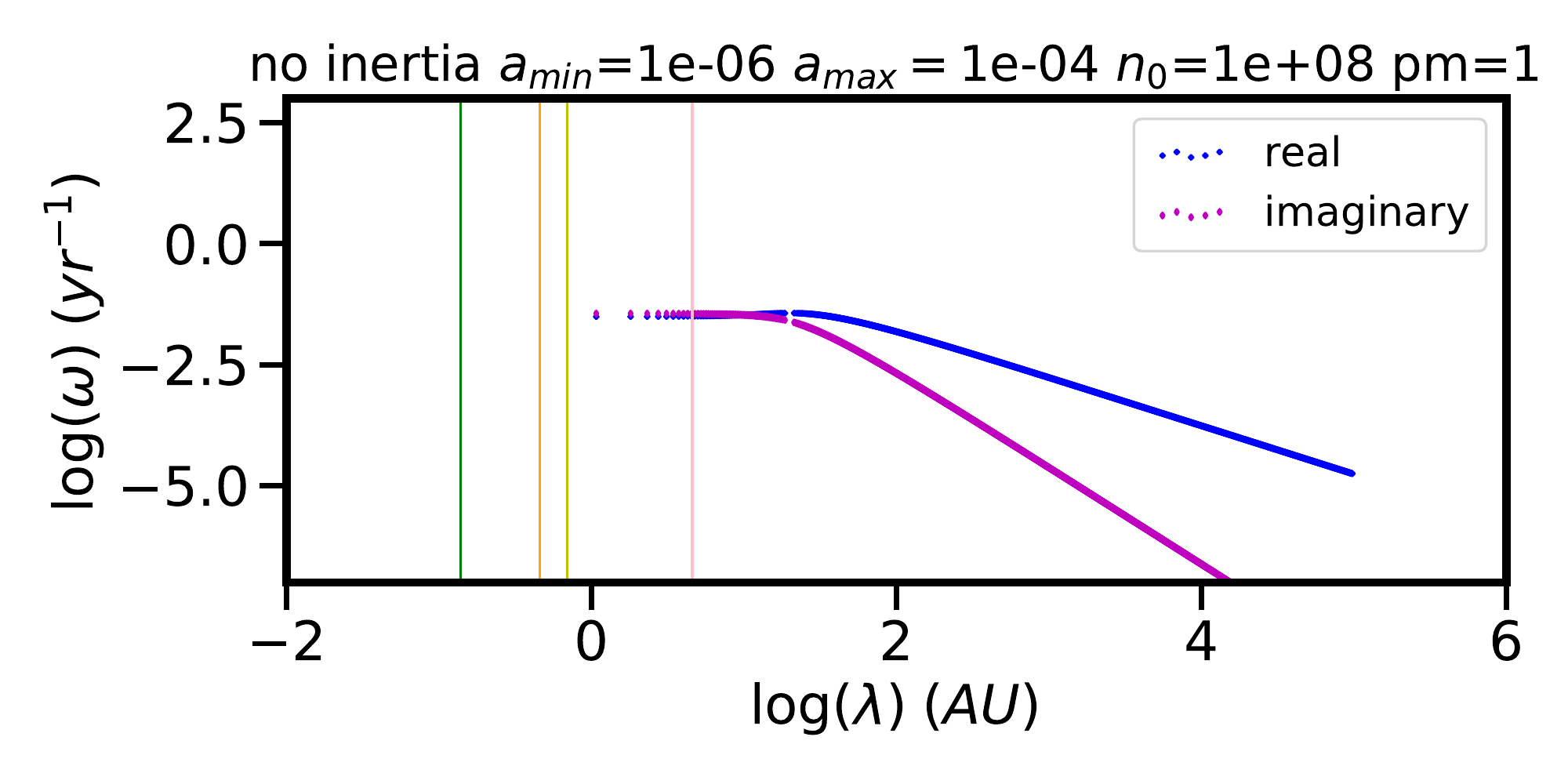}}
\put(0,3.7){\includegraphics[width=8cm]{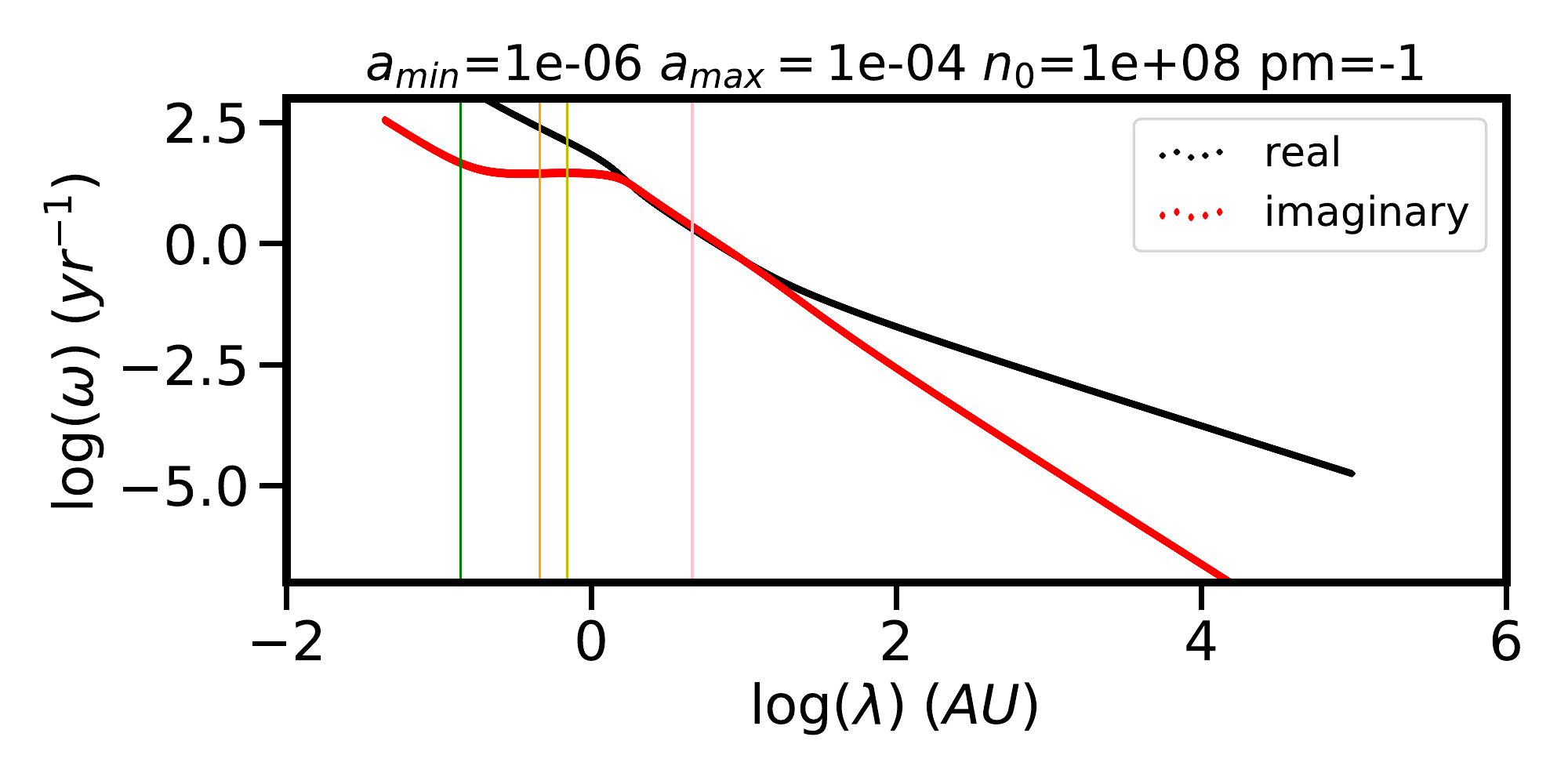}}
\put(0,0){\includegraphics[width=8cm]{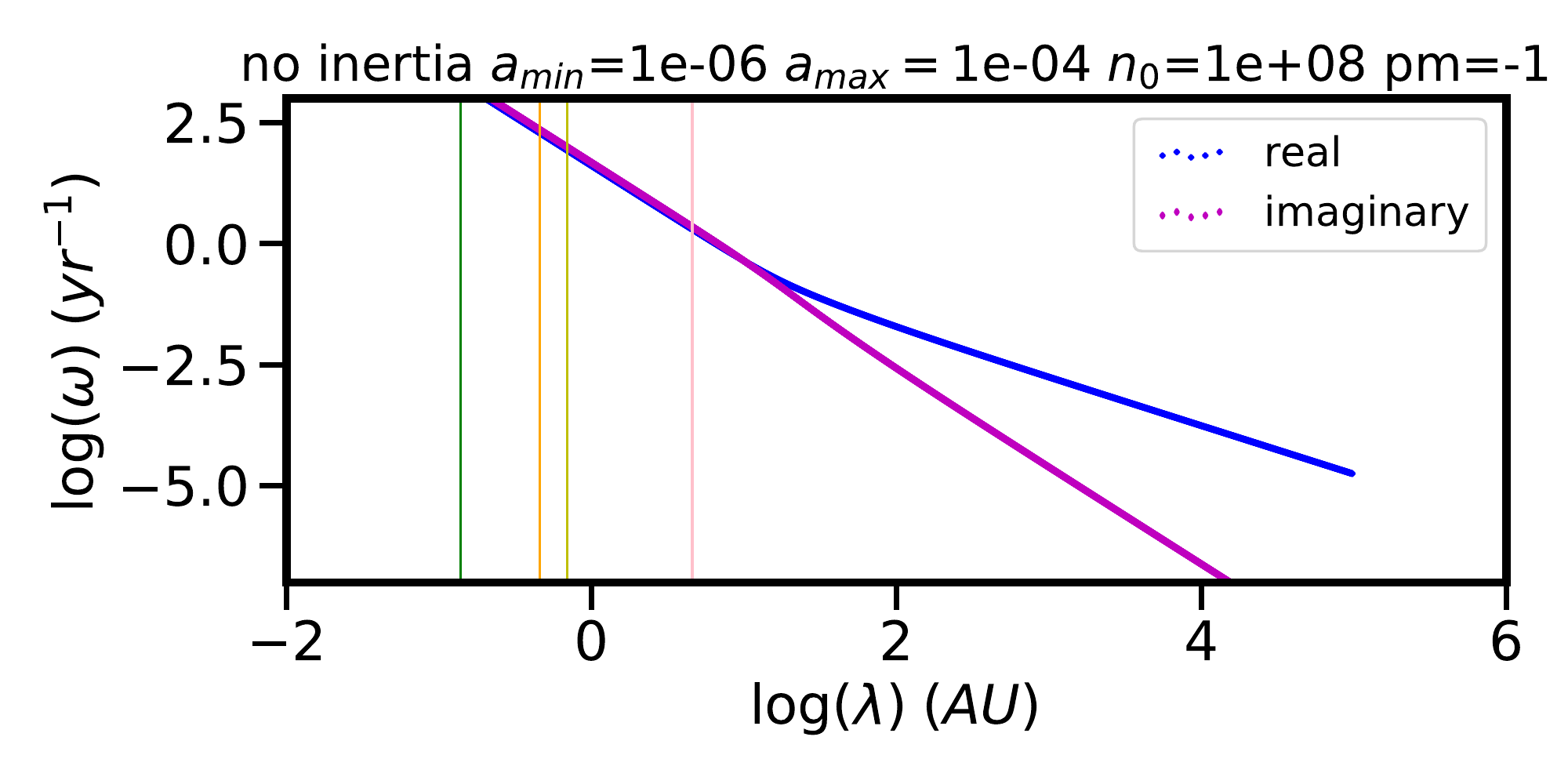}}
\end{picture}
\caption{  Same as Fig.~\ref{multn1e8}
 for $n_0=10^8$ cm$^{-3}$ and $a_{min}=1 \times 10^{-6}$ cm $a_{max}=1 \times 10^{-4}$ cm. }
\label{multn1e8}
\end{figure}

We now discuss the dependence of the waves on gas density. We recall that from Eqs.~(\ref{AWdissn_num})-(\ref{AWdissd_num}), 
we expect the non-ideal MHD behaviour to occur at wavelengths that scale as $\simeq n^{-1}$.
Figures~\ref{multn1e3} and \ref{multn1e8} portray the dispersion relation for $n_0=10^3$ cm$^{-3}$ and $n_0=10^8$ cm$^{-3}$
respectively. Note that for $n_0=10^8$ cm$^{-3}$, we have used $a_{min}=10^{-6}$ cm and  $a_{max}=10^{-4}$ cm instead 
of $a_{min}=10^{-7}$ cm and  $a_{max}=10^{-5}$ cm because it is very likely that due to the differential motions between grains induced by magnetic drift, small grains disappear 
as the collapse proceed \citep{guillet2020,lebreuilly2023}. 

The two dispersion relations appear to be similar to the one of Fig.~\ref{multn1e6} once a rescaling is performed (as suggested by Eq.~(\ref{AWdissn_num})).
For $n_0=10^3$ cm$^{-3}$, Alfv\'en waves would barely propagate for $\lambda < 3 \times 10^4$ au without grain inertia (as seen from second-top and bottom panel). 
On the contrary, first and third-top panels show that the waves do propagate except in a relatively narrow range of wavelengths. The 
"+" and "-" modes present the same characteristic than in Fig.~\ref{multn1e6}, that is to say the "+" mode has a distribution of $\omega$ and its mean value
appears to be close to the dispersion relation obtained for the "-" mode. At short wavelengths, 
the dependence of ${\cal R}_e(\omega)$ is again given by $\omega \propto \lambda^{-1.2}$. 

For  $n_0=10^8$ cm$^{-3}$, the agreement between the case with and without inertia is very good for $\lambda > 2-3$ au. 
Interestingly, since $\lambda _{AW,hall,n} > \lambda _{AW,diss,n}$, which are represented respectively by the pink and yellow lines, 
a clear whistler regime develops for 2 au $< \lambda <$ 20 au.

\subsection{Effect of grain size and ionisation}

\setlength{\unitlength}{1cm}
\begin{figure}
\begin{picture} (0,8)
\put(0,4){\includegraphics[width=8cm]{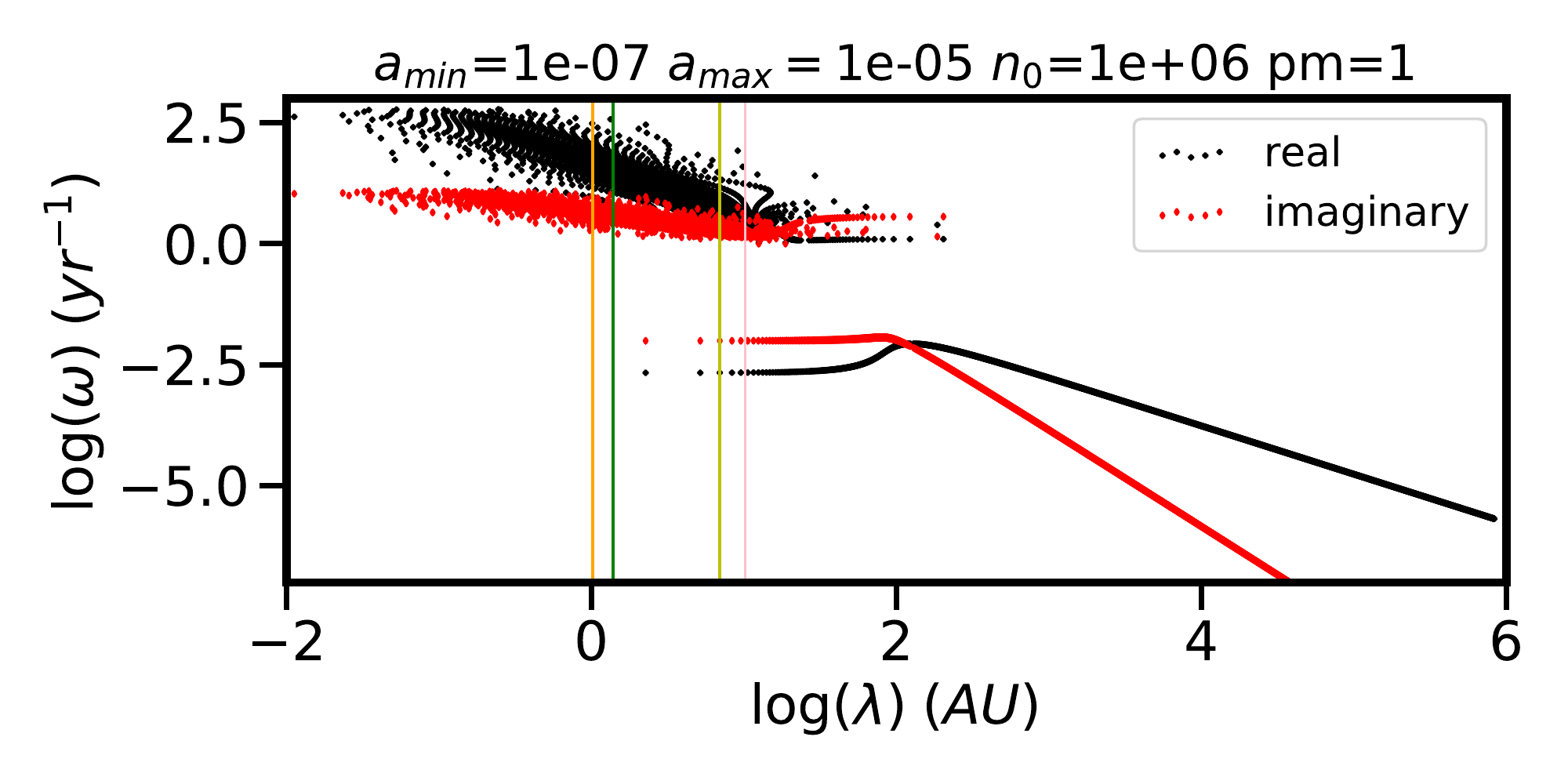}}
\put(0,0){\includegraphics[width=8cm]{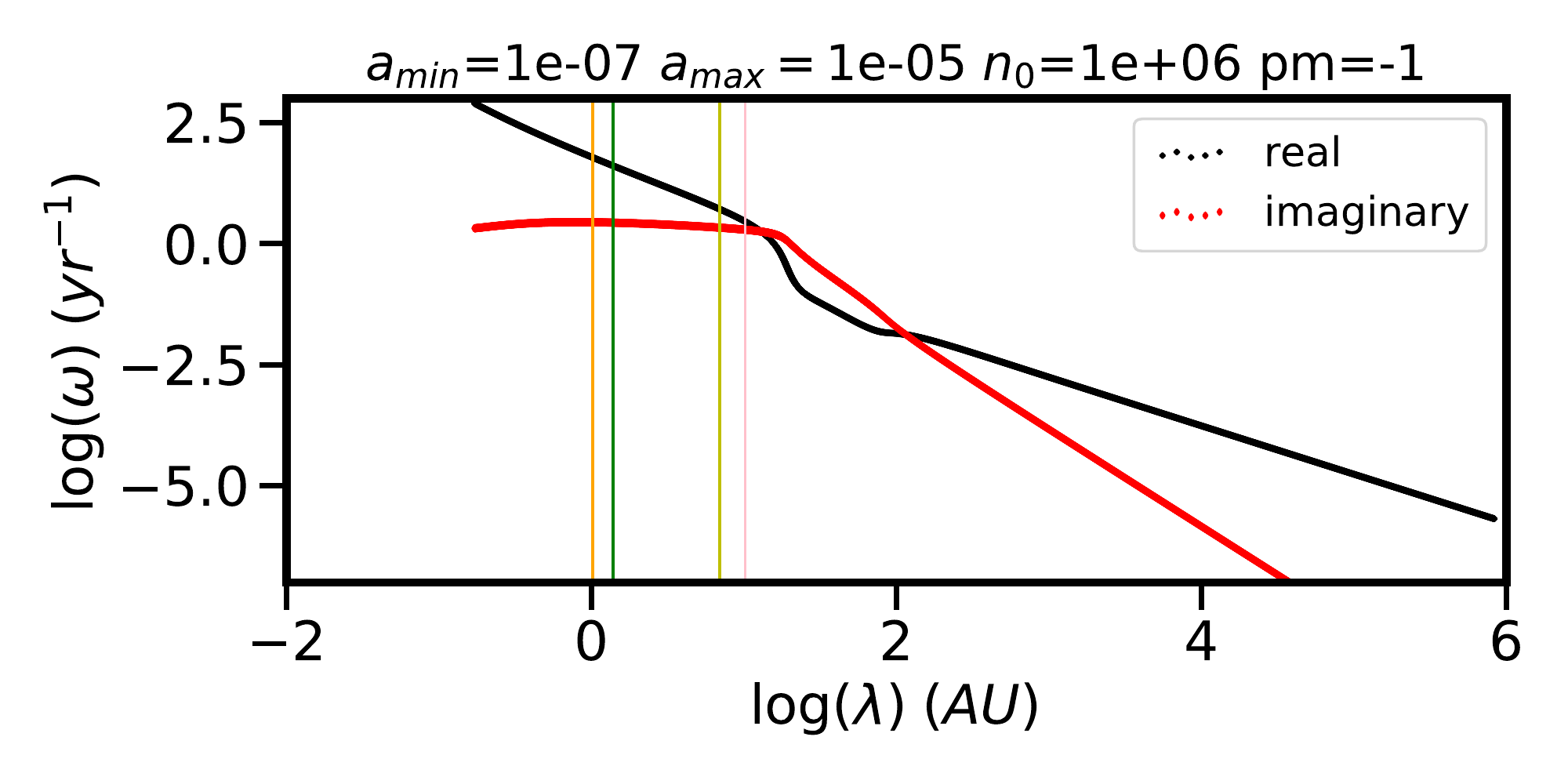}}
\end{picture}
\caption{  Wave frequency as a function of wavelengths for $n_0=10^6$ cm$^{-3}$ and $a_{min}=1 \times 10^{-7}$ cm $a_{max}=1 \times 10^{-5}$ cm as well as a low ionisation rate of $10^{-17}$ s$^{-1}$.  The yellow, green, pink and orange vertical lines represent $\lambda _{AW,diss,n}$, $\lambda _{AW,diss,d}$, $\lambda _{AW,hall,n}$, $\lambda _{AW,hall,d}$ respectively
(see Eqs.~\ref{AWdissn_num}-\ref{AWdhalld_num}). }
\label{multn1e6low}
\end{figure}

\setlength{\unitlength}{1cm}
\begin{figure}
\begin{picture} (0,8)
\put(0,4){\includegraphics[width=8cm]{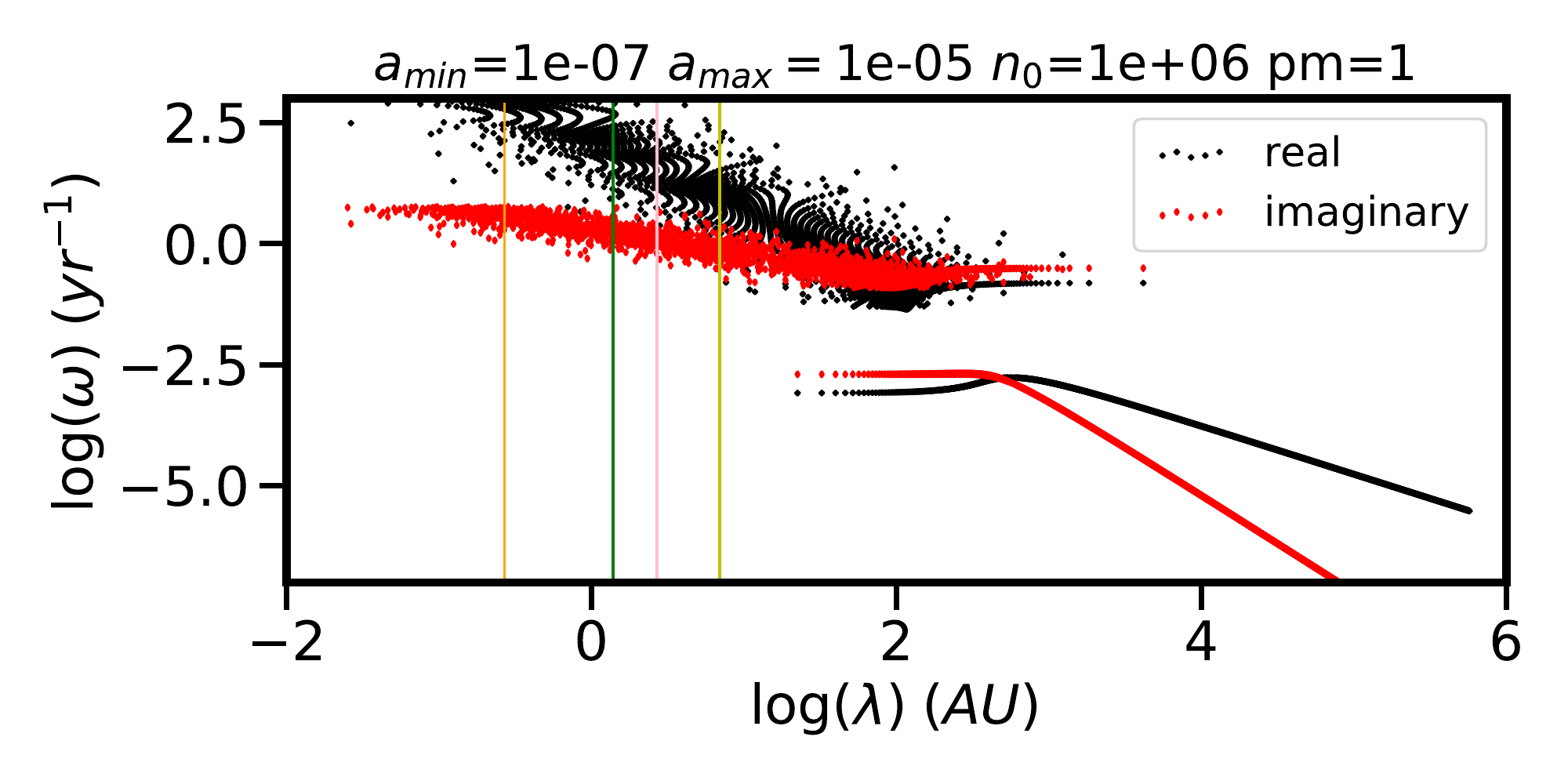}}
\put(0,0){\includegraphics[width=8cm]{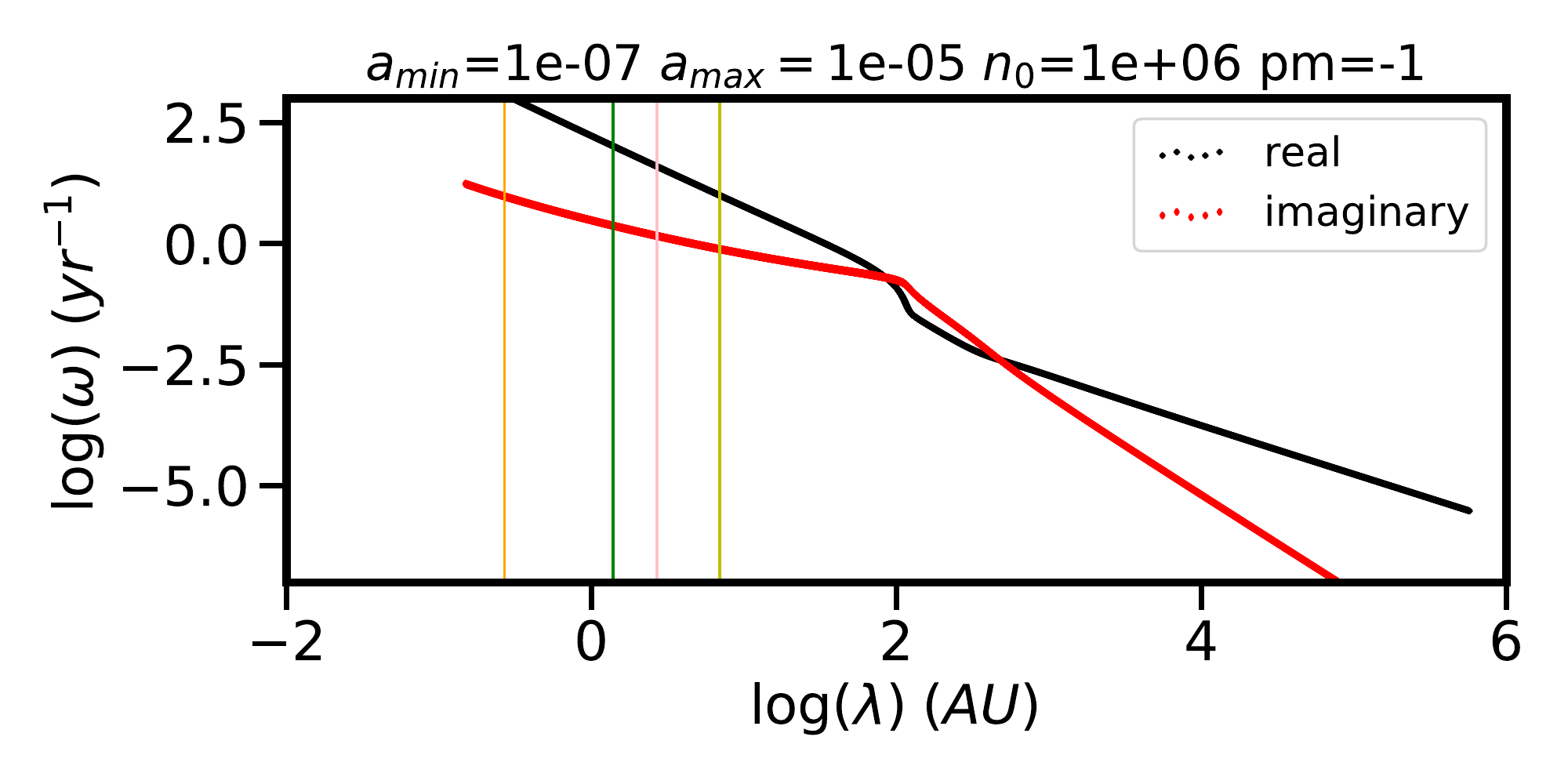}}
\end{picture}
\caption{  Same as Fig.~\ref{multn1e6low}  for $n_0=10^6$ cm$^{-3}$ and $a_{min}=1 \times 10^{-7}$ cm $a_{max}=1 \times 10^{-5}$ cm as well as with 
$\lambda_{mrn}=-2.1.$ }
\label{multn1e6mu2.1}
\end{figure} 

To investigate the influence that grain distribution and ionisation may have on the dispersion relation, we present two cases one with an ionisation rate 
5 times lower than the one used in Fig.~\ref{multn1e6} and one with  $\lambda_{mrn}=-2.1$ instead of $\lambda_{mrn}=-3.5$  usually assumed. With 
$\lambda_{mrn}=-2.1$ there are more large grains and less small grains. 

The dispersion relation seen in Fig.~\ref{multn1e6low}  is very similar to the one presented in Fig.~\ref{multn1e6}, in spite of the fact that the model 
has a lower ionisation rate. The main difference appears to be the imaginary part of $\omega$, which is inversely proportional to the dissipation rate, that 
is a factor of about 2 times higher. 

Figure~\ref{multn1e6mu2.1} presents more difference with  Fig.~\ref{multn1e6}. First, the dissipation rate is also larger than 
for $\lambda_{mrn} = -3.5$  (Fig.~\ref{multn1e6}). Second, regarding the "+" mode, the distribution of $\omega$ seen for $\lambda < 100$ au is 
significantly broader. This is expected since bigger grains have more inertia. Third, the wavelength at which classical 
Alfv\'en waves stop propagating, is about 5 times higher in Fig.~\ref{multn1e6mu2.1} than in Fig.~\ref{multn1e6}.

\subsection{Grain velocities}

\setlength{\unitlength}{1cm}
\begin{figure}
\begin{picture} (0,8)
\put(0,4){\includegraphics[width=8cm]{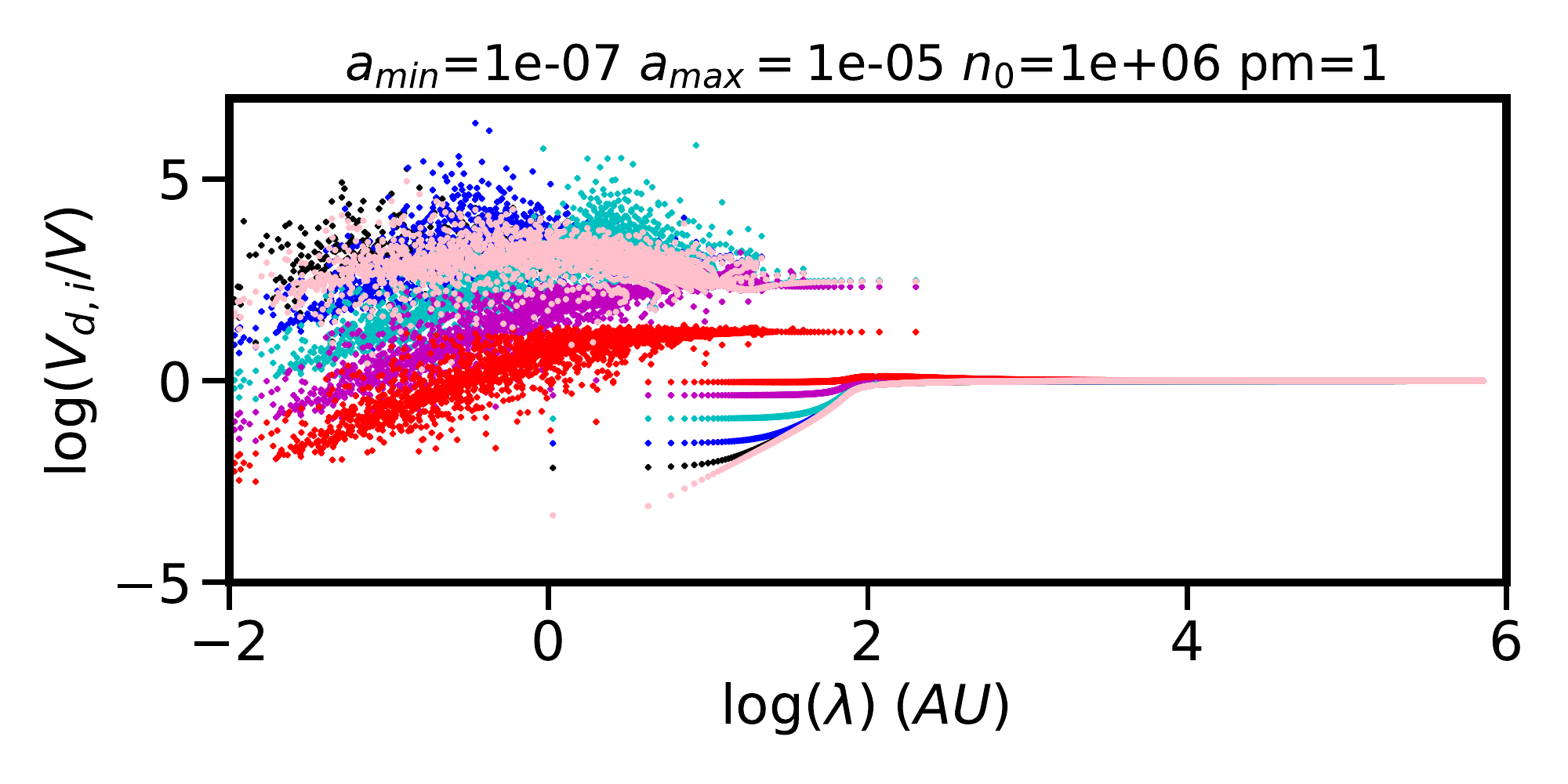}}
\put(0,0){\includegraphics[width=8cm]{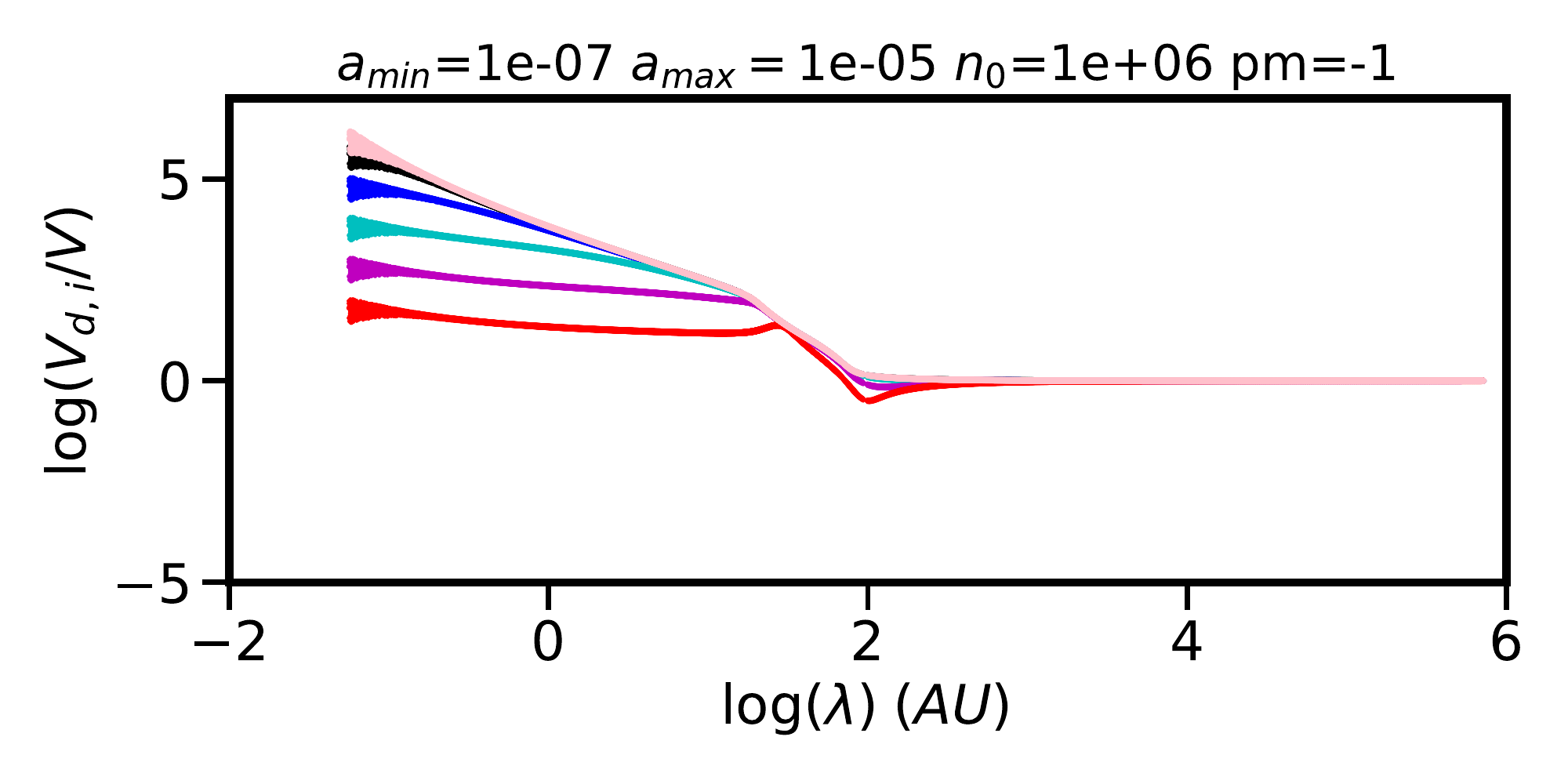}}
\end{picture}
\caption{  Velocities of ions and various grains  relative to the amplitude of the neutral velocity for $n_0=10^6$ cm$^{-3}$. The pink points correspond to the ion velocities, the dark points to the smallest grains, i.e. of size 
$a_{min}$ whereas the red points correspond to grain of size $a_{max}$. The blue, cyan and purple points correspond to grain of 
intermediate sizes evenly distributed (logarithmically) between $a_{min}$ and $a_{max}$.
   }
\label{Vd1e6}
\end{figure} 

Figure~\ref{Vd1e6} portrays the velocity range of the ions (pink points) and of the grains for various size from $a_{min}$ (dark points) to 
 $a_{max}$ (red points). The blue, cyan and purple points correspond to grains of intermediate size (in increasing order). 
 As expected at long wavelengths, the grains and the neutrals have the same velocities because they are dynamically coupled.
At short wavelengths,  the smaller the grains, the larger their velocities. As for the single size grain case, we therefore expect that 
at short wavelengths, the small grains and the ions present a velocity range that is larger than the one of the neutrals and of the 
big grains.


\section{Discussion}
Although the results obtained in this paper are restricted to Alfv\'en waves only, they represent 
nevertheless  a very fundamental mode of magnetized flows and it is worth discussing 
the effect this may have.

\subsection{Consequences for magnetized turbulence in molecular clouds}
It is sometimes speculated that the ion-neutral friction prevent the propagation of waves 
below  $\lambda_{diss,n}$  ($\simeq$0.1 pc for $n_0=10^3$ cm$^{-3}$) possibly even suppressing the turbulent cascade
at $\lambda < \lambda_{diss,n}$. 
Our results suggest 
that the waves  are likely not cut at  $\lambda_{diss,n}$ because  the grains lead to the propagation 
of Alfv\'en waves at the Alfv\'en speed of the dust, $B _0 / \sqrt{4 \pi \rho_d }$  (i.e. roughly ten times the 
Alfv\'en speed of the neutrals). Our results suggest that through the grain-neutral friction, 
the grains may transfer some momentum to the neutrals often more efficiently (depending of densities and wavelengths) that if 
the grains had no inertia, though this momentum diminishes
with wavelengths and tends toward 0 in the short wavelength limit. 
Very importantly, the Alfv\'en wave dissipation is considerably reduced when grain inertia is properly accounted for. 

Qualitatively this suggests that in molecular gas of density $\simeq 10^3$ cm$^{-3}$ at scale below $\simeq 0.1$ pc (see Fig.~\ref{multn1e3}),
the ion velocity,  could be larger than the velocity of the neutrals at least if Alfv\'en modes are sufficiently representative of motions 
in molecular clouds. Interestingly, \citet{pineda2021} have recently reported that the velocity dispersion of NH$_2^+$ are 
systematically larger by $\simeq 0.015$ km s$^{-1}$ than the one of  NH$_3$. They discuss the possibility that 
this may be due to the injection of Alfv\'en waves although they stress that the waves are quickly damped and therefore would need to 
be continuously injected. While it is likely that the motions in molecular clouds are fully turbulent, therefore implying energy cascade, it 
is likely that at scale below the dissipative scale, $\lambda_{diss,n}$, the charged dust and the ions acquire velocities larger 
than the ones of the neutrals. 

Indeed while it is now largely admitted that molecular clouds are both turbulent and magnetized
\citep[e.g.][]{HF12}, the nature of this turbulence at small scales is 
presently unclear precisely because it is expected that below 
$\simeq 0.1$ pc,  due to ambipolar diffusion, dissipation may  become important. 
The various attempts to study this turbulence have either used the so-called 
strong coupling approximation which consists in neglecting entirely the ion inertia \citep[e.g.][]{ntormousi2016} or the so-called heavy ion approximation 
\citep[e.g.][]{meyer2014,burkhart2015}, where it is assumed that the ions have a mass that is several orders 
of magnitude larger than their actual values, allowing then to limit the high Alfv\'en velocity that they would 
otherwise present. In principle, our results show that, even when the charges are carried by the ions and electrons 
(bottom panel of Fig.~\ref{ionisation}), the Alfv\'en speed  at wavelengths shorter than $\lambda_{diss,n}$, corresponds
to the one of the dust grain and not the one of the ions. Since the latter is typically three orders of magnitude larger 
than the Alfv\'en speed of the neutral, this suggests that a full modelling of molecular clouds at small scales, may be performed 
without requiring the use of heavy ion  approximation though the size distribution remains a serious difficulty.
In this respect, we note nevertheless that the single size grain and the multi-size grain cases studied in 
this paper, present strong similarities suggesting that the former may constitute a fruitful approach.

\subsection{Consequences for magnetic braking and disk formation}
\label{disk_form}
In astrophysical problems where rotation is a dominant process, magnetic field is usually an important mechanism 
through which angular momentum can be transported. This is particularly true for the formation 
of planet-forming disks where magnetic field is playing a crucial role \citep{zhao2020,maury2022}.

Magnetic braking is basically a flux of angular momentum carried away by torsional Alfv\'en waves. It operates at all 
densities as the collapse proceeds. For instance Fig.~4 of \citet{joos2012} reveals that at all times, the angular momentum 
of the gas above $10^8$ cm$^{-3}$ is significantly higher for a low magnetic field ($\mu=17$) than 
for a strong field ($\mu=2$). This implies that the magnetic braking has already occurred before the gas reaches 
this density. 

To determine whether or not grain inertia may influence magnetic braking in collapsing cores, let us estimate the scale at which magnetic braking applies.
We assume that the gas density is given by $\rho_{sis} = C_s^2 / (2 \pi G r^2) $, i.e. the density of the singular  isothermal sphere. This leads to 
$r_{sis} = C_s / ( \rho  2 \pi G )^{1/2} $. The relevant scale for magnetic braking is the thickness of the pseudo-disk, which depending of the magnetisation
has an aspect ratio of $\simeq$5-10 \citep{lishu1996,hf2008}.

From Fig.~\ref{multn1e6}, we see that at $10^6$ cm$^{-3}$, the wave propagation for wavelengths below 100 au 
is not correctly described if inertia is neglected. We also see that this scale is about 10 $\lambda_{diss,n}$ (yellow lines).
To estimate the typical density below which grain inertia may affect magnetic braking, we thus write 
$r_{sis} / 10 \simeq   10 \lambda_{diss,n}$. This leads to $n  \simeq 5 \times 10^5 \beta _{mag}^{-1} (a / 10^{-7} {\rm cm}) ^{-1}$ {\rm cm}$^{-3}$.
Since in dense cores, $\beta_{mag}$ is generally small (about 0.1 or less) and that grain growth is expected, we see  that grain inertia could 
possibly affect magnetic braking up to relevant densities during collapse. 
From the conclusions reached in Sect.~\ref{msg}, it seems possible that magnetic braking is enhanced and more 
symmetrical between the aligned and anti-aligned configurations which have been found different because 
of the Hall effect \citep{tsukamoto2015,wurster2016,leeyn2021}. Obviously this remains a rough estimate at this stage but it advocates 
for considering grain dynamics during collapse.

\subsection{Consequences for propagation of cosmic rays in the dense gas}

\setlength{\unitlength}{1cm}
\begin{figure}
\begin{picture} (0,12)
\put(0,0){\includegraphics[width=8cm]{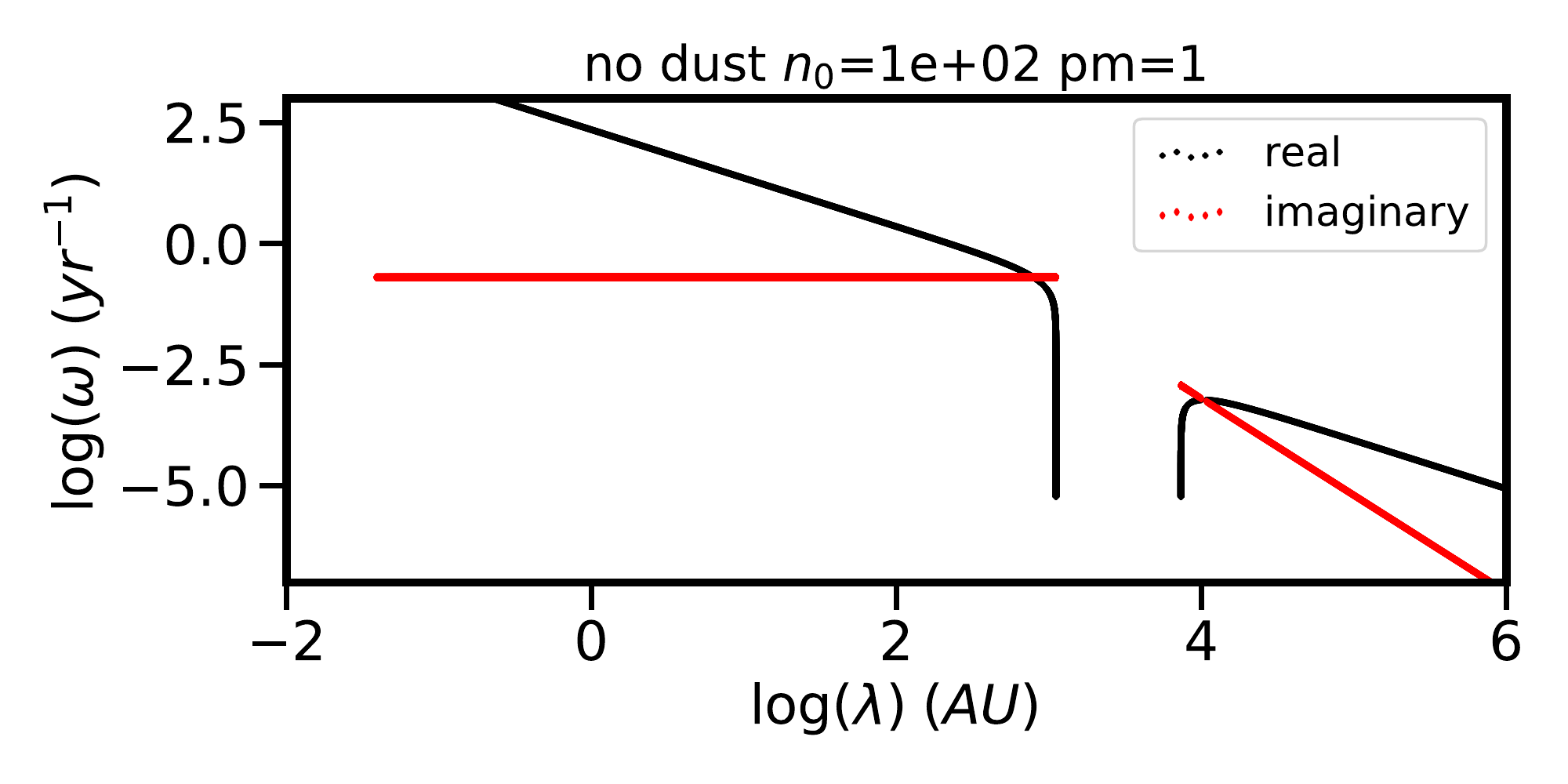}}
\put(0,4){\includegraphics[width=8cm]{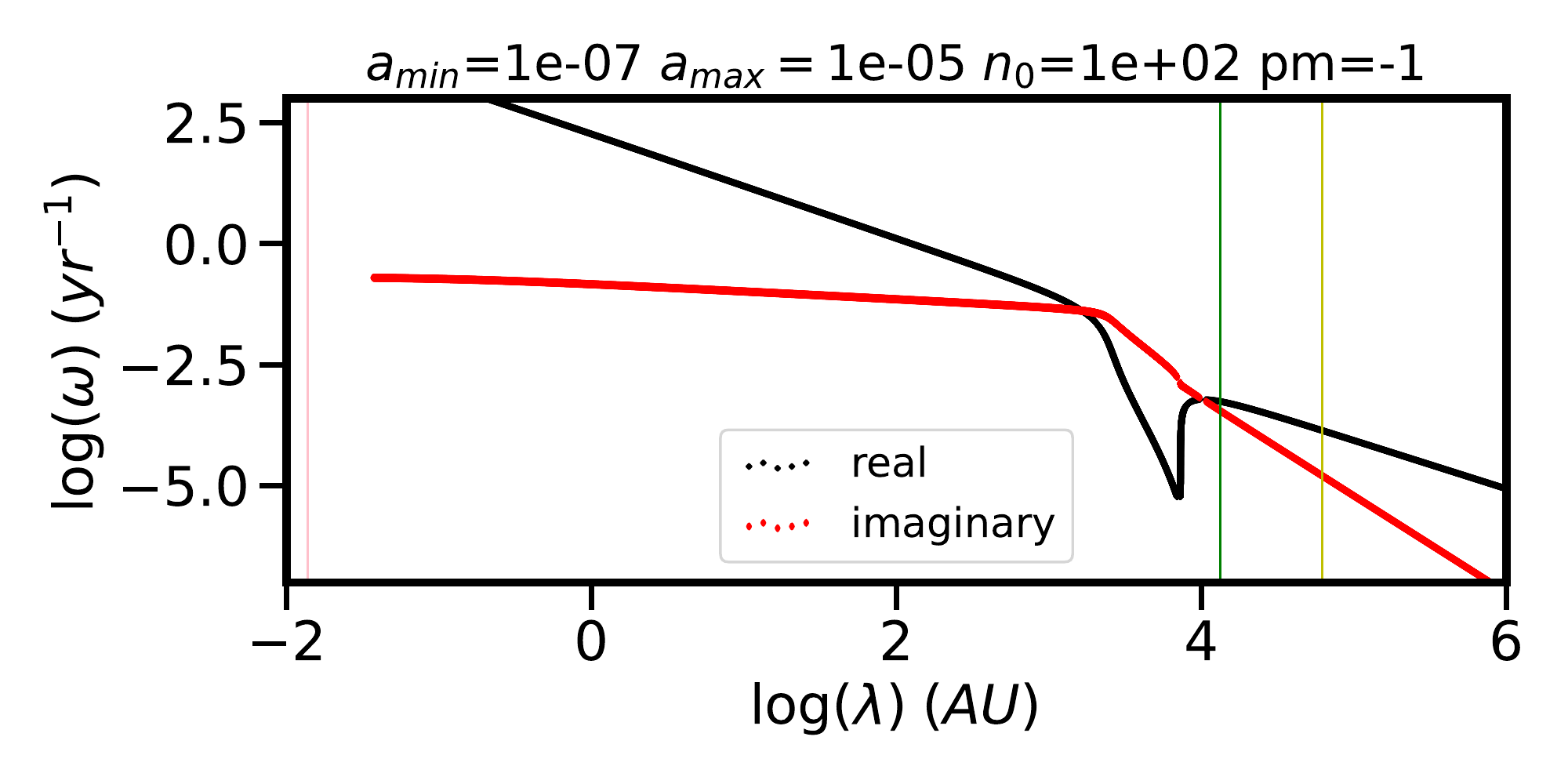}}
\put(0,8){\includegraphics[width=8cm]{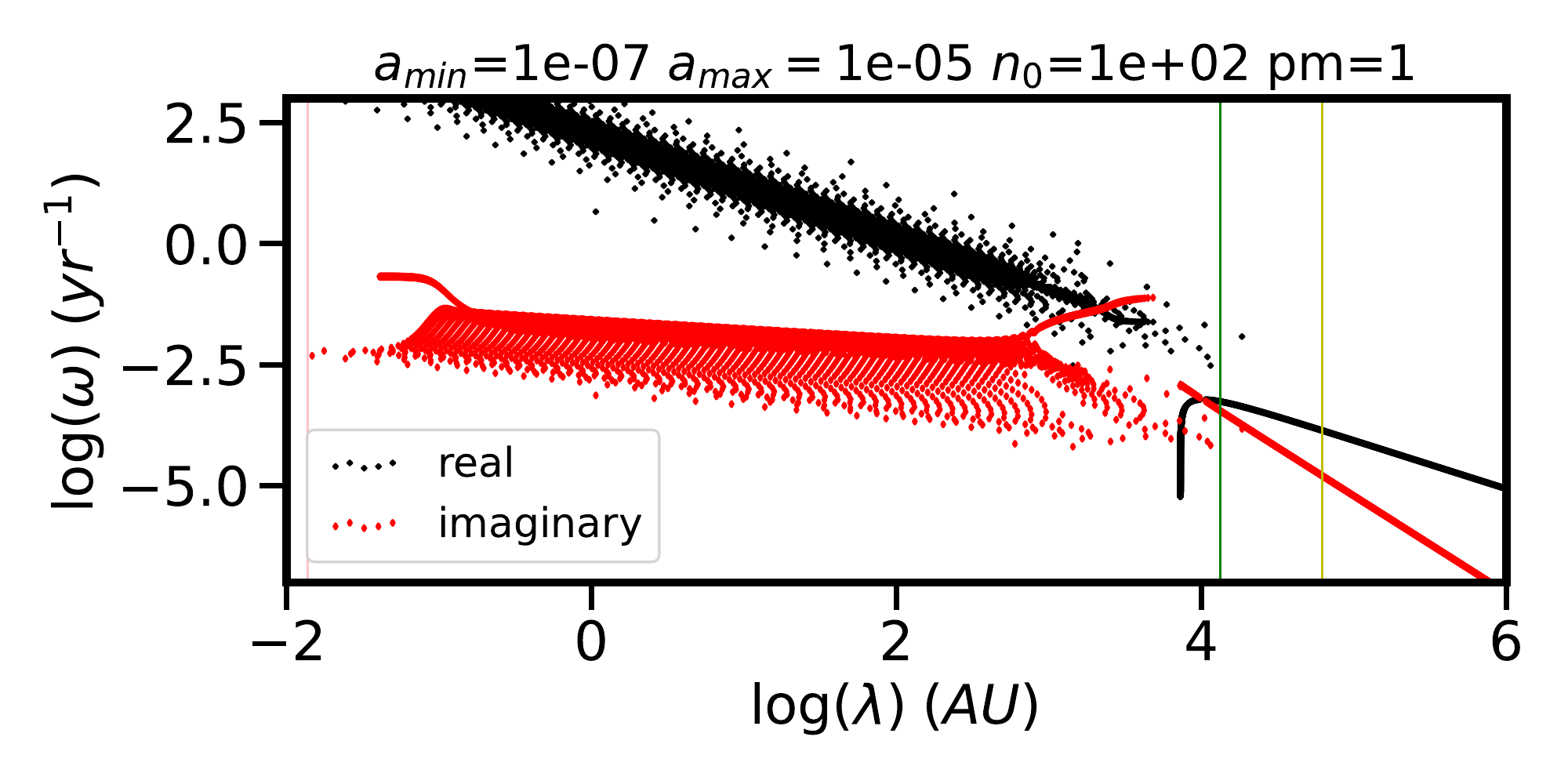}}
\end{picture}
\caption{  Wave frequency as a function of wavelengths for $n_0=10^2$ cm$^{-3}$. An ionisation of $10^{-4}$ is assumed. 
Bottom panel corresponds to the case with no dust, i.e. only ions and electrons (the "+" and "-" modes are indistinguishable). For  top and middle panels (respectively "-" and "+" modes), dust is included a standard 
MRN distribution with  $a_{min}=1 \times 10^{-7}$ cm $a_{max}=1 \times 10^{-5}$ cm is assumed.
   }
\label{n0100}
\end{figure}

 Finally, we discuss the consequences that our results may have on the propagation of cosmic rays within the 
ISM. Several recent studies  have included GeV cosmic rays in ISM and galaxy simulations \citep[e.g.][]{pfrommer2017,commercon2019,dashyan2020,girichidis2022} showing that as expected 
cosmic rays have an impact on galaxy evolution.
Cosmic rays having a bulk velocity larger than the Alfv\'en speed, excite Alfv\'en waves through gyroresonance 
and tend to be well-coupled to these waves \citep{kulsrud2005,everett2011,thomas2019,hopkins2021}. Since the streaming velocity is proportional to the Alfv\'en speed of 
the charge carriers, the influence of charged dust may change the picture significantly in the dense gas.  Indeed since the ionisation 
ranges from $10^{-4}$ to $10^{-10}$ depending of the gas density, the Alfv\'en speed would range in the absence of charged dust, from 30 to 10$^5$ the Alfv\'en speed 
within the neutral gas. Instead since the gas-to-dust ratio is about 1/100, the Alfv\'en speed is typically 10 times the Alfv\'en speed 
within the neutral gas. The other fundamental consequence is obviously the presence and the intensity of the Alfv\'en waves, which depends on 
the damping that they experiment. The latter is much reduced in the presence of charge carriers. 
The consequences that cosmic rays propagation in the dense gas may have in ISM clouds have been recently investigated by 
\citet{bustard2021}. For instance, their figures 3 and 4 illustrate how the variations of ion abundance may affect a cloud evolution. 

To illustrate further the effect that charged grains may have in this context, we have computed the dispersion relation for gas more typical of the cold HI. 
We adopt $n_0=100$ cm$^{-3}$ and an ionisation of $x_i \simeq x_e =10^{-4}$. For the  abundance of charged grains, we consider two cases. 
For the first, we adopt $n_g / n_0 = 10^{-7}$
which is close to the value given by Eq. C14 of \citet{wolfire2003}.  For the second, we assume that there is no grain, that is to say all charges are carried just 
by ions and electrons. 

Figure~\ref{n0100} shows the results. Top and middle panel display the dispersion relation for "+" and "-" modes when grains are accounted for whereas
bottom panel shows the dispersion relation when grains are ignored (the "+" and "-" modes are almost identical and only the "+" is shown).
As can be seen the dispersion relations present significant differences. In particular, the dissipation rate of the Alfv\'en waves is lower when grains are included
therefore the wave intensity excited by the cosmic ray propagation may be larger in the presence of grains. Note that the differences 
between the dispersion relations with and without grains nevertheless present similarities. This is because with an ionisation of $10^{-4}$ and given that 
the mass of the ions is about 10 $M_p$ the mass of the charged fluids differ by a factor of about 10 only. In the molecular gas where the ionisation rate 
is significantly lower, the differences would be far more pronounced.

\section{Conclusion}

In this paper we have calculated how Alfv\'en waves propagate in a weakly ionised and dusty medium like the ISM. 
We obtain the exact dispersion relation taking into account a continuous grain size distribution. For the sake of simplicity
and to interpret our results with grain size distribution, we also compute the single size grain case for which 
we obtain and discuss the wave behaviour in the short and long wavelength limit. From these asymptotic 
behaviour, we obtain simple estimates for the relevant wavelengths which separate the various propagation regimes of 
the waves. 

Importantly, we account for the grain inertia and we systematically compare the results with what is inferred 
when inertia is neglected. 

At long wavelengths, it is generally found that neglecting grain inertia is always a valid approximation. 
This remains correct up to the scale at which grain-neutral friction is so strong that the wave propagation 
is considerably reduced or even suppressed. 
At short wavelengths, the situation is however different. When grain inertia is neglected, it is found that due 
to the Hall effect, the two circularly polarised Alfv\'en waves behave differently. Whereas one has
a zero group velocity implying no wave propagation, the other presents a dispersion relation typical 
of whistler modes. When grain inertia  is accounted for and at density lower than $\simeq 10^8$ cm$^{-3}$,
 the two modes tend to be more symmetrical and  are essentially standard Alfv\'en waves propagating 
 at the Alfv\'en velocity of the grains although with ${\cal R} _e (\omega) \propto k^{\simeq 1.2}$.
 Interestingly we found that one of the two polarisation modes ("+") present a distribution of $\omega$
 rather than a single value. 

While at long wavelengths, the grains, the ions and the neutrals have nearly identical velocities, at short wavelengths, 
the ions and the small grains present velocities far higher than the neutrals.

\begin{acknowledgements}
We thank the anonymous referee for constructive and helpful comments that have improved the manuscript. 
PH acknowledge discussions with Jaime Pineda and Alexei Ivlev. 
   This research has received funding from the European Research Council
synergy grant ECOGAL (Grant : 855130).
\end{acknowledgements}

\bibliography{lars}{}
\bibliographystyle{aa} 

\appendix

\setlength{\unitlength}{1cm}
\begin{figure}
\begin{picture} (0,15)
\put(0,11.1){\includegraphics[width=8cm]{FIGURE/dustamin1e-07amax1e-05lam-3.5e+00n01e+06bet1e-01ir5.0e-17nb1e+04nbin_ge1e+03new_1.pdf}}
\put(0,7.4){\includegraphics[width=8cm]{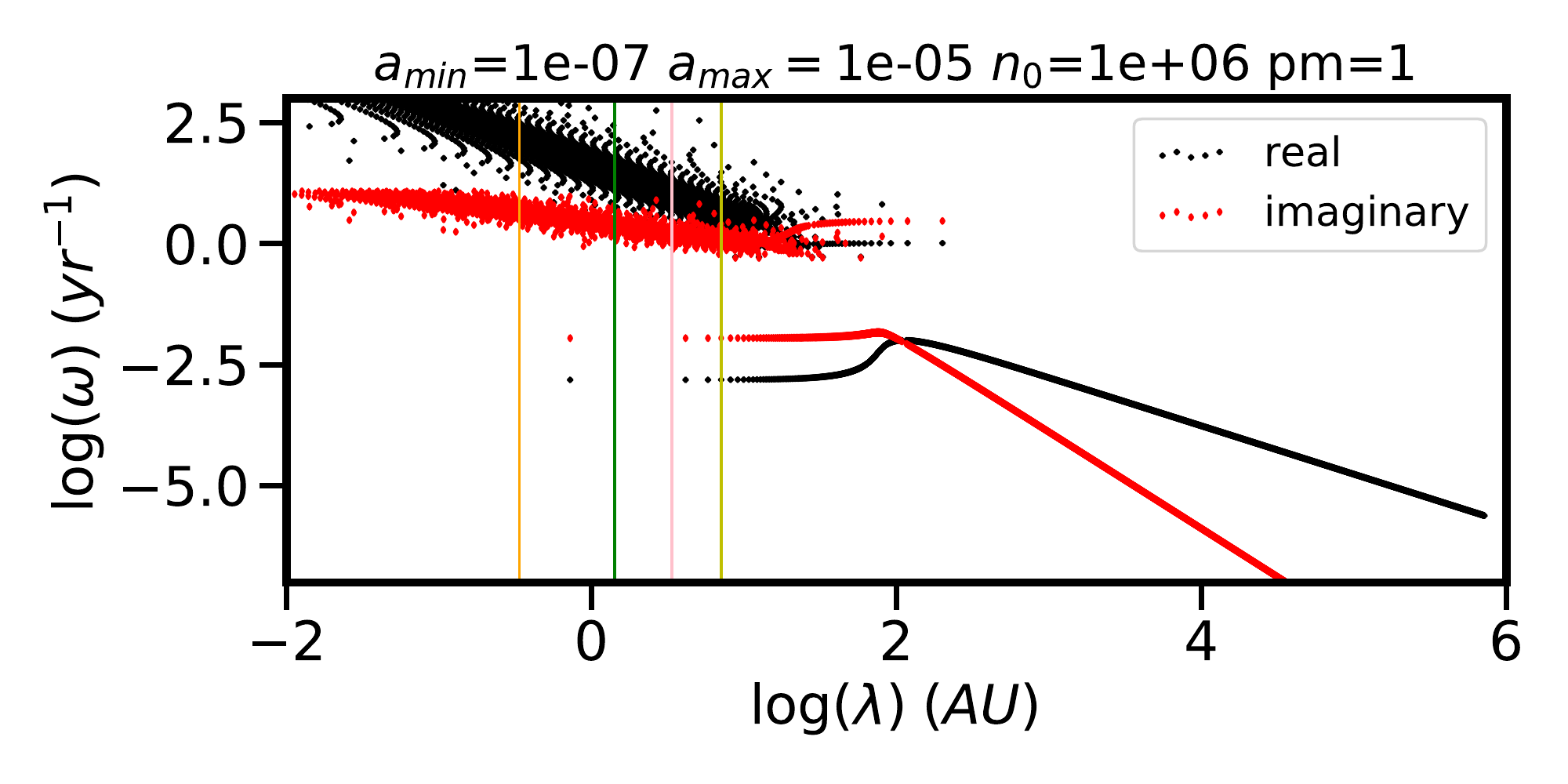}}
\put(0,3.7){\includegraphics[width=8cm]{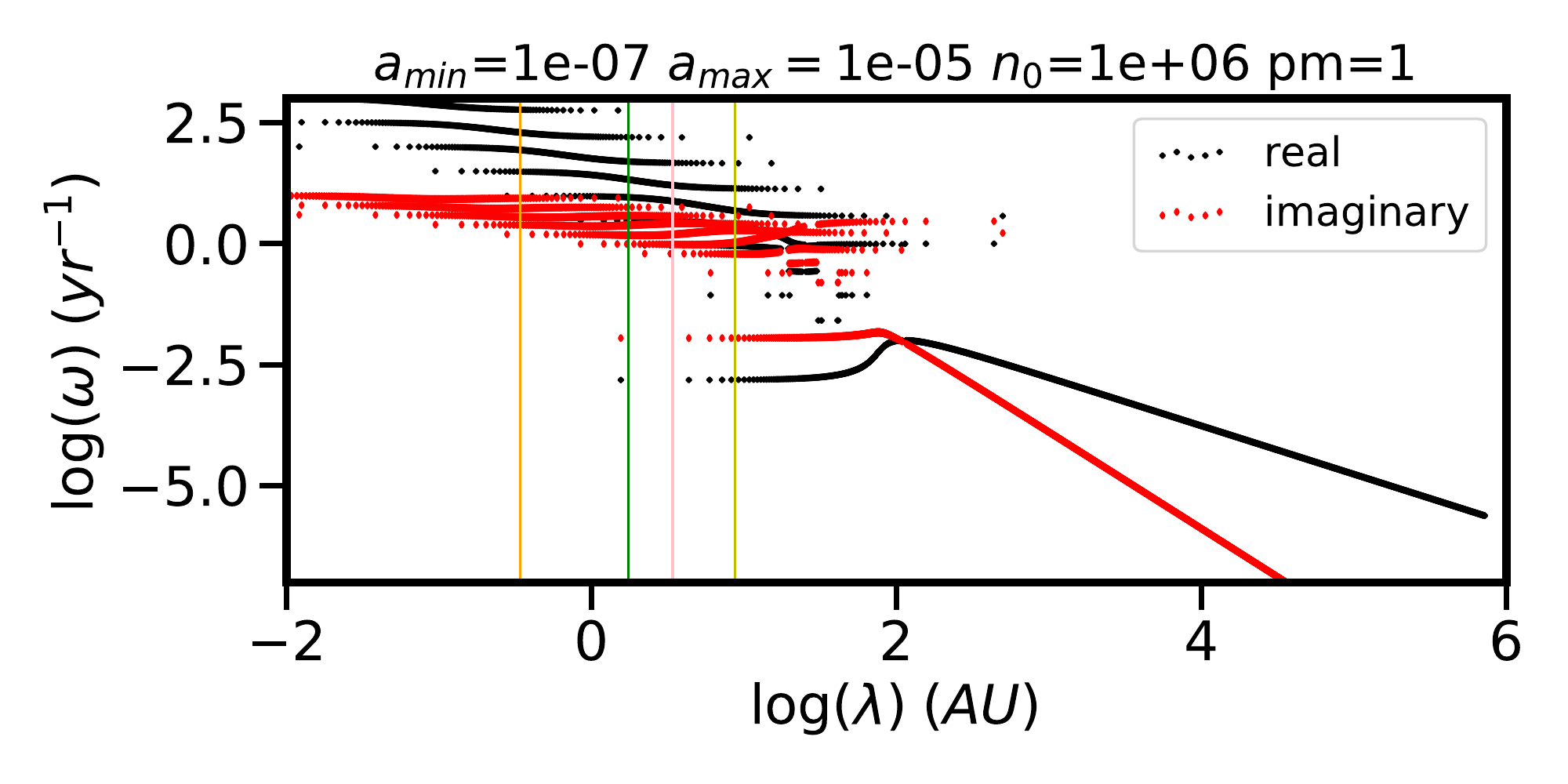}}
\put(0,0){\includegraphics[width=8cm]{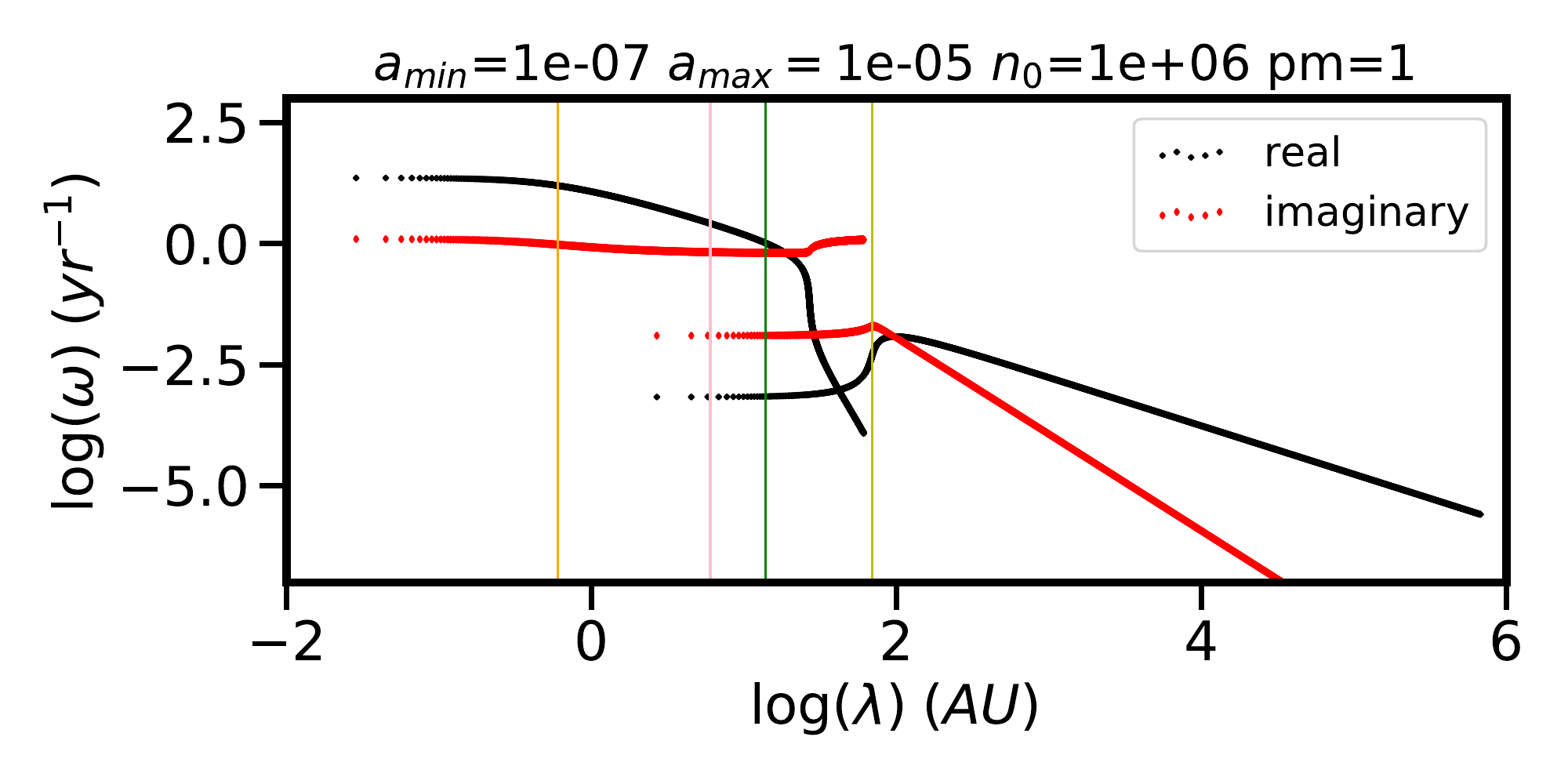}}
\end{picture}
\caption{  Wave frequency as a function of wavelengths for $n_0=10^6$ cm$^{-3}$ and $a_{min}=1 \times 10^{-7}$ cm $a_{max}=1 \times 10^{-5}$ cm.
All panels display the real and imaginary part of the frequency  for the "+" circularly polarised mode. 
Top panel uses $10^3$ bins of grains, second-top panel 100, third-top panel 10 and bottom panel only 1 bin of grains. }
\label{multn1e6_ngrain}
\end{figure} 

\section{Grain and ion properties}
\label{grainion}
Here we give a detailed description of the model and assumptions that are made in the paper.

$K_a$, the friction coefficient, is generally written as $K_a= 1/ ( (\rho+\rho_a)  \tau _{a} ) $ \citep[e.g.][]{marchand2016,lebreuilly2019}, where

\begin{eqnarray}
\tau_{a} = \sqrt{{\pi \gamma \over 8} } { \rho_g \over \rho + \rho_a} { a \over c_s}
\end{eqnarray}
where $\rho_g = M_a / (4 \pi / 3 a ^3)$ is the grain density assumed to be $\rho_g = 2.4 $g cm$^{-3}$.

\begin{eqnarray}
\tau_{a} = {3 \over 4} \sqrt{{\pi \gamma \over 8} } { M_a \over (\rho + \rho_a) c_s \pi a ^2} 
\end{eqnarray}
where $\gamma$ is the adiabatic index, $a$ is the grain radius and $M_a$ its mass.

Thus

\begin{eqnarray}
K _{a} = {4 \over 3} \sqrt{{8 \over \pi \gamma} } {  c_s \pi a^2 \over  M_a} 
\end{eqnarray}

The grain distribution is given by
\begin{eqnarray}
{dn \over da} = C a^ {\lambda _{mrn} }, 
\end{eqnarray}
where typically $\lambda _{mrn}=-3.5$.

The normalisation constant $C$ is determined through the dust density

\begin{eqnarray}
\rho_d = \int _{a_{min}} ^{a_{max}} M_a {dn \over da} da = {4 \pi \over 3} 
\rho_g
\int _{a_{min}} ^{a_{max}} C a^{\lambda _{mrn} +3} da \\
= {4 \pi \over 3} {C \rho_g \over 4 + \lambda _{mrn} } \left( a_{max}^{4+\lambda _{mrn}} -  a_{min}^{4+\lambda _{mrn}} \right), 
\nonumber
\end{eqnarray}

Leading to 

\begin{eqnarray}
C = { 3 (4 + \lambda _{mrn}) \over 4 \pi \left( a_{max}^{4+\lambda _{mrn}} -  a_{min}^{4+\lambda _{mrn}} \right) }  { \rho_a \over \rho_g} , 
\end{eqnarray}

The total number of grains is given by

\begin{eqnarray}
n_d = \int _{a_{min}} ^{a_{max}} {dn \over da} da = \int _{a_{min}} ^{a_{max}} C a^{ \lambda _{mrn} } da = {C \over 1 + \lambda _{mrn} } \left( a_{max}^{1+\lambda _{mrn}} -  a_{min}^{1+\lambda _{mrn}} \right), 
\end{eqnarray}


We consider $N_b$ logarithmic bins of grains given by
\begin{eqnarray}
a_j = a_{min} \left( { a_{max} \over a_{min} } \right) ^{j/N_b}
\end{eqnarray}

The mean size, number, density, mass  of grains in bin $j$ is

\begin{eqnarray}
n_{d,j} = \int _{a_{j}} ^{a_{j+1}} {dn \over da} da = {C \over 1 + \lambda _{mrn} } \left( a_{j+1}^{1+\lambda _{mrn}} -  a_{j}^{1+\lambda _{mrn}} \right), 
\end{eqnarray}

\begin{eqnarray}
\rho_{d,j} = \int _{a_{j}} ^{a_{j+1}} M_a {dn \over da} da 
 = {4 \pi \over 3} {C \rho_g \over 4 + \lambda _{mrn} } \left( a_{j+1}^{4+\lambda _{mrn}} -  a_{j}^{4+\lambda _{mrn}} \right), 
\end{eqnarray}


\begin{eqnarray}
<a_{j}> = {\int _{a_{j}} ^{a_{j+1}} a {dn \over da} da \over \int _{a_{j}} ^{a_{j+1}}  {dn \over da} da }
= {1 + \lambda _{mrn} \over 2 + \lambda _{mrn} } { a_{j+1}^{2+\lambda _{mrn}} -  a_{j}^{2+\lambda _{mrn}} \over
a_{j+1}^{1+\lambda _{mrn}} -  a_{j}^{1+\lambda _{mrn}} }, 
\end{eqnarray}

While the friction coefficient is given by
\begin{eqnarray}
\rho_{d,j}  K _{j} = \int _{a_{j}} ^{a_{j+1}} K_a  M_a {dn \over da} da 
 = {4 \pi c_s \over 3} \sqrt{ { 8 \over \pi \gamma} } {C  \over 3 + \lambda _{mrn} } \left( a_{j+1}^{3+\lambda _{mrn}} -  a_{j}^{3+\lambda _{mrn}} \right), 
\end{eqnarray}

For the ions we consider \citep{draine1983}

\begin{eqnarray}
\tau_{i} = {1 \over a _{i,he} } { m_i + m_{h2} \over m_{h2} n_{h2} } {1 \over  < \sigma_{coll}  w > _i},
\end{eqnarray}
and

$K_i = \gamma_{ad} = 3.5 \times 10^{13}$ g$^{-1}$ cm$^3$ s$^{-1}$.

For the electrons, we have

\begin{eqnarray}
 < \sigma_{coll}  w > _e = 10^{-15} \left( {128 k_B T \over 9 \pi m_e} \right) = 8.3 \, 10^{-10} T^{1/2} {\rm cm^3 s^{-1}} ,
\end{eqnarray}
and

\begin{eqnarray}
\gamma _e = m_n^{-1} < \sigma_{coll}  w > _e = {8.3 \, 10^{-10} \over 4 10^{-24} } T^{1/2} {\rm cm^3 s^{-1}} 
= 2 \, 10^{14}  T^{1/2} {\rm cm^3 s^{-1}}.
\nonumber
\end{eqnarray}

\section{Neglecting the ion inertia }
Here we provide the dispersion relation when the ion inertia is neglected. 
Since the ionisation is very small (typically below $10^{-7}$) in the dense ISM, the inertia of ions is negligible and thus
  we have
\begin{eqnarray}
  {\bf E} =  -{ {\bf  V_i} \over c } \times {\bf B}
 - {1 \over Z_i n_i e}  \rho \rho_i K_i ({\bf V} - {\bf V_i} ).
\label{E_mg}
\end{eqnarray}

The momentum equation for the neutrals leads to
\begin{eqnarray}
\nonumber
 \left(  i \omega  + \rho_i K_i  + \int da {d \rho_{d,a} \over d a}   K_{d,a} \right)  V^\pm  =   \rho_i K_i  V_i^\pm \\
  + \int da {d \rho_{d,a} \over d a}   K_{d,a}  V_{d,a} ^\pm ,
\label{V_mg_lin}
\end{eqnarray}

The dynamical equation of the grains gives $V_{d,a}^\pm$ as a function of $V^\pm$ and $V_i^\pm$ and can be rewritten as

\begin{eqnarray}
\label{Vd_lin_mg2}
 V_{d,a}^\pm  
 =  C_{n,a}  V ^\pm + C_{i,a}  V_i ^\pm, \\
 = 
\left( { \rho K_{d,a} - {\rho \rho_i K_i \over Z_i n_i e} {Z_{d,a}  e \over  m_{d,a} } \over i \omega + \rho K_{d,a}  \pm i {B_0 Z_{d,a} e \over c m_{d,a}}  } \right) V ^\pm 
 + \left(  {  {\rho \rho_i K_i \over Z_i n_i e} {Z_{d,a}  e \over m_{d,a} }  \pm i {B_0 Z_{d,a} e \over c m_{d,a}}  \over  i \omega + \rho K_{d,a}  \pm i {B_0 Z_{d,a} e \over c m_{d,a}}   } \right) V_i ^\pm.
\nonumber
\end{eqnarray}

With Eq.~(\ref{j_mg}), we obtain:
\begin{eqnarray}
 V_i^\pm =      \pm k  {c \over 4 \pi n_i Z_i e } b^\pm  - {1 \over n_i Z_i e } \int da {d \rho _{d,a} \over da} { Z_{d,a} e \over m_{d,a} } V_{d,a} ^\pm, 
\label{j_mg_lin}
\end{eqnarray}

Finally with Eq.~(\ref{E_mg}), we obtain 
\begin{eqnarray}
 \omega  b ^\pm = -  k B_0 V_i^\pm \pm i k { c \rho \rho_i K_i \over Z_i n_i e } (V ^\pm - V_i ^\pm).
\label{b_mg_lin}
\end{eqnarray}
which can be recast as
\begin{eqnarray}
 b ^\pm =  \left( -  {k  \over \omega } B_0 - \pm i {k \over \omega} { c \rho \rho_i K_i \over Z_i n_i e }  \right) V_i^\pm 
 \pm i {k \over \omega} { c \rho \rho_i K_i \over Z_i n_i e } V ^\pm.
\label{b_mg_lin2}
\end{eqnarray}

\begin{widetext}

With Eq.~(\ref{j_mg_lin}), we obtain 

\begin{eqnarray}
\left( 1 \pm  {k^2  \over \omega }   {c B_0 \over 4 \pi n_i Z_i e }  +  i {k^2 \over \omega} { c^2 \rho \rho_i K_i \over 4 \pi (Z_i n_i e)^2 }  \right) V_i^\pm = 
+  i {k^2 \over \omega} { c^2 \rho \rho_i K_i \over 4 \pi (Z_i n_i e)^2 } V ^\pm
  - {1 \over n_i Z_i e } \int da {d \rho _{d,a} \over da} { Z_{d,a} e \over m_{d,a} } V_{d,a} ^\pm, 
\label{ViVVp}
\end{eqnarray}

With Eq.~(\ref{Vd_lin_mg2}), Eq.~(\ref{V_mg_lin}) and Eq.~(\ref{ViVVp}) become

\begin{eqnarray}
 \left(  i \omega  + \rho_i K_i  + \int da {d \rho_{d,a} \over d a}   K_{d,a} (1 - C_{n,a}) \right)  V^\pm  =  
  \left(  \rho_i K_i  +   \int da {d \rho_{d,a} \over d a}    K_{d,a} C_{i,a}  \right)  V_i^\pm,  
\label{VVi1}
\end{eqnarray}

\begin{eqnarray}
\left( 1 \pm  {k^2  \over \omega }   {c B_0 \over 4 \pi n_i Z_i e }  +  i {k^2 \over \omega} {c^2 \rho \rho_i K_i \over 4 \pi (Z_i n_i e)^2 } 
+   {1 \over n_i Z_i e } \int da {d \rho _{d,a} \over da} { Z_{d,a} e \over m_{d,a} } C_{i,a}
 \right) V_i^\pm = 
\left(   i {k^2 \over \omega} { c^2 \rho \rho_i K_i \over 4 \pi (Z_i n_i e)^2 }    - {1 \over n_i Z_i e } \int da {d \rho _{d,a} \over da} { Z_{d,a} e \over m_{d,a} } C_{n,a}  
\right) V ^\pm,
 \label{VVi2}
\end{eqnarray}

The dispersion relation follows


\begin{eqnarray}
\nonumber
\left( 1 \pm  {k^2  \over \omega }   {c B_0 \over 4 \pi n_i Z_i e }  +  i {k^2 \over \omega} { c^2 \rho \rho_i K_i \over 4 \pi (Z_i n_i e)^2 } 
+   {1 \over n_i Z_i e } \int da {d \rho _{d,a} \over da} { Z_{d,a} e \over m_{d,a} } C_{i,a}
 \right)
 \left(  i \omega  + \rho_i K_i  + \int da {d \rho_{d,a} \over d a}   K_{d,a} (1 - C_{n,a}) \right)   =  \\
  \left(  \rho_i K_i  +   \int da {d \rho_{d,a} \over d a}    K_{d,a} C_{i,a}  \right) \left(   i {k^2 \over \omega} { c^2 \rho \rho_i K_i \over 4 \pi (Z_i n_i e)^2 }    - {1 \over n_i Z_i e } \int da {d \rho _{d,a} \over da} { Z_{d,a} e \over m_{d,a} } C_{n,a}  
\right),  
\label{rel_disp}  
\end{eqnarray}

where 

\begin{eqnarray}
\nonumber
 C_{n,a}  &=&
\left( { \rho K_{d,a} - {\rho \rho_i K_i \over Z_i n_i e} {Z_{d,a}  e \over  m_{d,a} } \over i \omega + \rho K_{d,a}  \pm i {B_0 Z_{d,a} e \over c m_{d,a}}  } \right) = { C_{n,num} \over i \omega + C_{det}} , \\
C_{i,a} &=& 
  \left(  {  {\rho \rho_i K_i \over Z_i n_i e} {Z_{d,a}  e \over m_{d,a} }  \pm i {B_0 Z_{d,a} e \over c m_{d,a}}  \over  i \omega + \rho K_{d,a} \pm i {B_0 Z_{d,a} e \over c m_{d,a}}   } \right) = { C_{i,num} \over i \omega + C_{det}}.
\label{CCnia}
\end{eqnarray}


\section{The single grain case: derivation}
\label{singlesizegrain}
The dispersion relation as stated by Eq.~(\ref{rel_disp}) is not a polynomials. 
Here we infer  the dispersion relation for a single size grain fluid for which we have the usual  polynomial form.
Although it is straightforward, the calculation is a bit 
cumbersome and some details  are given below. 

\begin{eqnarray}
\nonumber
\left( n_i Z_i e  \pm  {k^2  \over \omega }   {c B_0 \over 4 \pi}  +  
+   \rho _{a}  { Z_{a} e \over m_{a} } C_{i,a}
 \right)
 \left(  i \omega   + \rho_{a}   K_{a} (1 - C_{n,a}) \right)   =  \\
  \left(    \rho_{a}   K_{a} C_{i,a}  \right) \left( -  \rho _{a}  { Z_{a} e \over m_{a} } C_{n,a}  
\right),  
\label{rel_disp_simpl_ap}
\end{eqnarray}

\begin{eqnarray}
\nonumber
 C_{n,a}  &=&
\left( { \rho K_{a} \over i \omega + \rho K_{a}  \pm i {B_0 Z_{a} e \over c m_{a}}  } \right) = { C_{n,num} \over i \omega + C_{det}} , \\
C_{i,a} &=& 
  \left(  {   \pm i {B_0 Z_{a} e \over c m_{a}}  \over  i \omega + \rho K_{a}  \pm i {B_0 Z_{a} e \over c m_{a}}   } \right) = { C_{i,num} \over i \omega + C_{det}}.
\label{CCnia2_ap}
\end{eqnarray}

\begin{eqnarray}
\nonumber
\left( n_i Z_i e  \pm  {k^2  \over \omega }   {c B_0 \over 4 \pi}  
+   \rho _{a}  { Z_{a} e \over m_{a} } C_{i,a}
 \right)
   i \omega   + \rho_{a}   \left( n_i Z_i e  \pm  {k^2  \over \omega }   {c B_0 \over 4 \pi}  \right)
  K_{a} (1 - C_{n,a}) +   \rho _{a}^2  { Z_{a} e \over m_{a} } C_{i,a}  K_{a}  =  0
\label{rel_disp_simpl2}
\end{eqnarray}

\begin{eqnarray}
\nonumber
\left( n_i Z_i e  \pm  {k^2  \over \omega }   {c B_0 \over 4 \pi}  
+   \rho _{a}  { Z_{a} e \over m_{a} } \left(  {   \pm i {B_0 Z_{a} e \over c m_{a}}  \over  i \omega + \rho K_{a}  \pm i {B_0 Z_{a} e \over c m_{a}}   } \right)
 \right)
   i \omega   + \rho_{a}   \left( n_i Z_i e  \pm  {k^2  \over \omega }   {c B_0 \over 4 \pi}  \right)
  K_{a} \left( {  i \omega   \pm i {B_0 Z_{a} e \over c m_{a}}   \over i \omega + \rho K_{a}  \pm i {B_0 Z_{a} e \over c m_{a}} }  \right) +  \\
   \rho _{a}^2  { Z_{a} e \over m_{a} }   K_{a} \left(  {   \pm i {B_0 Z_{a} e \over c m_{a}}  \over  i \omega + \rho K_{a}  \pm i {B_0 Z_{a} e \over c m_{a}}   } \right) =  0
\label{rel_disp_simpl3}
\end{eqnarray}

\begin{eqnarray}
\nonumber
 \pm  i k^2     {c B_0 \over 4 \pi} + 
  i \omega   n_i Z_i e  
  {   i \omega + \rho K_{a} \over  i \omega + \rho K_{a}  \pm i {B_0 Z_{a} e \over c m_{a}}   } 
   + \rho_{a} K_{a}  \left(  \pm  {k^2  \over \omega }   {c B_0 \over 4 \pi}  \right)
   \left( {  i \omega   \pm i {B_0 Z_{a} e \over c m_{a}}   \over i \omega + \rho K_{a}  \pm i {B_0 Z_{a} e \over c m_{a}} }  \right) +  \\
  n_i Z_i e  \rho _{a}   K_{a} \left(  {  i \omega  \over  i \omega + \rho K_{a}  \pm i {B_0 Z_{a} e \over c m_{a}}   } \right) =  0
\label{rel_disp_simpl4}
\end{eqnarray}

\begin{eqnarray}
 \pm  i k^2     {c B_0 \over 4 \pi} + 
  i \omega   n_i Z_i e  
  {   i \omega + (\rho + \rho_{a}) K_{a} \over  i \omega + \rho  K_{a}  \pm i {B_0 Z_{a} e \over c m_{a}}   } 
   +   \left(  \pm  {k^2  \over \omega }   {c B_0 \over 4 \pi}  \right)
   \left( {  i \omega \rho_{a}  K_{a}  - \pm i {B_0 n_i Z_i e \over c }  K_{a}   \over i \omega + \rho K_{a}  \pm i {B_0 Z_{a} e \over c m_{a}} }  \right)  =  0
\label{rel_disp_simpl5}
\end{eqnarray}

\begin{eqnarray}
 \pm  i k^2    \omega {c B_0 \over 4 \pi}  
\left(  i \omega + \rho  K_{a}  \pm i {B_0 Z_{a} e \over c m_{a}}  \right)
  + 
  i \omega^2   n_i Z_i e  
  \left(   i \omega + (\rho + \rho_{a}) K_{a}  \right)
   +   \left(  \pm  k^2    {c B_0 \over 4 \pi}  \right)
   \left(   i \omega \rho_{a}  K_{a}  - \pm i {B_0 n_i Z_i e \over c }  K_{a}      \right)  =  0
\label{rel_disp_simpl6}
\end{eqnarray}

where we used the fact that $n_i Z_i = - n_{a} Z_{a}$.

\begin{eqnarray}
-    n_i Z_i e   \omega^3 
+ \left(  - \pm  k^2     {c B_0 \over 4 \pi} +
  i  n_i Z_i e (\rho + \rho_{a}) K_{a}  \right)   \omega^2  
    + \left( 
     k^2     {c B_0 \over 4 \pi} \left( \pm  i (\rho + \rho_{a})  K_{a} - {B_0 Z_{a} e \over c m_{a}} \right)  \right) \omega
     -i k ^2 {B_0^2 \over 4 \pi} n_i Z_i e K_{a} = 0 
\label{rel_disp_simpl7}
\end{eqnarray}

\begin{eqnarray}
   \omega^3
+ \left(   \pm  k^2     {c B_0 \over 4 \pi   n_i Z_i e} 
  - i  (\rho + \rho_{a}) K_{a}  \right)   \omega ^2 
    -  
     k^2   \left( \pm  i (\rho + \rho_{a})  K_{a} {c B_0 \over 4 \pi  n_i Z_i e } + {B_0^2  \over 4 \pi \rho _{a}} \right) \omega 
      + i k ^2 {B_0^2 \over 4 \pi}  K_{a} = 0
\label{rel_disp_simpl8_ap}
\end{eqnarray}

This relation can also be expressed as

\begin{eqnarray}
k^2  = {   \omega^3  - i  (\rho + \rho_{a}) K_{a} \omega^2  \over \left(  -\pm  \omega^2     {c B_0 \over 4 \pi   n_i Z_i e} + \left( \pm  i (\rho + \rho_{a})  K_{a} {c B_0 \over 4 \pi  n_i Z_i e } + {B_0^2  \over 4 \pi \rho _{a}} \right) \omega 
      - i {B_0^2 \over 4 \pi}  K_{a}   \right) }
\label{k_omeg}
\end{eqnarray}

In the limit $k \rightarrow \infty$, we see that if $\omega$
remains finite, it must satisfy a second order equation that is independent of 
$k$, namely

\begin{eqnarray}
 \left(   \pm     {c B_0 \over 4 \pi   n_i Z_i e} 
   \right)   \omega ^2 
    -  
     \left( \pm  i (\rho + \rho_{a})  K_{a} {c B_0 \over 4 \pi  n_i Z_i e } + {B_0^2  \over 4 \pi \rho _{a} } \right) \omega 
      + i  {B_0^2 \over 4 \pi}  K_{a} = 0
\label{mode_omeg}
\end{eqnarray}

\end{widetext}

\section{Influence of numerical parameters on the dispersion relation in the multi-size grain case}
\label{num_methode}

Here we investigate the influence of the number of bins of grains on our results, particularly on the "+" mode as we saw that it presents
a distribution of $\omega$ instead of a single value. 
Figure~\ref{multn1e6_ngrain} displays the dispersion relation for 1000 (top panel), 100 (second-top panel), 
10 (third-top panel) and 1 (bottom panel) bins of grains. The long wavelengths are barely affected by a change of bin number. 
However at short wavelengths, clearly the bin distribution evolves from a single value (bottom panel) to discrete bands of $\omega$
to an apparent continuum (top and second-top panels).

\end{document}